\documentclass[review,12pt]{elsarticle}

\usepackage{amssymb}
\usepackage{amsthm}
\usepackage{amsmath}

\usepackage{mathrsfs}
\usepackage{graphicx}
\usepackage{epstopdf}
\usepackage{float}
\usepackage{caption}
\usepackage{subcaption}
\usepackage{bm}
\usepackage{bbm}
\usepackage{mathrsfs}
\usepackage{hyperref}
\usepackage{cleveref}
\usepackage{soul}
\usepackage{accents}
\usepackage{graphicx}
\usepackage{xcolor}
\usepackage[margin=2cm]{geometry}
\usepackage{courier} 
\usepackage{nomencl}
\usepackage{framed}
\usepackage{listings} 
\usepackage{tabu} 
\usepackage{longtable}
\usepackage{changepage} 
\biboptions{sort&compress} 
\usepackage{url} 

\hypersetup{
    colorlinks=true,
    linkcolor=blue,
    filecolor=magenta,      
    urlcolor=cyan,
    }
    
\journal{ }

\makeatletter
\def\@author#1{\g@addto@macro\elsauthors{\normalsize%
    \def\baselinestretch{1}%
    \upshape\authorsep#1\unskip\textsuperscript{%
      \ifx\@fnmark\@empty\else\unskip\sep\@fnmark\let\sep=,\fi
      \ifx\@corref\@empty\else\unskip\sep\@corref\let\sep=,\fi
      }%
    \def\authorsep{\unskip,\space}%
    \global\let\@fnmark\@empty
    \global\let\@corref\@empty  
    \global\let\sep\@empty}%
    \@eadauthor={#1}
}
\makeatother

\begin{document}

\begin{frontmatter}



\title{A COMSOL framework for predicting hydrogen embrittlement - Part II: phase field fracture}


\author[UBU]{Andrés Díaz\corref{cor1}}
\ead{adportugal@ubu.es}

\author[UBU]{Jesús Manuel Alegre}
\author[UBU]{Isidoro Iván Cuesta}

\author{Emilio Mart\'{\i}nez-Pa\~neda\corref{cor1}\fnref{Ox,IC}}
\ead{emilio.martinez-paneda@eng.ox.ac.uk}

\address[UBU]{University of Burgos, Escuela Politécnica Superior, 09006 Burgos, Spain}
\address[Ox]{Department of Engineering Science, University of Oxford, Oxford OX1 3PJ, UK}
\cortext[cor1]{Corresponding author.}

\begin{abstract}
Prediction of hydrogen embrittlement requires a robust modelling approach and this will foster the safe adoption of hydrogen as a clean energy vector. A generalised computational model for hydrogen embrittlement is here presented, based on a phase field description of fracture. In combination with Part I of this work, which describes the process of hydrogen uptake and transport, this allows simulating a wide range of hydrogen transport and embrittlement phenomena. The material toughness is defined as a function of the hydrogen content and both elastic and elastic-plastic material behaviour are incorporated, enabling to capture both ductile and brittle fractures, and the transition from one to the other. The accumulation of hydrogen near a crack tip and subsequent embrittlement is numerically evaluated in a single-edge cracked plate, a boundary layer model and a 3D vessel case study, demonstrating the potential of the framework. Emphasis is placed on the numerical implementation, which is carried out in the finite element package COMSOL Multiphysics, and the models are made freely available.\\

\end{abstract}


\end{frontmatter}


\makenomenclature

\begin{framed}

\nomenclature{\(\Pi\)}{Potential energy}
\nomenclature{\(A\)}{Crack surface area}
\nomenclature{\(\Psi_b\)}{Bulk strain energy}
\nomenclature{\(\Psi_f\)}{Fracture energy}
\nomenclature{\(G\)}{Strain energy release rate}
\nomenclature{\(G_c\)}{Critical strain energy release rate}
\nomenclature{\(\psi_e\)}{Elastic strain energy density}
\nomenclature{\(\gamma_s\)}{Crack surface density function}
\nomenclature{\(\phi\)}{Phase field damage variable}
\nomenclature{\(\Pi_{\ell}\)}{Regularised potential energy}
\nomenclature{\(\ell\)}{Phase field length scale}
\nomenclature{\(\psi\)}{Potential energy density}
\nomenclature{\(\psi_s\)}{Stored strain energy density}
\nomenclature{\(\psi_e^0\)}{Undamaged elastic strain energy density}
\nomenclature{\(\psi_p^0\)}{Undamaged plastic strain energy density}
\nomenclature{\(g(\phi)\)}{Elastic degradation function}
\nomenclature{\(h(\phi)\)}{Plastic degradation function}
\nomenclature{\(c_{\omega}\)}{Regularisation parameter associated to the functional $\omega(\phi)$}
\nomenclature{\(\omega(\phi)\)}{Ambrosio-Tortorelli functional}
\nomenclature{\(k\)}{Numerical parameter to prevent instabilities}
\nomenclature{\(\beta_p\)}{Weight factor for the plastic contribution to damage}
\nomenclature{\(\mathcal{H}_e\)}{History variable to prevent damage reversibility considering the elastic strain energy density}
\nomenclature{\(\mathcal{H}_e^+\)}{History variable to prevent damage reversibility including only the positive part of $\psi_e^0$}
\nomenclature{\(\mathcal{H}\)}{History variable to prevent damage reversibility including elastic and plastic contributions}
\nomenclature{\(\sigma_{f0}\)}{Undamaged flow stress}
\nomenclature{\(\sigma_{y0}\)}{Undamaged initial yield stress}
\nomenclature{\(H_0\)}{Undamaged hardening modulus}
\nomenclature{\(\varepsilon_p\)}{Equivalent plastic strain}
\nomenclature{\(N\)}{Hardening exponent}
\nomenclature{\(E_0\)}{Undamaged elastic modulus}
\nomenclature{\(\nu\)}{Poisson's ratio}
\nomenclature{\(\boldsymbol{\sigma}\)}{Damaged stress tensor}
\nomenclature{\(\boldsymbol{\sigma}_0\)}{Undamaged stress tensor}
\nomenclature{\(\mathbf{C}_0\)}{Undamaged elasticity matrix}
\nomenclature{\(\pmb{\varepsilon}_e\)}{Elastic strain tensor}
\nomenclature{\(K_0\)}{Undamaged bulk modulus}
\nomenclature{\(\mu_0\)}{Undamaged shear modulus}
\nomenclature{\(|\boldsymbol{\sigma}|\)}{Equivalent stress}

\nomenclature{\(t\)}{Time}
\nomenclature{\(C\)}{Total hydrogen concentration}
\nomenclature{\(C_L\)}{Hydrogen concentration in lattice sites}
\nomenclature{\(C_T\)}{Hydrogen concentration in trapping sites}
\nomenclature{\(D_L\)}{Ideal diffusivity through lattice sites}
\nomenclature{\(R\)}{Constant of gases}
\nomenclature{\(T\)}{Temperature}
\nomenclature{\(\sigma_h\)}{Hydrostatic stress}
\nomenclature{\(\bar{V}_H\)}{Partial molar volume of hydrogen in the host metal}
\nomenclature{\(\mathbf{v}\)}{Convection velocity vector to model stress-assisted diffusion}

\nomenclature{\(D_L^{mov}\)}{Artificial diffusivity to capture hydrogen advancing through cracks}
\nomenclature{\(k_{mov}\)}{Constant to control artificial diffusivity}
\nomenclature{\(\phi_{th}\)}{Damage threshold when artificial diffusivity is activated}

\nomenclature{\(N_L\)}{Number of lattice sites per unit volume}
\nomenclature{\(N_T\)}{Number of trapping sites per unit volume (trap density)}
\nomenclature{\(K_T\)}{Equilibrium constant for trapping}
\nomenclature{\(E_B\)}{Binding energy for trapping}

\nomenclature{\(\rho\)}{Dislocation density}
\nomenclature{\(a\)}{Lattice parameter}
\nomenclature{\(\gamma\)}{Geometric parameter correlating dislocation and trap densities}

\nomenclature{\(\theta_L\)}{Occupancy of lattice sites defined as $C_L/N_L$}
\nomenclature{\(\theta_T\)}{Occupancy of trapping sites defined as $C_T/N_T$}

\nomenclature{\(G_c^0\)}{Critical strain energy release rate in the absence of hydrogen}
\nomenclature{\(f(C)\)}{Hydrogen degradation function as a function of local concentration}
\nomenclature{\(\theta_s\)}{Hydrogen coverage on the incipient crack surface}
\nomenclature{\(\chi\)}{Hydrogen degradation coefficient}
\nomenclature{\(c\)}{Hydrogen concentration expressed as an impurity factor}
\nomenclature{\(N_M\)}{Number of host metal atoms per unit volume}
\nomenclature{\(\rho_M\)}{Density of host metal}
\nomenclature{\(A_M\)}{Atomic weight of host metal}
\nomenclature{\(\beta\)}{Number of interstitial lattice sites per metal atom}
\nomenclature{\(f_{\infty}\)}{Hydrogen degradation at high concentrations}
\nomenclature{\(\xi\)}{Shape parameter for a hydrogen degradation function}

\nomenclature{\(c(\phi)\)}{Degradation function in the built-in phase field model in COMSOL}


\nomenclature{\(C_{env}\)}{Concentration of lattice hydrogen for a given environment}
\nomenclature{\(C_L^0\)}{Initial hydrogen concentration in lattice sites}

\nomenclature{\(K_c^0\)}{Reference fracture toughness corresponding to $G_c^0$}
\nomenclature{\(R_0\)}{Reference length of the fracture process zone in the absence of hydrogen}
\nomenclature{\(\Delta a\)}{Computed crack advance}

\nomenclature{\(D_{eff}\)}{Local effective diffusivity due to trapping effects}

\nomenclature{\(c_d\)}{Diffusion term in the Helmholtz equation}
\nomenclature{\(f_s\)}{Source term in the Helmholtz equation}
\nomenclature{\(a_c\)}{Absorption term in the Helmholtz equation}

\nomenclature{\(K_I\)}{Loading stress intensity factor for the boundary layer}
\nomenclature{\(R_b\)}{Radius of the remote boundary layer}
\nomenclature{\(\theta\)}{Polar coordinate of a system centred at the crack tip}

\nomenclature{\(u\)}{Displacement applied remotely}
\nomenclature{\(M_H\)}{Atomic weight of hydrogen}

\printnomenclature
\end{framed}

\section{Introduction}
\label{Introduction}

The prediction of hydrogen-assisted fracture or hydrogen embrittlement is crucial to optimise material selection and design for components that operate in the presence of hydrogen environments. The potential use of green hydrogen as a sustainable energy carrier can smooth the way to a low-carbon economy, but one of the major challenges in hydrogen transport and storage is material compatibility; many alloys, e.g. steels \cite{malheiros2022local}, nickel alloys \cite{Lu2020On725}, titanium alloys \cite{Tal-Gutelmacher2005TheAlloys} or aluminium alloys \cite{Scully2012HydrogenAlloys} suffer from a critical reduction in fracture resistance after hydrogen ingress in the bulk material. This is especially critical in high-strength steels \cite{Murakami2013HydrogenInclusions} and a trade-off between strength and susceptibility to hydrogen embrittlement is faced, limiting the possible thickness and weight reduction of components. Due to the complex phenomena involved, e.g. hydrogen uptake \cite{Martinez-Paneda2020GeneralisedTips}, diffusion and trapping \cite{Chen2020ObservationPrecipitates}, interaction with dislocations \cite{Ferreira1998HydrogenDislocations}, vacancies \cite{Tehranchi2019TheMetals} and grain boundaries \cite{Ding2022Hydrogen-enhancedTransition} or fracture energy reduction \cite{Alvaro2015HydrogenTesting}, a physically-based and robust modelling framework is required for the study of hydrogen-assisted fracture. Additionally, hydrogen generation takes place in corrosion-related phenomena \cite{Petroyiannis2004Corrosion-inducedAlloys} and other industrial processes such as welding \cite{Zafra2021FractureMicromechanisms} or plating \cite{Figueroa2008TheSteels}. \\

Prediction of hydrogen-assisted failures requires developing models capable of resolving the mechanisms of hydrogen uptake, transport and embrittlement \cite{martinez2021progress,djukic2019synergistic}. An overview of existing models to predict hydrogen uptake and transport was covered in part I of this work \cite{PartI}, along with their detailed numerical implementation. Simulating the embrittlement stage is arguably the most complicated step, due to its complex underpinning material physics, not yet fully understood \cite{chen2024hydrogen}, and because of the computational challenges associated with tracking evolving discontinuities (cracks) in a coupled chemo-mechanical (or electro-chemo-mechanical) setting. The first coupled deformation-diffusion-fracture formulations that explicitly resolved crack evolution in hydrogenated metals used the cohesive zone model \cite{Serebrinsky2004AEmbrittlement,scheider2008simulation,Olden2008ApplicationSteel,moriconi2014cohesive,delBusto2017AFatigue,jemblie2017review,yu2017cohesive}. Cohesive zone formulations enabled a rigorous treatment of hydrogen-assisted fractures, based on well-established physical variables: fracture energy and strength, including their sensitivity to hydrogen content. However, these models require prior knowledge of the crack propagation path and often lead to convergence issues \cite{yu2016viscous,martinez2017non}. The eXtended Finite Element Method (X-FEM) has also been used to model hydrogen-assisted fracture \cite{kim2021hydrogen}, yet X-FEM-based approaches struggle to capture complex cracking phenomena such as crack coalescence, particularly in 3D. Hydrogen-assisted cracking has also been predicted by means of local continuum damage mechanics models \cite{seo2022fracture,negi2024coupled}, including Gurson-like models which are based on ductile fracture processes \cite{yu2019hydrogen,depraetere2021fully}. However, these models take as input numerous fitting parameters with no physical connection to the embrittlement process and are inherently mesh-dependent, hindering their use beyond simple academic exercises. Very recently, Pinto and co-workers \cite{pinto2024simulation} developed a non-local continuum damage model for hydrogen embrittlement, which removes the mesh objectivity issues. The challenges of (simultaneously) ensuring numerical robustness, mesh-independence, and physical rigour were addressed in 2018, with the proposal of the first phase field models for hydrogen-assisted fracture \cite{Martinez-Paneda2018ACracking,Duda2018ASolids}. Phase field modelling has emerged as a powerful approach to predict fracture that is attracting growing interest due to its numerical and physical advantages: (i) the underlying principles are consistent with the thermodynamics of fracture, taking as input two well-defined physical parameters (strength and toughness, akin to cohesive zone models) \cite{Kuhn2010AFracture,kristensen2021assessment}; (ii) a scalar phase field represents the smeared crack without the need for explicit simulation of discrete surfaces \cite{bourdin2000numerical,Linse2017AFracture}; (iii) the crack trajectory does not need to be predefined a priori \cite{Francfort1998RevisitingProblem}; (iv) interactions between deformation, fracture and chemical processes, e.g. solute migration, naturally emerge in the framework of continuum mechanics and thermodynamics \cite{kristensen2020applications,kimura2021irreversible,Chen2022Phase-fieldMultiphysics,anand2019modeling}.\\  

Since the earlier developments by Mart\'{\i}nez-Pa\~neda and co-workers \cite{Martinez-Paneda2018ACracking,isfandbod2021mechanism}, phase field models for hydrogen embrittlement have been extended in multiple directions, building upon their unprecedented ease of implementation and versatility. Kristensen \textit{et al.} \cite{Kristensen2020AEmbrittlement} used strain gradient plasticity theory to incorporate the role of plastic strain gradients and geometrically necessary dislocations (GNDs) in enhancing crack tip stresses and hydrogen concentrations ahead of evolving cracks. Huang and Gao \cite{Huang2020PhaseEmbrittlement} extended the model to account for both decohesion and HELP-based degradation mechanisms. Mandal and co-workers \cite{wu2020phase,mandal2021comparative} proposed a phase field cohesive zone model (PF-CZM) for hydrogen embrittlement that enables a direct definition of the material strength. Golahmar \textit{et al.} \cite{Golahmar2022AFatigue} and Cui \textit{et al.} \cite{cui2024computational} incorporated cyclic damage, to predict the role of hydrogen on fatigue crack nucleation and fatigue crack growth. Cui \textit{et al.} \cite{Cui2022AEmbrittlement} presented a generalised stress corrosion cracking model that could capture corrosion (anodic dissolution) and hydrogen-assisted cracking, through a multi-phase-field formulation. Dinachandra and Alankar \cite{dinachandra2022adaptive} developed a phase field framework for hydrogen embrittlement that included adaptive mesh refinement. Yang \textit{et al.} \cite{yang2023phase} investigated hydrogen-assisted failures using the length-scale insensitive degradation function recently proposed by Lo \textit{et al.} \cite{lo2023phase}. Si \textit{et al.} \cite{si2024adaptive} presented a multi-patch isogeometric phase field model for hydrogen-assisted failures. Suvin \textit{et al.} \cite{suvin2024adaptive} combined phase field fracture and the adaptive scaled boundary finite element method to predict hydrogen-assisted failures while taking advantage of re-meshing algorithms. Liu \textit{et al.} \cite{liu2024hydrogen} proposed a phase field formulation for hydrogen embrittlement based on the virtual element method (VEM). Phase field methods for hydrogen embrittlement have also been used to gain complementary insight to that of experiments and to answer critical technological questions. For example, Cupertino-Malheiros \textit{et al.} \cite{cupertino2024suitability} used a coupled elastic-plastic phase field fracture model to determine the suitability of SENT testing and its optimal characteristics (e.g., test duration). Negi \textit{et al.} \cite{negi2024phase} conducted 3D phase field simulations to understand the role of hydrogen in DCB specimens undergoing sulfide stress cracking. Li and Zhang \cite{li2022analysis} analysed crack growth in CT samples of 45CrNiMoVA steel. Valverde \textit{et al.} \cite{valverde2022computational} and Grant \textit{et al.} \cite{grant2024simulating} developed a microstructurally-sensitive phase field fracture model to predict hydrogen-induced intergranular fracture. Mart\'{\i}nez-Pa\~neda \textit{et al.} used phase field to assess the suitability of slow strain rate tensile testing (SSRT) for assessing hydrogen embrittlement susceptibility. Zhang \textit{et al.} \cite{zhang2024phase} used phase field to investigate multiple crack interaction in hydrogenated metals. Zhao and Cheng \cite{zhao2024phase} used phase field to determine the conditions under which dents can lead to hydrogen-assisted failures. Mandal \textit{et al.} \cite{mandal2024computational} and  Wijnen \textit{et al.} \cite{wijnen2024computational} combined weld process modelling with coupled deformation-diffusion-fracture phase field simulations to predict the critical pressure at which hydrogen transport pipelines will fail. Thus, recent years have seen remarkable interest in phase field fracture modelling as a tool to predict hydrogen-assisted failures and multiple developments have arisen, independently, from this. The goal of this paper is to combine the main developments into a novel, generalised phase field fracture formulation and to provide a robust implementation framework into \texttt{COMSOL Multiphysics}, which is made freely available to the community. \\

We present a phase field model for hydrogen-assisted fracture that combines advanced hydrogen uptake and transport modelling, as discussed below and in Part I \cite{PartI}, with a generalised description of fracture that includes both elastic and plastic contributions and can handle brittle and ductile fractures. In Section \ref{Sec:Theory}, the governing equations and constitutive relationships are presented for both the phase field fracture and the hydrogen transport model. The focus is on phase field fracture features, such as the elastic and plastic strain energy densities, the energy split, the degradation functions employed and the treatment of damage irreversibility. Theoretical details of hydrogen transport are kept to a minimum, with the reader referred to Part I \cite{PartI} of this work. The numerical implementation of the generalised model presented is detailed in Section \ref{Sec:Numerical implementation}, considering a user-oriented philosophy to guide the reader. Then, representative case studies are presented in Section \ref{Sec:num_examples}, spanning a wide range of boundary value problems, from a single-edge cracked plate, to a boundary layer model and to a 3D high-pressure vessel. Hydrogen concentration levels, loading rates and trapping parameters are varied to assess modelling capabilities, and the numerical instability and convergence are extensively discussed. The manuscript ends with concluding remarks in Section \ref{Sec:Concluding remarks}.

\section{Theory: a coupled phase field-based formulation to predict hydrogen embrittlement}
\label{Sec:Theory}

The basic principles behind the governing equations modelling phase field evolution and hydrogen transport are briefly presented here. For the sake of brevity, the formulation of kinematics is omitted and only the local balances and constitutive expressions of interest are presented. 

\subsection{Phase field and the thermodynamics of fracture}
\label{Sec:Theory fracture}

The phase field criterion for crack propagation is in accordance with Griffith's energy balance. A new crack surface $A$ is created when the strain energy release rate, $G$, is equal to a critical value, defined as $G_c$: 
\begin{equation}
\label{Eq: Griffith's balance}
    \frac{\text{d}\Pi}{\text{d}A}=\frac{\text{d}\Psi_b}{\text{d}A}+\frac{\text{d}\Psi_f}{\text{d}A} = G-G_c= 0
\end{equation}
where the potential energy  $\Pi$ can be expressed as the sum of a bulk strain energy $\Psi_b$ and a fracture energy $\Psi_f$ \cite{Li2023AFatigue}, according to the variational principle by Francfort and Marigo \cite{Francfort1998RevisitingProblem}: 
\begin{equation}
    \Pi=\Psi_b + \Psi_f = \int_{\Omega}{\psi_s \, \text{d}V} +  \int_{\Gamma}{G_c \, \text{d}A}
\end{equation}
where $\psi_s$ is the stored strain energy density, the sum of the elastic and the fraction of the plastic energy that is not dissipated into heat. The potential energy and therefore Griffith's criterion for fracture can be regularized to a functional $\Pi_{\ell}$ considering a crack surface density functional, $\gamma_s(\phi, \nabla \phi)$.
\begin{equation}
    \Pi_{\ell}= \int_{\Omega}{ \left( \psi_s + G_c\gamma_s(\phi, \nabla \phi) \right) \, \text{d}V} 
\end{equation}
The minimization problem of $\Pi_{\ell}$ therefore governs crack initiation and propagation. The total internal energy density $\psi$ is defined as the sum of the degraded elastic and plastic strain energy densities and the dissipated crack surface energy density. The latter is regularised through the scalar phase field $\phi$ and the length scale $\ell$, such that
\begin{equation}
\label{Eq: free energy}
    \psi = g(\phi)\psi_e^0 + h(\phi)\psi_p^0 + \frac{G_c}{4c_w\ell}\left(w(\phi)+\ell^2|\nabla\phi|^2\right)
\end{equation}
where $\psi_e^0$ and $\psi_p^0$ represent the undamaged elastic and plastic strain energy densities, respectively. Two different degradation functions are here defined: $g(\phi)$ for the elastic and $h(\phi)$ for the plastic term. Alessi et al. \cite{Alessi2018ComparisonPlasticity} establish a general framework with different degradation functions for the isotropic hardening contribution and for the plastic dissipated work; however, in the present work, both terms are included in $\psi_p^0$. The contribution from the interaction between the solute atom and the host lattice is omitted from the free energy expression (\ref{Eq: free energy}) for two reasons: the lattice dilation corresponding to hydrogen in low-solubility alloys is negligible \cite{Taha2001AEmbrittlement} and the diffusion driving force is derived directly from known expressions for the stress-dependent chemical potential \cite{Li1966TheSolids}. The last term of Eq. (\ref{Eq: free energy}) recovers the fracture energy over a discontinuous crack surface when $\ell \rightarrow 0$ and $G_c$ is the critical energy release rate, also referred to as the material toughness. A critical strength can be defined from $G_c$, Young's modulus and $\ell$, providing a physical meaning to the length scale and a strategy to obtain experimental-numerical correlations. The function $w(\phi)$ and the associated parameter $c_w$ are derived from the Ambrosio-Tortorelli functional choice (AT1 or AT2) \cite{Ambrosio1990ApproximationT-convergence, Alessi2018ComparisonPlasticity}. In the present study the AT2 model is chosen, therefore $w(\phi)=\phi^2$ and $c_w=1/2$. This results in a material strength equal to \cite{kristensen2021assessment}:
\begin{equation}\label{eq:sigmaC}
\sigma_c = \frac{9}{16} \sqrt{\frac{E G_c}{3 \ell}}
\end{equation}

Damage is captured by the phase field scalar $\phi$ where $\phi = 0$ represents undamaged condition and $\phi=1$ totally broken material. To reproduce the consequent loss of stiffness, the degradation function $g(\phi)$ must comply with the conditions $g(0)=1$ and $g(1)=0$. The degradation function for the elastic contribution and stress degradation is taken as:
\begin{equation}
\label{Eq: g degrad}
    g(\phi) = (1-\phi)^2+k
\end{equation}
where $k$ is a parameter to prevent numerical instabilities due to zero stiffness when $\phi \to 1$. In the present study: $k=10^{-6}$. The plastic degradation function $h(\phi)$ is assumed to follow the same quadratic function only when the contribution is not weighted ($\beta_p=1$). However, when the dissipation of plastic work as heat is considered, i.e. $\beta_p<1$, the plastic degradation function is modified for consistency:
\begin{equation}
\label{Eq: g degrad Bp}
    h(\phi)=\beta_p (1-\phi)^2 +1-\beta_p
\end{equation}

From the principle of virtual work, a local force balance can be established for the phase field problem \cite{Kristensen2020AEmbrittlement}:
\begin{equation}
    \nabla \cdot \frac{\partial \psi}{\partial \nabla \phi} -\frac{\partial \psi}{\partial \phi} = 0
\end{equation}

Deriving hence the free energy density expression (\ref{Eq: free energy}), a yield surface for damage conditions \cite{Alessi2018ComparisonPlasticity} can be generalised as follows:
\begin{equation}
    g'(\phi)\psi_e^0+h'(\phi)\psi_p^0 + \frac{G_c}{2\ell c_w}\left( \frac{w'(\phi)}{2}-\ell^2\nabla^2 \phi\right)=0
\end{equation}
where $(\cdot)' = \partial (\cdot)/\partial \phi$. All the chosen functions are then derived and grouped into a Helmholtz's equation form so that the damage governing equation reads:
\begin{equation}
    -G_c\ell\nabla^2 \phi+ \left( 2(\psi_e^0+\beta_p\psi_p^0) +\frac{G_c}{\ell} \right) \phi= 2(\psi_e^0+\beta_p\psi_p^0)
\end{equation}

In the present implementation, the governing equation is rearranged to obtain non-dimensional terms:
\begin{equation}
    -\ell^2\nabla^2 \phi+ \left( \frac{2\ell(\psi_e^0+\beta_p\psi_p^0)}{G_c} + 1 \right) \phi= \frac{2\ell(\psi_e^0+\beta_p\psi_p^0)}{G_c}
\end{equation}

The weight factor for the plastic contribution, $\beta_p$, emerges here from the choice of the plastic degradation function $h(\phi)$ and its derivation. Some authors have also introduced a plastic threshold, $W_0$, with the aim of controlling the initiation of plastic-enhanced damage \cite{Borden2016AEffects}, substituting the term $\psi_p^0$ by $\langle\psi_p^0-W_0\rangle=\max(0,\psi_p^0-W_0)$. However, this threshold is not here analysed.\\

Two isotropic hardening laws are considered: linear and power-law hardening. When linear hardening is assumed, the evolution of the undamaged flow stress $\sigma_{f0}$ reads
\begin{equation}
\label{Eq: linear hardening law}
   \sigma_{f0} = \sigma_{y0}+H_0\varepsilon_p
\end{equation}
where $\varepsilon_p$ represents the equivalent plastic strain, $\sigma_{y0}$ the initial and undamaged yield stress and $H_0$ the hardening modulus. It must be highlighted that, in order to define a consistent nomenclature, undamaged strain energy densities are expressed with a 0 superscript whereas a 0 subscript is added for undamaged stress values or elastic-plastic moduli. The plastic strain energy density for the linear isotropic hardening case is: 
\begin{equation}
\label{Eq: psi_p^0 linear hardening}
   \psi_p^0 = \sigma_{y0}\varepsilon_p+\frac{1}{2}H_0\varepsilon_p^2
\end{equation}

The power-law hardening relationship is implemented following the Swift hardening model, which is already in-built into COMSOL:
\begin{equation}
\label{Eq: power law hardening law}
    \sigma_{f0} = \sigma_{y0} \left( 1+ \frac{\varepsilon_p}{\varepsilon_0}\right)^{N}
\end{equation}
where $\varepsilon_0$ is the yield strain, here fixed as $\sigma_{y0}/E_0$, and $N$ is the strain hardening exponent. In this case, the plastic strain energy density is given by
\begin{equation}
\label{Eq: psi_p^0 power law}
   \psi_p^0 = \frac{\sigma_{y0}\varepsilon_0}{1+N}\left[\left(1+\frac{\varepsilon_p}{\varepsilon_0}\right)^{1+N}-1\right]
\end{equation}

The phase field fracture formulation is often altered to ensure that damage only nucleates in tensile regions, as common in metals, and to enforce damage irreversibility. This is here achieved following the so-called hybrid approach by Ambati \textit{et al.} \cite{Ambati2015AFormulation}, whereby the tensile part of the elastic strain energy density $\psi_e^{0+}$ is assumed to drive fracture but the total strain energy density is used in the balance of linear momentum, retaining the linearity of the problem. Such that, using a history variable ($\mathcal{H}$) to ensure damage irreversibility,
\begin{equation}
    \boldsymbol{\sigma} := g(\phi)\frac{\partial \psi_e^0}{\partial \pmb{\varepsilon}_e} 
\end{equation}
\begin{equation}
    \mathcal{H}_e^+(t) := \max\limits_{\tau \in [0,t]}\psi_e^{0+}(\tau)
\end{equation}

\noindent where $\boldsymbol{\sigma}$ is the stress tensor, $\bm{\varepsilon}_e$ is the elastic strain tensor, and $t$ denotes time. Therefore, the decomposition of the strain tensor is not necessary and one just has to degrade the elasticity matrix to find the damaged stress. The relationship between damaged and undamaged (or effective) stress values can also be expressed as follows:
\begin{equation}
\label{Eq: stress elast matrix}
    \boldsymbol{\sigma} := g(\phi)\frac{\partial \psi_e^0}{\partial \pmb{\varepsilon}_e} = g(\phi)\boldsymbol{\sigma}_0 = g(\phi) \mathbf{C}_0: \pmb{\varepsilon}_e 
\end{equation}
where $\mathbf{C}_0$ is the undamaged elasticity matrix. Since every component of that matrix is proportional to Young's modulus, the substitution of $E_0$ by $g(\phi)E_0$ imposes the stress degradation shown in Eq. (\ref{Eq: stress elast matrix}). It remains to define $\psi_e^0$, for which we follow the volumetric-deviatoric split by Amor \textit{et al.} \cite{Amor2009RegularizedExperiments}. Hence, for an undamaged bulk modulus $K_0$ and an undamaged shear modulus $\mu_0$, the tensile and compressive parts of the elastic strain energy density read,
\begin{equation}
    \psi_e^{0+} = \frac{1}{2}K_0 \langle \text{tr}(\pmb{\varepsilon}_e) \rangle_+^2 + \mu_0 (\pmb{\varepsilon}_e':\pmb{\varepsilon}_e')
\end{equation}
\begin{equation}
    \psi_e^{0-} = \frac{1}{2}K_0 \langle \text{tr}(\pmb{\varepsilon}_e) \rangle_-^2 
\end{equation}
where $\langle x \rangle_{\pm} = (x\pm|x|)/2$, and $\pmb{\varepsilon}_e'=\pmb{\varepsilon}_e-\text{tr}(\pmb{\varepsilon}_e)\mathbf{I}/3$. Since $\psi_p^0$ is expected to monotonically increase, the phase field driving force $\mathcal{H}$ is here defined and implemented as:
\begin{equation}\label{Eq:epDrivingForce}
    \mathcal{H}(t) =\mathcal{H}_e^+(t) + \beta_p\psi_p^0(t)
\end{equation}

Therefore, the phase field governing equation in a non-dimensional Helmholtz's form reads:
\begin{equation}
\label{Eq: Helmholtz governing}
    -\ell^2\nabla^2 \phi+ \left[ \frac{2\ell\mathcal{H}}{G_c} + 1 \right]\phi= \frac{2\ell\mathcal{H}}{G_c}
\end{equation}

Finally, the yielding criterion is established using the damaged equivalent stress and thus the undamaged flow stress $\sigma_{f0}$ is multiplied by the plastic degradation function $h(\phi)$:
\begin{equation}
\label{Eq: Yield criterion}
    |\boldsymbol{\sigma}|-h(\phi)\sigma_{f0}= |g(\phi)\boldsymbol{\sigma}_0|-h(\phi)\sigma_{f0}
    =0
\end{equation}

\begin{figure}[H]
  \makebox[\textwidth][c]{\includegraphics[width=0.8\textwidth]{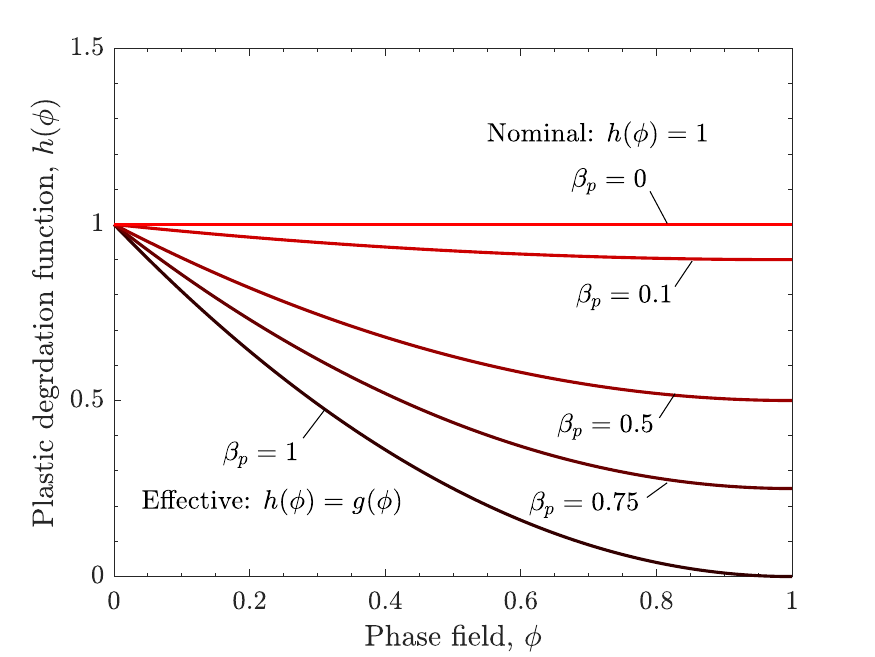}}%
  \caption{Comparison of different degradation functions.}
  \label{Fig: plastic degradation}
\end{figure}

The choice of $\beta_p$, as shown in Fig. \ref{Fig: plastic degradation}, determines not only the plastic contribution to fracture, but also the plastic flow behaviour after damage onset. For $\beta_p = 0$, $h(\phi) = 1$ and the nominal (undamaged) flow (yield) stress is considered in the yielding criterion.  As discussed in Ref. \cite{Marengo2023AFracture}, in the nominal approach, also called weak plasticity-damage coupling, plastic flow stops after the onset of damage, as $\bm{\sigma}$ decreases but $\sigma_{f0}$ remains constant, and some degree of elastic unloading is observed. In contrast, $\beta_p = 1$ results in a strong coupling and continuous plastic flow after damage onset because the yield stress $\sigma_{f0}$ is reduced according to $h(\phi) = g(\phi)$. This effective approach circumvents potentially non-physical elastic unloading phenomena but produces plastic localisation and mesh dependency. In this work, the choice $\beta_p=0.1$ is favoured, as it provides a rigorous, thermodynamically-consistent description of the energy balance during the fracture process. As demonstrated in the seminal work by Taylor and Quinney \cite{taylor1934latent}, the amount of plastic work that is stored in the material (and hence available to contribute to the fracture process) is about 10\% under quasi-static loading, with the remaining being dissipated into heat. As shown in Fig. \ref{Fig: plastic degradation}, the $\beta_p = 0.1$ choice delivers a plastic degradation function that is close to the nominal one, implying that plastic flow is hindered in damaged regions. In the context of hydrogen-assisted failures, one could encompass localised plasticity mechanisms such as HELP by making $\beta_p$ dependent on the concentration of hydrogen, increasing the role of plasticity in the fracture process.

\subsection{Stress-assisted hydrogen diffusion}
\label{Sec:theory hydrogen transport}

A two-level modelling approach is considered for hydrogen transport following the seminal works of Sofronis and McMeeking \cite{Sofronis1989NumericalTip} and Krom \textit{et al.} \cite{Krom1999HydrogenTip}. Thus, hydrogen total concentration is split into two terms: $C_L$ for hydrogen atoms in lattice sites and $C_T$ for hydrogen atoms trapped in defects such as carbides, dislocations and grain boundaries. The total hydrogen content at each material point then reads $C=C_L+C_T$. The role of trapping in delaying diffusion and influencing crack growth is investigated in Section \ref{Sec: R-curves trapping effects}. Modelling insights for other phenomena involved in hydrogen transport, such as hydrogen transport by dislocations, electrochemical adsorption/absorption, kinetic modelling of trapping, and hydrogen-induced softening, are discussed and implemented in the first part of this work \cite{PartI}; these can be straightforwardly incorporated into the present deformation-diffusion-fracture framework.\\

The flux term in the mass balance equation is modified when taking into account the hydrostatic stress influence on the chemical potential \cite{Sofronis1989NumericalTip}:
\begin{equation}
\label{Eq. governing eq}
    \frac{\partial C_L}{\partial t}+\nabla \cdot \left(-D_L\nabla C_L+\frac{D_L \Bar{V}_H}{RT} C_L\nabla \sigma_h \right) = 0
\end{equation}
where $D_L$ represents the hydrogen diffusion coefficient through the ideal lattice, $\bar{V}_H$ is the partial molar volume of hydrogen in the host metal, $T$ the temperature and $R$ the constant of gases. The hydrostatic stress is defined from the first stress tensor invariant: $\sigma_h = \text{tr}(\boldsymbol{\sigma})/3$, considering the damaged value for hydrogen drifted diffusion as in Ref. \cite{Cui2022AEmbrittlement}.
For implementation purposes, the drifted diffusion is modelled through a convection velocity, $\textbf{v}$, within a conservative convective term:
\begin{equation}
\label{Eq. governing convec}
    \frac{\partial C_L}{\partial t}+\nabla \cdot \left(-D_L\nabla C_L+\textbf{v} C_L \right) = 0
\end{equation}

\begin{figure}[H]
  \makebox[\textwidth][c]{\includegraphics[width=0.8\textwidth]{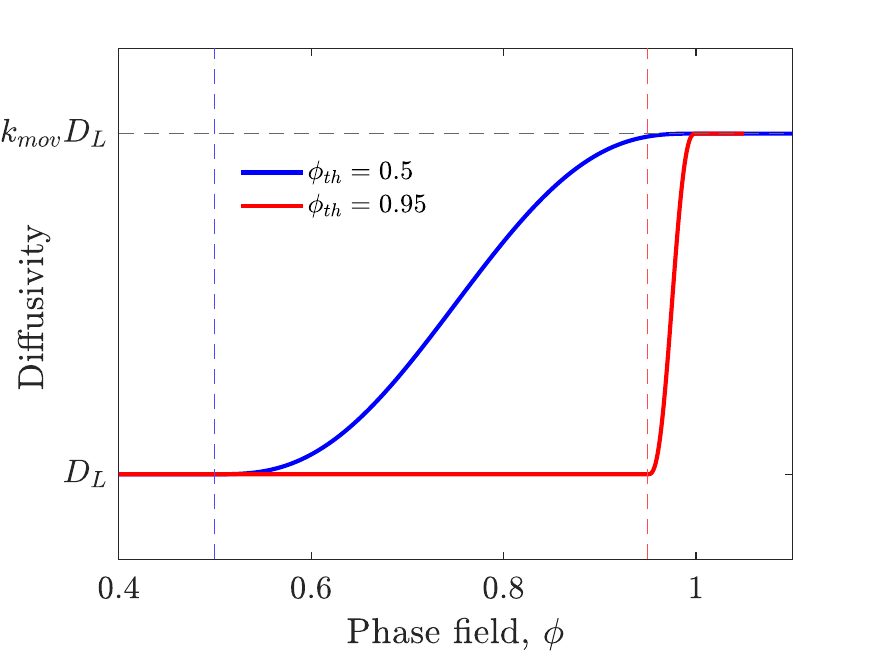}}%
  \caption{Enhancing diffusivity through a step function to capture how the rapid fluid (gaseous or electrochemical) hydrogen-containing environment progresses with crack advanced.}
  \label{Fig: step function}
\end{figure}

An important modelling challenge for hydrogen transport coupled to fracture is to reproduce fluid advance as the crack propagates. Kristensen et al. \cite{Kristensen2020AEmbrittlement} proposed a penalty-based method to enforce the condition $C_L = C_{env}$ in the cracked material, where $C_{env}$ is the hydrogen concentration produced by the environment in the crack surface. The possible implementation strategies of this condition in COMSOL are discussed in Section \ref{Sec:Hydrogen implementation}. In the present study, an alternative method based on artificially increasing lattice diffusivity is presented and followed. The lattice diffusion coefficient $D_L$ is therefore replaced by $D_L^{mov}$ in the complete domain:  
\begin{equation}
\label{Eq: DL movBCs}
    D_L^{mov}=D_L[1+k_{mov} \text{step}(\phi-\phi_{th})]
\end{equation}
where $\phi_{th}$ is a threshold parameter that controls the damage level required to consider that the fluid is moving through the cracked material and $k_{mov}\gg 1$ is the numerical parameter multiplying $D_L$ when the material is totally broken, i.e. $D_L^{mov}(\phi=1)=(1+k_{mov})D_L$. The step expression represents a smooth function with the conditions step(0) = 0, i.e. step($\phi=\phi_{th}$) = 0, and step($\phi=$ 1) = 1. The scheme of this function is shown in Fig. \ref{Fig: step function} for the two threshold values that are assessed in Section \ref{Sec:num_examples}, $\phi_{th}=0.5$ and 0.95.

To account for trapping effects, a reaction term is included in the governing equation reproducing the sink effect of hydrogen atoms hopping from lattice to trapping sites.
\begin{equation}
\label{Eq. governing trap}
    \frac{\partial C_L}{\partial t}+\frac{\partial C_T}{\partial t}+\nabla \cdot \left(-D_L\nabla C_L+\textbf{v} C_L \right) = 0
\end{equation}

Assuming thermodynamic equilibrium and following the derivation chain rule, the reaction term can be expressed as \cite{Dadfarnia2011HydrogenEmbrittlement}:
\begin{equation}
\label{Eq. reaction term}
    \frac{\partial C_T}{\partial t}=\frac{K_T N_T/N_L}{[1+(K_T-1)C_L/N_L]^2}\frac{\partial C_L}{\partial t}
\end{equation}
where the trapping equilibrium constant is a function of the defect binding energy ($E_B$) as $K_T=\exp[E_B/(RT)]$. Here, $N_L$ is the number of lattice sites per unit volume and $N_T$ represents the density of trapping sites. If multiple traps are present, the governing equation includes the sum of all trapping contributions:
\begin{equation}
\label{Eq. governing trap multitrap}
    \frac{\partial C_L}{\partial t}+\sum_i\frac{\partial C_T^i}{\partial t}+\nabla \cdot \left(-D_L\nabla C_L+\textbf{v} C_L \right) = 0
\end{equation}
where $C_T^i$ represents the hydrogen concentration in the trap site $i$. Each term is determined through the expression (\ref{Eq. reaction term}) considering the corresponding $N_T^i$ and $K_T^i$, or $E_B^i$. When the trapping site represents hydrogen retention in dislocations, i.e. $N_T^i=N_T^d$, the evolution of this value with plastic straining is here modelled following a geometrical relationship with the dislocation density, $\rho$, for a bcc lattice \cite{Lufrano1998HydrogenX-750}:
\begin{equation}
\label{Eq: N_T function of rho}
    N_T^d = \frac{\sqrt{2}\rho}{a}
\end{equation}
where $a$ is the lattice parameter ($a=$2.866$\times 10^{-10}$ m for iron), and the dislocation density evolution is modelled through a piece-wise function as in \cite{GilmanJJ.1969MicromechanicsSolids}:
\begin{equation}
\label{Eq: rho evolution}
    \rho=
    \begin{cases}
        \rho_0 + 2\gamma\varepsilon_p & \text{if } \varepsilon_p \le 0.5\\
        \rho_0+\gamma & \text{if } \varepsilon_p > 0.5\\
    \end{cases}
\end{equation}
where $\rho_0$ is the dislocation density in the unstrained condition and $\gamma$ must be experimentally fitted. Other $N_T(\varepsilon_p)$ expressions fitted from permeation testing can also be included in this modelling framework; a common one is the fit to the experiments by Kumnick and Johnson \cite{Kumnick1980DeepIron} on pure iron. In any case, the dependence $N_T(\varepsilon_p)$ results in the need for an extra term for the trapping reaction in Equation (\ref{Eq. reaction term}). Here, only dislocation sites are assumed to depend on plastic strain: 
\begin{equation}
\label{Eq. reaction term Krom}
    \frac{\partial C_T^d}{\partial t} = \frac{K_T^d N_T^d/N_L}{[1+(K_T^d-1)C_L/N_L]^2}\frac{\partial C_L}{\partial t}+ \theta_T^d\frac{dN_T^d}{d\varepsilon_p}\frac{\partial\varepsilon_p}{\partial t}
\end{equation}
The last term is referred as Krom's term after Ref. \cite{Krom1999HydrogenTip} and is proportional to the occupancy of hydrogen in traps, $\theta_T^d$, which can be obtained assuming thermodynamic equilibrium as $\theta_T^d = K_T^d\theta_L/(1+K_T^d\theta_L)$, to  the rate of trap creation and to the plastic strain rate. The rate of trap creation, $dN_T^d/d\varepsilon_p$, is found considering Eq. (\ref{Eq: N_T function of rho}) and the dislocation density evolution (\ref{Eq: rho evolution}).

\subsection{Fracture energy degradation}
\label{Sec. Fractue energy degradation}

The coupled nature of the present framework raises from the dependence of hydrogen transport on $\sigma_h$ and $\varepsilon_p$ but also on the influence of the local hydrogen concentration $C$ on the fracture toughness, or critical strain energy release $G_c$:
\begin{equation}
\label{Eq: Gc damaged}
    G_c = G_c^0 f(C)
\end{equation}
where $G_c^0$ represents the material toughness in the absence of hydrogen and $f(C)$ is a hydrogen-enhanced damage expression. Martínez-Pañeda \textit{et al.} \cite{Martinez-Paneda2018ACracking} developed a hydrogen-informed phase field framework where they assumed a coverage-based fracture energy reduction following atomistic calculations. This resulted in a linear degradation law with a single hydrogen degradation parameter $\chi$, such that
\begin{equation}
\label{Eq: coverage-based degradation}
    f(C) = 1-\chi\theta_s = 1-\chi\frac{c}{c+\exp\left(-\Delta g_b^0/(RT)\right)}
\end{equation}
where $\theta_s$ represents the hydrogen coverage, i.e. the number of hydrogen atoms per surface site ``on each slowly formed crack surface" \cite{Jiang2004FirstMetals}, which is assumed to be in equilibrium with the bulk hydrogen, i.e. with the impurity fraction $c$. The equilibrium between bulk and surface states is modelled by a Langmuir-McLean isotherm that includes a segregation energy or Gibb's energy difference $\Delta g_b^0$ \cite{Serebrinsky2004AEmbrittlement}. The impurity fraction can be calculated as the ratio between hydrogen concentration $C$ and the number of host metal atoms per unit volume $N_M$:
\begin{equation}
\label{Eq: c impurity}
    c = \frac{C}{N_M}= \frac{C[\text{mol}/\text{m}^3] A_M}{\rho_M} 
\end{equation}
where $A_M$ and $\rho_M$ represent the atomic weight and the density of the metal, respectively. In the present paper the concentration reducing the fracture energy is taken as the lattice hydrogen concentration and thus $c$ can be expressed as a function of the lattice occupancy, $\theta_L$, and the number of interstitial sites per metal atom, $\beta$ \cite{Krom2000HydrogenSteel}:
\begin{equation}
\label{Eq. impurity fraction}
    c = \frac{C_L}{N_M}=\frac{\beta C_L}{N_L} = \beta \theta_L
\end{equation}

\begin{figure}[H]
    \centering
    \makebox[\textwidth][c]{
        \begin{subfigure}[t]{0.55\textwidth}  
            \centering
            \includegraphics[width=1.1\textwidth]{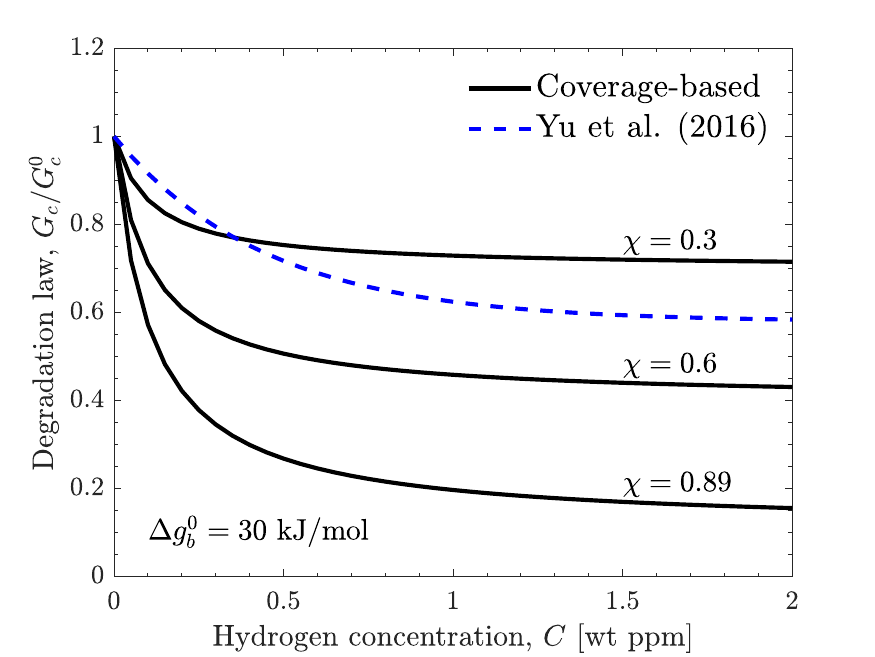}
            \subcaption{}
            \label{Fig: degradation}
        \end{subfigure}
        \begin{subfigure}[t]{0.55\textwidth} 
            \centering
            \includegraphics[width=1.1\textwidth]{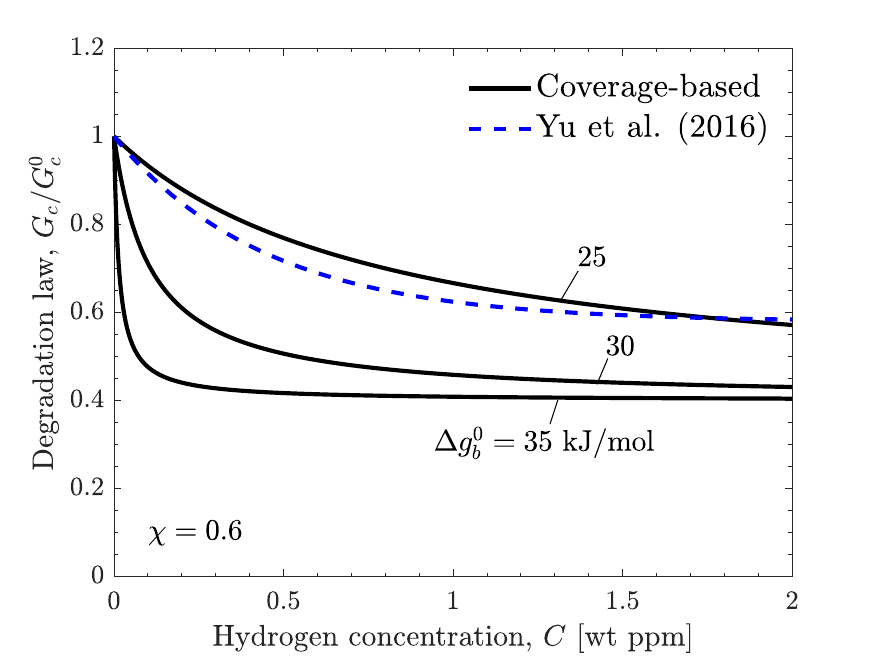}
            \subcaption{}
            \label{Fig: degradation dG}
        \end{subfigure}
    }
    \caption{Atomistically-informed hydrogen degradation laws and comparison with the phenomenological law by Yu et al. \cite{Yu2016ASteels}: (a) influence of the hydrogen degradation coefficient $\chi$ for a segregation energy of $\Delta g_b^0 = 30$ kJ/mol, and (b) influence of the segregation energy for a hydrogen degradation coefficient $\chi=0.6$.}
    \label{fig:Xi_effect}
\end{figure}

Despite the relevance of atomistic-informed hydrogen degradation expressions, one has to make a choice in regards to the magnitude of the segregation energies to the emerging crack surfaces; i.e., determine what interfaces are most likely to decohere and conduct a trapping analysis for that trap type. Other bottom-up approaches can be considered to reproduce hydrogen-modified fracture from continuum models at lower scales. For example, Ahn et al. \cite{Ahn2007ModelingAlloys} performed unit cell simulations of pre-charged void cells and considered hydrogen-enhanced local softening to obtain a reduction in the maximum stress for different triaxialities. These unit cell results were translated into hydrogen-informed traction-separation laws for cohesive zone models (CZMs). On the other hand, Yu \textit{et al.} \cite{Yu2016ASteels} use a hydrogen-informed CZM in the analysis of notched specimens with different notch radii to fit the experimental tensile curves. The reduction in cohesive strength is then fitted as $\sigma_c = \sigma_c^0f(C)$ using a \emph{uniform degradation law} which is independent of the triaxiality:
\begin{equation}
\label{Eq: Yu degradation}
    f(C) = (1-f_{\infty})\exp(-\xi C_L)+ f_{\infty}
\end{equation}
where $f_{\infty}$, the asymptotic strength reduction at high concentrations, and the shape parameter $\xi$ need to be experimentally calibrated. For a high-strength AISI 4135 steel (denoted as B15 in the experimental work by Wang \textit{et al.} \cite{Wang2005EffectSteel}), the degradation law was fitted by Yu \textit{et al.} \cite{Yu2016ASteels} as $f_{\infty}=0.579$ and $\xi=2.227$ (wt ppm)$^{-1}$. The coverage-based degradation law for three different $\chi$ values and the empirical expression proposed by Yu \textit{et al.} \cite{Yu2016ASteels} are compared in Fig. \ref{Fig: degradation}. The initial shape of the coverage-based curves is determined by the segregation energy; in Fig. \ref{Fig: degradation}, a magnitude of $\Delta g_b^0 = 30$ kJ/mol is assumed, which is a typical value for grain boundary interfaces. The role of the segregation energy in governing hydrogen uptake for a given hydrogen concentration is shown in Fig. \ref{Fig: degradation dG}. Similar to the equilibrium relationship between $C_L$, binding energy $E_B$ and trap occupancy $\theta_T$ \cite{Fernandez-Sousa2020AnalysisFatigue}, there is a notable increase in coverage with increasing segregation energy, and this brings a noticeable reduction in fracture resistance at low hydrogen concentrations.

In the present work, the hydrogen degradation law is assumed to be a function of $C_L$, while $C_T$ is not explicitly included. However, the coverage-based fracture energy reduction considers that hydrogen segregation, i.e. trapping at an interface, dominates embrittlement. Segregated hydrogen or any other concentration of trapped hydrogen can be expressed as a function of $C_L$ using Langmuir-McLean isotherm or thermodynamic equilibrium. Therefore, any other phenomenological expression could be fitted considering only $C_L$. Engineering approaches can also be considered, where the reduction if fracture energy, e.g. $G_c$ obtained from fracture experiments, is fitted as a function of experimental hydrogen content. As an example, a consistent procedure assuming an exponential degradation law has been proposed by Mandal \textit{et al.} \cite{mandal2024computational}.

\section{Numerical implementation}
\label{Sec:Numerical implementation}

We proceed to describe the implementation of the generalised phase field fracture theory for hydrogen embrittlement presented in Section \ref{Sec:Theory}, which includes multi-trapping, elastic and plastic contributions to the fracture driving force and a new formulation for capturing enhanced hydrogen transport through cracks. In addition, particular emphasis is placed on its numerical implementation within COMSOL Multiphysics. Various implementation approaches are considered, with the most general form considering the following physics nodes and sub-nodes: \texttt{Solid Mechanics}, \texttt{Helmholtz Equation} and \texttt{Transport of Diluted Species}. The codes developed can be freely downloaded from \url{https://mechmat.web.ox.ac.uk/codes}\\ 

In the \texttt{Solid Mechanics} Physics interface, the dependent variable is the displacement vector $\textbf{u}$. The set of equations includes the governing equation based on the balance of linear momentum ($\nabla \cdot \boldsymbol{\sigma} = 0$ in the absence of body forces and for quasi-static conditions), compatibility equations (to relate $\textbf{u}$ and the strain field $\pmb{\varepsilon}$) and a constitutive model which by default is based on linear elasticity ($\boldsymbol{\sigma}=\mathbf{C}: \pmb{\varepsilon}_e$).  
Within the \texttt{Linear Elastic Material} node, the \texttt{Plasticity} sub-node is governed by the flow rule and the yield criterion. The present study is restricted to small strains\footnote{While the extension to large strains is straightforward, as discussed in Part I \cite{PartI}, these play a negligible role in crack growth studies for the conditions relevant to hydrogen-assisted fracture.} and a Von Mises yield function. Both linear and power law isotropic hardening behaviours are considered. Except for the degradation of the elastic modulus, the default \texttt{Solid Mechanics} equations are not modified and thus are not presented here; these are standard and details can be found in the COMSOL documentation. Only the novel features are discussed below; we begin by presenting the phase field fracture implementation in Section \ref{Sec:Phase field implementation}, covering the in-built and user-defined options, proceed to discuss the hydrogen transport implementation (Section \ref{Sec:Hydrogen implementation}), and finish with a discussion on solution strategies (Section \ref{Sec:Solution strategies}).  

\subsection{Phase field fracture implementation}
\label{Sec:Phase field implementation}

\subsubsection{Built-in damage modelling}
\label{Sec:Damage implementation}

COMSOL includes, since version 6.0, an in-built implementation of phase field fracture. From version 6.2, this in-built implementation is also compatible with the plastic material models. To activate this, one should access the sub-node \texttt{Damage} under the \texttt{Linear Elastic Material} sub-node and select \texttt{Phase field damage}. The following features must be highlighted in this in-built implementation.\\

\noindent \textbf{\textit{Crack driving force}}. The crack driving force can be defined to be based on the elastic strain energy density, on the total strain energy density, or user-defined. These two choices, with or without plastic contribution, are assessed in \ref{Appendix: comparison with built-in} and compared to the (user) PDE-based implementation. It must be noted that only the AT2 model is considered for the in-built phase-field balance equation and therefore the elastic stage before the onset of damage typical of the AT1 formulation \cite{Alessi2018ComparisonPlasticity} cannot be reproduced. However, an elastic domain before damage can be introduced by activating the fracture energy threshold. Other alternative definitions of the crack driving force that are based on principal stress or strain criteria are also available yet not used here.\\

\noindent \textbf{\textit{Strain energy split}}. Both so-called volumetric-deviatoric or spectral decompositions of $\psi_e^0$ are available, in addition to a stress-based spectral decomposition. The results presented in Section \ref{Sec:num_examples} are based on the volumetric split.\\

\noindent \textbf{\textit{User-defined degradation}}. Both power law, equivalent to Eq. (\ref{Eq: g degrad}), and user-defined damage functions $d(\phi)$ can be defined. This is related to the more frequently used degradation function $g(\phi)$ as $d(\phi) = 1- g(\phi)$. Here, we choose to focus on the widely used quadratic function.\\

\noindent \textbf{\textit{Viscous regularisation}}. A characteristic time $\tau$ can be fixed, significantly smaller than the time increment, i.e. $\tau \ll \Delta t$, to stabilise the phase-field equation by introducing an artificial delay in damage evolution. The relationship between this viscous time and the numerical viscosity $\eta$ [kN$\cdot$s/mm$^2$] as defined by Miehe et al.  \cite{Miehe2010ASplits} is $\tau=\eta \ell/G_c$. In the present work, viscous regularisation is not considered, as no convergence issues were observed.\\

One disadvantage of the in-built phase field fracture implementation is that Dirichlet boundary conditions or initial values cannot be assigned to the phase field variable, $\phi$, and thus the cracks must be geometrically introduced or considering a previous step. Additionally, it must be taken into account that when the plasticity model is combined with the built-in damage, the corresponding yielding criterion considers the undamaged stress values.

\subsubsection{User-defined: PDE modelling}
\label{Sec:PDE implementation}

As shown in Section \ref{Sec:Theory fracture}, the phase field balance can be expressed as a Helmholtz-type equation - see Eq. (\ref{Eq: Helmholtz governing}). In COMSOL, PDEs resembling the Helmholtz equation can be readily defined by the user and are found under the \texttt{Mathematics} module and the \texttt{Classical PDEs} category. With that structure, the modelled terms are easily identified as a \texttt{Diffusion Coefficient} ($c_d$), an \texttt{Absorption coefficient} ($a_c$) and a \texttt{Source term} ($f_s$):
\begin{equation}
\label{Eq: Helmholtz Comsol}
    \nabla\cdot(-\underbrace{\ell^2}_{c_d}\nabla\phi)+ \underbrace{\left( \frac{2\ell\mathcal{H}}{G_c} + 1 \right)}_{a_c}\phi = \underbrace{\frac{2\ell\mathcal{H}}{G_c}}_{f_s}
\end{equation}

The rate-independent Helmholtz form (\ref{Eq: Helmholtz Comsol}) is not appropriate if one wishes to incorporate viscous regularisation; in that case, the more general \texttt{Coefficient Form PDE} or the \texttt{Stabilized Convection-Diffusion Equation} can be used to implement a damping term. However, no convergence issues were found in the computations conducted here and as such there was no need to consider viscosity.\\

To store the history field $\mathcal{H}$, a variable is created under the component definitions, and updated with the help of an auxiliary \texttt{State Variable} $H_{sv}$ to avoid circular dependency. This is achieved using the conditional expression:
\begin{equation}
\label{Eq: H update}
    \mathcal{H}(t+\Delta t)=
    \begin{cases}
         (\psi_e^{0+}+\psi_p^0)[t+\Delta t] & \text{if } (\psi_e^{0+}+\psi_p^0)[t+\Delta t] > H_{sv}(t)\\
         H_{sv}(t) & \text{else } \\
    \end{cases}
\end{equation}
It is important to update the state variable \texttt{After step} and follow the scheme in Eq. (\ref{Eq: H update}) to avoid a staggered storage of $\mathcal{H}$ that is expected to accumulate errors in fatigue or rate-dependent analysis \cite{Ai2022AParticles}.\\ 

In addition, the user-defined Young's modulus is implemented as $g(\phi)E_0$, where $E_0$ is the undamaged material property; this definition captures the appropriate coupling, as demonstrated in Eq. (\ref{Eq: stress elast matrix}). The plastic degradation is introduced in the yield criterion by a user-defined hardening $h(\phi)\sigma_{f0}$, where $\sigma_{f0}$ follows linear (\ref{Eq: linear hardening law}) or power-law (\ref{Eq: power law hardening law}) hardening laws. It must be highlighted that, after the degradation of the elasticity matrix, the COMSOL variable \texttt{solid.Ws} corresponds to the damaged value of the elastic strain energy density, i.e. $g(\phi)\psi_e^0$, and thus cannot be directly used to define $\psi_e^0$. Similarly, the internally calculated dissipated energy, \texttt{solid.Wp} is not used and the plastic strain energy density, $\psi_p^0$ is defined from the theoretical expressions provided above; i.e., Eq. (\ref{Eq: psi_p^0 linear hardening}) for linear hardening or Eq. (\ref{Eq: psi_p^0 power law}) for a power-law hardening.

\subsection{Hydrogen transport implementation}
\label{Sec:Hydrogen implementation}

Like other transport phenomena (e.g. heat transfer), a hydrogen diffusion model with the convection and reaction terms presented in Section \ref{Sec:theory hydrogen transport} can be modelled using a \texttt{Coefficient Form PDE} or a \texttt{General Form PDE}. However, the \texttt{Transport of Diluted Species} module is proposed as a valid strategy for reproducing trapping phenomena and stress-drifted diffusion in the two-level approach described in Section \ref{Sec:theory hydrogen transport}. Additionally, this module includes built-in stabilisation schemes to avoid numerical noise. The \texttt{Convection} option and its \texttt{Conservative form} must be selected to model the Eq. (\ref{Eq. governing convec}). Implementation details can be found in Part I of this work \cite{PartI}. Nevertheless, several aspects must be highlighted that are intrinsic to the coupling of hydrogen transport and phase field fracture:\\ 

\noindent \textbf{\textit{Hydrostatic stress term}}. To introduce the hydrostatic stress gradient ($\nabla \sigma_h$) in the convection term, $\sigma_h$ is stored as a new variable \texttt{Sh=-solid.pm}, with \texttt{solid.pm} being the internal variable for the pressure. Following Ref. \cite{Cui2022AEmbrittlement}, the damaged stress is considered to drive hydrogen lattice diffusion. The built-in differentiation operator is then used to find the gradient; for instance, in a 2D model, the components of the convection velocity vector (\textbf{v}) corresponding to $x$ and $y$ include the expressions \texttt{d(Sh,x)} and \texttt{d(Sh,y)}, respectively. The influence of discretization on the smoothness of $\sigma_h$ and $C_L$ distributions is discussed in Part I of this work \cite{PartI}, yet it should be noted that this influence is negligible under small strains, as relevant here and therefore all approaches are valid.\\ 
    
\noindent \textbf{\textit{Moving boundary conditions}}. Different options have been explored to capture how the environment advances with a propagating crack, i.e. to enforce $C_L(\phi>\phi_{th})=C_{env}$, where $C_{env}$ is the hydrogen concentration of the environment and $\phi_{th}$ is a damage threshold defining when the micro-crack network is dense enough for the fluid to progress through it. The options considered are: (i) weak contribution, (ii) weak constraint, (iii) pointwise constraint and (iv) artificial diffusivity. In the first approach, defining a weak contribution, a penalty term can be implemented equal to $k_p(C_L-C_{env})\text{test}(C_L)\text{step}(\phi-\phi_{th})$, where test() is a COMSOL built-in operator to express the test functions of the solution variable. However, the penalty method is too sensitive to the choice of penalty constant ($k_p$) and for some conditions results in oscillating concentrations near the crack. The weak constraint, in contrast, enforces the condition $(C_L-C_{env})\text{step}(\phi-\phi_{th})=0$ but introduces a Lagrange multiplier as an additional dependent variable resulting in a high computational cost and convergence deterioration. A pointwise constraint is similar in nature, since a constraint and the corresponding reaction force, i.e. $\text{test}(C_L)\text{step}(\phi-\phi_{th})$, are applied, but the Lagrange multipliers are implicitly eliminated and this results in more stable solutions than the penalty method. A final option that circumvents the need for constraints is based on the definition of an artificial diffusivity, as presented in Eq.  (\ref{Eq: DL movBCs}). This last option was found to be the most robust and efficient strategy to model the aforementioned moving chemical boundary conditions in COMSOL.\\ 
    
\noindent \textbf{\textit{Degradation of fracture energy}}. To capture its susceptibility to the hydrogen content, the material toughness ($G_c$) is defined as a variable in the COMSOL interface (as opposed to a parameter), following Eq. (\ref{Eq: Gc damaged}). The intermediate variables, e.g. the impurity fraction $c$ or the coverage $\theta_s$, are defined also in the variable list as a function of the dependent variable $C_L$ (\texttt{CL}).\\ 
    
\noindent \textbf{\textit{Trapping term}}. A \texttt{Reaction} node is selected within the \texttt{Transport of Diluted Species} module to implement the expression in Eq. (\ref{Eq. reaction term}). It must be noted that this reaction term is proportional to $\partial C_L/\partial t$, which can be accessed in COMSOL using the time-derivative operator (\texttt{CLt}).

\subsection{Solution strategies}
\label{Sec:Solution strategies}

Two solution strategies have been initially considered: the backward differentiation formula (BDF) and the generalised alpha method. Despite the generalised alpha method demonstrating second-order accuracy and robustness in phase field fracture studies \cite{Borden2012AFracture, Zhou2018PhaseStudies, Liu2020Micro-crackingModeling}, some instabilities have been found during crack propagation. BDF is more stable, resulting in smoother $\phi$, $\sigma_h$ and $C_L$ distributions for the same tolerance and time steps but longer computational times. It must be noted that the chosen BDF solver in COMSOL has a variable order between 2 and 5, where the order is automatically reduced when a higher stability is needed. Time steps are automatically taken by the solver (\texttt{Time stepping: Free}) but the maximum increment size has been controlled to ensure that all crack propagation phenomena are resolved; this becomes particularly relevant for conditionally stable staggered solution approaches (so-called single-pass staggered), as discussed below. It is also important to consider a sensitivity study with the value of the tolerance factor to ensure the accuracy of the results.\\

Considering a predefined maximum displacement $u_{\text{max}}$ to be applied in a remote boundary, an upper bound increment $\Delta u$ is fixed to avoid error accumulations, especially in staggered schemes:   
\begin{equation}
    \Delta u = \Delta\bar{u} \ u_{\text{max}} 
\end{equation}
where $\Delta\bar{u}$ is the non-dimensional displacement increment. The choice of this upper bound limits the time increment but does not fix its value, which can be lower when the convergence is poor. The maximum time increment is defined accordingly as:
\begin{equation}
    \Delta t = \Delta\bar{u} \frac{u_{\text{max}}}{\dot{u}} = \Delta\bar{u} \ t_{\text{max}} 
\end{equation}
This upper bound for $\Delta t$ is implemented as the \texttt{Maximum step} in the \texttt{Time Stepping} options. A value of $\Delta\bar{u} = 10^{-3}$ has been used in the present study. As an alternative to a fixed $\Delta\bar{u}$ value, an adaptive time increment is proposed for some cases showing unstable crack propagation, especially for elastic-brittle cases. The upper limit of the time increment
depends here on the rate of the phase field evolution:
\begin{equation}
    \Delta t (t)=\frac{1}{k_s \dot{\phi}_g(t)} 
\label{Eq: adaptive time step}    
\end{equation}
where a value of $k_s$ equal to 10$^2$ is a reasonable choice to capture damage nucleation and crack propagation, and $\dot{\phi}_g$ is tracked as the maximum rate of $\phi$ using the \texttt{Maximum} operator within the in-built \texttt{Nonlocal couplings} in COMSOL:
\begin{equation}
    \dot{\phi}_g(t) = \max\limits_{\textbf{x} \in \Omega} \dot{\phi}(\textbf{x},t)
\end{equation}
For single-pass staggered schemes and unstable crack propagation, this adaptive time stepping improves the accuracy of results, as discussed in \ref{Appendix: time stepping and discretization}.

All fields in a hydrogen-informed phase field problem are coupled: hydrogen transport is influenced by the stress-strain state, phase field evolution is governed by the strain energy density and the hydrogen content, and the material stiffness degrades with $\phi$. Therefore, the nodal solution variables, displacement field $\textbf{u}$, phase field $\phi$ and hydrogen lattice concentration $C_L$ should be solved simultaneously in a single system of equations. This approach is termed as \emph{monolithic} in phase field studies or \texttt{Fully Coupled} in COMSOL. However, taking only the displacement and phase field problem, the free energy functional is non-convex with respect to ($\textbf{u}$, $\phi$) and the solution usually diverges in the post-peak loading regime \cite{Gerasimov2016AFracture}. This deterioration in convergence is usually tackled by solving the unknown fields sequentially; i.e., the phase field is fixed during the iterative solution of the displacement problem and vice versa. This so-called \emph{staggered} approach is less efficient and thus slower than the monolithic scheme, but more robust due to the convexity of the problem in $\textbf{u}$ and $\phi$, separately. To improve this trade-off between efficiency and robustness \cite{Navidtehrani2021AMethod}, modifications of monolithic schemes have been proposed such as quasi-Newton schemes \cite{Wu2020OnTheory,Kristensen2020PhaseScheme} or line search procedures \cite{Gerasimov2016AFracture}, but these are not available in COMSOL and thus are not considered here. Consequently, the present work implements a staggered scheme by defining \texttt{Segregated Steps} in COMSOL. We also choose to solve the hydrogen transport problem separately, even though the robustness of the coupled solution with $\textbf{u}$ is similar, as shown in Part I of this work \cite{PartI}. It is worth noting that it has been reported that the inefficiency of the staggered approach can be overcome by using acceleration techniques \cite{Storvik2021AnFracture}. COMSOL implements an Anderson acceleration algorithm that uses information from a defined number of previous iterations to improve convergence. In the present work, the \texttt{Dimension of iteration space} is fixed as 10 since considering more iterations might cause instabilities \cite{Schapira2023PerformanceSolutions}. Finally, it is worth emphasising that COMSOL allows for both single and multi-pass staggered schemes; that is, the user can define the number of recursive iterations over the segregated step as an option in the \texttt{Termination technique} list. Differences between single-pass and multi-pass segregated schemes are assessed in \ref{Appendix: time stepping and discretization}. In each individual step for the corresponding dependent variable, $\textbf{u}$, $\phi$ and $C_L$, the termination criterion is based on a tolerance value which is here fixed as $R_{tol} = 0.001$ for relative tolerance and $0.1R_{tol}$ for absolute tolerance. An additional segregated step is added to update the state variable $\mathcal{H}$, when using the PDE-based implementation.

\section{Numerical experiments}
\label{Sec:num_examples}

The capabilities of the present framework are shown in three case studies, and their analysis also reveals new insight into the interactions between crack growth and hydrogen transport and trapping. The same mechanical properties are considered in all three case studies, as listed in Table \ref{Tab:mat_elastic_plastic}; despite the general validity of the model for the transport of solid impurities in a metal and the consequent modified fracture, this work evaluates a set of parameters typical of steels due to the relevance of hydrogen embrittlement in these alloys. One should note that, as discussed in Section \ref{Sec:Theory fracture}, the choice of $\ell$ determines the value of the material strength, via Eq. (\ref{eq:sigmaC}). This is relevant for crack nucleation or short crack analyses and also influences plastic dissipation during crack growth (and hence it can be benchmarked against experimental crack growth resistance curves, see Ref. \cite{mandal2024computational}).

\begin{table}[H]
\centering
\caption{Mechanical parameters for the simulated elastic-plastic material.}
\label{Tab:mat_elastic_plastic}
   {\tabulinesep=1.2mm
   \makebox[\textwidth][c]{\begin{tabu} {cccccccccccccc}
       \hline
$E_0$  & $\nu$ & $\sigma_{y0}$ & $H_0$ & $G_c^0$ & $\ell$ \\ \hline
210 & 0.3  & $0.003E_0$ & $0.03E_0$ & 10-50 & 0.05 \\
(GPa) & (-) &  (MPa) & (MPa) & (N/mm) & (mm)  \\ \hline
   \end{tabu}}}
\end{table}

The parameters employed in the hydrogen diffusion and trapping model are listed in Table \ref{Tab:mat_diffusion}. The lattice diffusivity and partial molar volume values, $D_L$ and $\bar{V}_H$ are chosen as in Ref. \cite{Sofronis1989NumericalTip}, and represent typical values for a ferritic steel. Most of the analyses 
conducted consider the case of hydrogen entry during loading in a pre-charged bulk material; this means that both boundary and initial conditions are equal to $C_L^0$, unless the absence of precharging is explicitly stated. Different lattice concentrations are simulated, from 0.1 to 1 wt ppm, even though trapping effects are more relevant in lower concentration regimes \cite{Diaz2019AnalysisIron}. The last three parameters in Table \ref{Tab:mat_diffusion}, $N_L$, $N_T$ and $E_B$, are required for the trapping-modified model, see Eq. (\ref{Eq. governing trap}); however, $N_L$ is also indirectly considered in the coverage-based degradation law to determine the impurity fraction $c$. 
The concentration variables are considered in mol/m$^3$ by default within the \texttt{Transport of Diluted Species} module. To convert wt ppm into mol/m$^3$, the following expression is used:
\begin{equation}
\label{Eq: wt ppm to mol/m3}
    C[\text{mol}/\text{m}^3] = C[\text{wt ppm}]10^{-6}\frac{ \rho_M}{M_H} 
\end{equation}
\noindent where $M_H$ is the atomic weight of hydrogen, i.e. 1.008 g/mol. The parameter $\beta$ in Eq. (\ref{Eq. impurity fraction}) is fixed as 6 interstitial sites per metal atom for tetrahedral occupancy in bbc iron \cite{Krom2000HydrogenSteel}, and the segregation energy equals $\Delta g_b^0 = 30$ kJ/mol, as in Ref. \cite{Serebrinsky2004AEmbrittlement}. The choice of this value is expected to capture hydrogen trapping in grain boundaries for intergranular embrittlement or the segregation in any newly created surface during fracture. The coverage-based linear degradation law proposed by Mart\'{\i}nez-Pa\~neda \textit{et al.} \cite{Martinez-Paneda2018ACracking} is followed with the fitted value of $\chi = 0.89$ for iron but the magnitude of $\chi$ is also varied to assess its effect.

\begin{table}[H]
\centering
\caption{Hydrogen transport parameters for the simulated steel considering trapping effects.}
\label{Tab:mat_diffusion}
   {\tabulinesep=1.2mm
   \makebox[\textwidth][c]{\begin{tabu} {cccccccccccccc}
       \hline
$D_L$  & $\bar{V}_H$ & $T$ & $C_{env}$, $C_L^0$  & $N_L$ & $N_T$ & $E_B$\\ \hline
1.27$\times$10$^{-8}$ & 2$\times$10$^{-6}$  & 293 & 0.1--1.0 & 5.1$\times$10$^{29}$ & 10$^{-4}$--10$^{-2}$ $N_L$ & 30--40\\
(m$^2$/s)  & (m$^3$/mol) &  (K) & (wt ppm) & (sites/m$^3$) & (traps/m$^3$) & (kJ/mol)\\ \hline
   \end{tabu}}}
\end{table}

\subsection{Notched square plate tests}
\label{Sec:Plate}

\begin{figure}[H]
\makebox[\linewidth][c]{%
        \begin{subfigure}[b]{0.4\textwidth}
                \centering
                \includegraphics[scale=0.7]{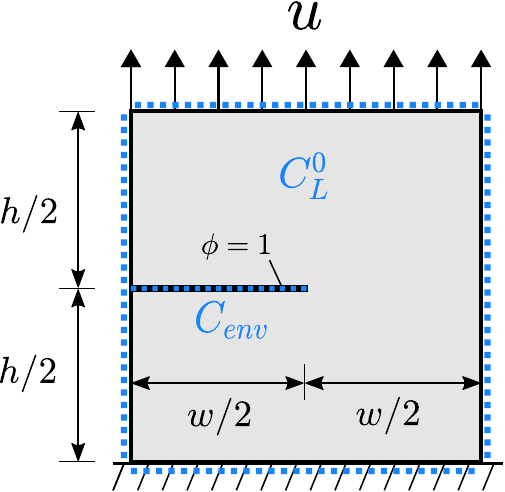}
                \caption{}
                \label{}
        \end{subfigure}
        \begin{subfigure}[b]{0.55\textwidth}
                \raggedleft
                \includegraphics[scale=0.18]{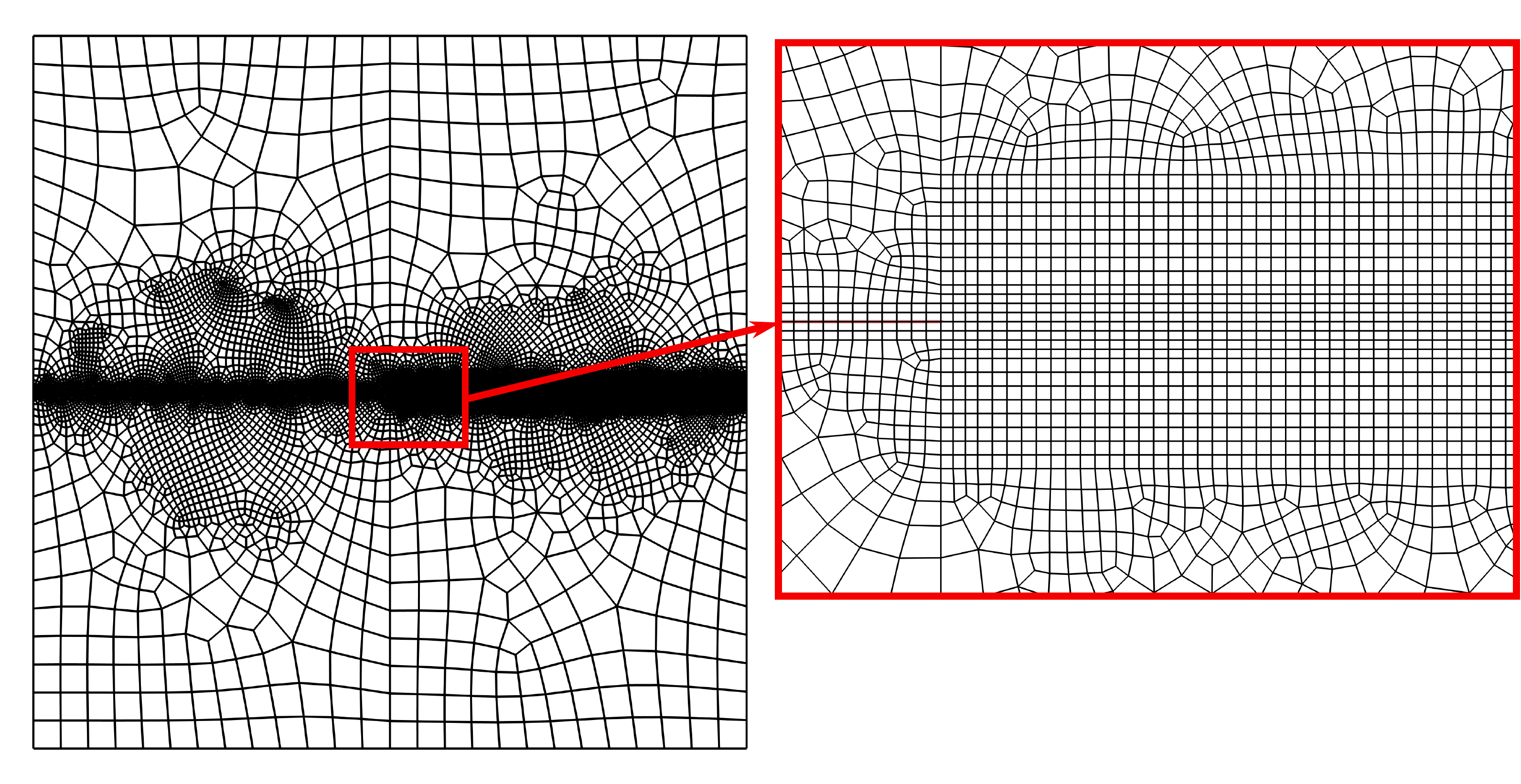}
                \caption{}
                \label{}
        \end{subfigure}}
        \caption{Numerical experiments on a single-edge cracked plate undergoing uniaxial loading: (a) scheme of the geometry and the boundary conditions, and (b) finite element mesh.}
        \label{Fig:schemeSENT}
\end{figure}

We begin our numerical experiments by considering the case of a notched square plate subjected to uniaxial loading, as depicted in Fig. \ref{Fig:schemeSENT}a. This is a paradigmatic benchmark in the phase field fracture and phase field hydrogen embrittlement communities and thus serves as a validation problem. The applied load is introduced through a prescribed displacement on the top edge, which is ramped in time according to $u=\dot{u}t$, where $\dot{u}$ denotes the fixed displacement rate. While typical loading rates in hydrogen-assisted cracking problems are not sufficiently high to induce inertia or rate-dependent material behaviour, the loading rate plays a role due to the inherent transient behaviour of hydrogen diffusion. 

As shown in Fig. \ref{Fig:schemeSENT}b, a fine mesh is used in the region ahead of the crack. To ensure mesh-insensitive results, the characteristic element size in this region is chosen to be $h_e \le \ell/5$. A total of 18,324 quadrilateral elements are used to discretise the square plate. Cubic discretisation is used for the displacement field and the phase field, while linear discretisation is used for the lattice hydrogen concentration.  It should be noted that quadratic discretisation is enough to capture the interplay between deformation and diffusion under small strains, as discussed in \ref{Appendix: time stepping and discretization} and detailed in Part I of this work \cite{PartI}. While COMSOL offers the possibility of using adaptive re-meshing, this feature has not been explored in the present work. Unless otherwise stated, the dimensions of the 2D square plate sample are $h=w=5$ mm. Plane strain conditions are assumed. As depicted in Fig. \ref{Fig:schemeSENT}a, hydrogen entry through the crack surface is simulated by fixing a Dirichlet boundary condition for the hydrogen transport problem, i.e. $C_L = C_{env}$ on the crack surface. This boundary condition is also applied to all the outer edges of the single-edge cracked plate. Also, unless otherwise stated, a pre-charged domain is simulated and thus the initial concentration is taken as $C_L^0$. Two types of material behaviour are considered, the one corresponding to a brittle elastic solid (Section \ref{Sec:elastic_plate}), which serves as a validation benchmark, and one where plasticity effects are considered in both deformation and fracture, as well as in hydrogen trapping, to gain new insight (Section \ref{Sec:plastic_plate}). 

\subsubsection{Brittle-elastic model}
\label{Sec:elastic_plate}

First, we consider the case of a linear elastic material, absent of plasticity effects. Therefore, the crack driving force only includes $\mathcal{H}_e^+$. Since there is no plastic dissipation and this is a mode I fracture problem, the failure process is brittle, with the crack propagating in an unstable fashion. Hence, if using a single-pass staggered scheme one must use a very small time step to capture the behaviour in the failure regime appropriately. To this end, the maximum increment is limited to $\Delta \bar{u} = 10^{-3}$ in the single-pass scheme considered in this section.

\begin{figure}[H]
  \makebox[\textwidth][c]{\includegraphics[width=0.8\textwidth]{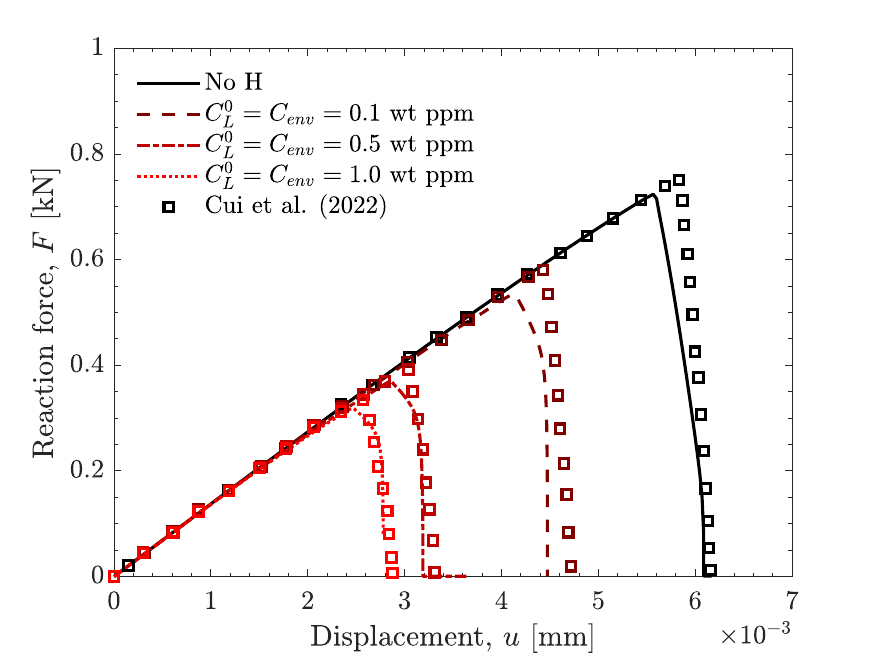}}%
  \caption{Validation of the numerical implementation against the behaviour of a linear elastic solid. Comparison against the results by Cui \textit{et al.} \cite{Cui2022AEmbrittlement} (symbols). A very good agreement is obtained across a wide range of hydrogen concentrations. The time increment and tolerance are chosen as $\Delta \bar{u} = 10^{-3}$ and $R_{tol}=0.005$.}
  \label{Fig: validate embrittlement}
\end{figure}

First, we validate our implementation against the results by Cui \textit{et al.} \cite{Cui2022AEmbrittlement}. Mimicking Ref. \cite{Cui2022AEmbrittlement}, and different from the other calculations in this work, we assume a smaller plate ($h=w= 1$ mm), take larger displacement increments and consider the following fracture parameters: $G_c = 2.7$ N/mm, $\ell=0.0075$ mm. Additionally, in contrast to all other examples, a sharp crack is initially introduced using COMSOL's \texttt{Crack} feature within the \texttt{Solid Mechanics} module and considering a \texttt{Slit} crack surface definition, which duplicates the nodes in the crack region. This is here referred to as a \emph{geometrically-induced} sharp crack, as opposed to a \emph{phase field-induced} crack, whereby the crack is defined by setting $\phi=1$ as an initial condition in a given region (Fig. \ref{Fig:schemeSENT}a). The moving chemical boundary conditions are applied with a threshold of $\phi_{th}=0.5$. The results obtained are shown in Fig. \ref{Fig: validate embrittlement}, where the load versus displacement curves are obtained for different hydrogen concentrations, including the hydrogen-free case ($C_L^0=C_{env}=0$ wt ppm). A very good agreement is attained with the results by Cui \textit{et al.} \cite{Cui2022AEmbrittlement}, validating the present implementation. It is important to note that in both the benchmark Ref. \cite{Cui2022AEmbrittlement} and the present results, the damaged hydrostatic stress $\sigma_h$ drives diffusion, in contrast to the approach adopted in Ref. \cite{Martinez-Paneda2018ACracking}. 
In addition, the loss of carrying capacity is strongly affected by the considered time stepping in this single-pass scheme. For that reason, a multi-pass staggered scheme or an adaptive time increment have been implemented to solve this sensitivity, as analysed in \ref{Appendix: time stepping and discretization}. The multi-pass scheme is adopted in following sections where multiple iterations are performed over the full segregated step until the tolerance criterion is verified. 

\subsubsection{Ductile-plastic model}
\label{Sec:plastic_plate}

\begin{figure}[H]
  \makebox[\textwidth][c]{\includegraphics[width=0.8\textwidth]{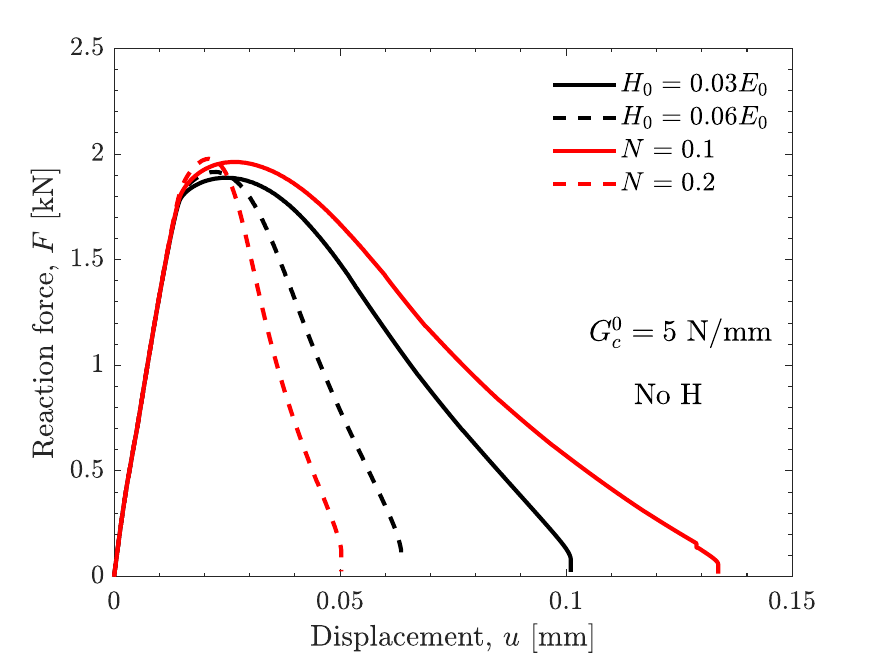}}%
  \caption{Force versus displacement behaviour of an elastic-plastic single-edge tension square plate for various degrees of work hardening, as characterised by a linear hardening law (with hardening coefficient $H_0=0.03E_0$ and $H_0=0.06E_0$) and a power-law (with strain hardening exponent $N=0.1$ and $N=0.2$). Results obtained for $G_c^0=5$ N/mm, $\ell=0.05$ mm and no hydrogen ($C_L^0=C_{env}=0$ wt ppm).}
\label{Fig: hardening comparison}  
\end{figure}

The validated model is then extended to incorporate plasticity effects in the deformation, diffusion and fracture processes, with the fracture driving force given in Eq. (\ref{Eq:epDrivingForce}). The chosen multi-pass staggered scheme does not show time step sensitivity, and the maximum increment is chosen as $\Delta u=7.5\times10^{-6}$ mm here. In this case study, the initial crack is introduced considering a prescribed $\phi$ value equal to 1 (i.e., a \textit{phase field-induced} crack). The influence of the work hardening behaviour is shown in Fig. \ref{Fig: hardening comparison}, where force versus displacement results are presented for $G_c^0 = 5$ N/mm, $\ell = 0.05$ mm, $\beta_p = 0.1$, and both linear and power-law hardening, with two choices of the linear hardening modulus $H_0$ and the strain hardening exponent $N$. The sample is assumed to be hydrogen-free. 

The results, presented in Fig. \ref{Fig: hardening comparison}, reveal distinct behaviour relative to the case of an elastic, brittle solid (Fig. \ref{Fig: validate embrittlement}). A more ductile response is observed, with crack growth being stable and thus the drop in load after the peak being less sharp. This smoother softening behaviour is seen to be more ductile with decreasing work hardening, as this results in a higher degree of plastic dissipation. The calculations with a higher degree of work hardening (dashed lines, $H_0=0.06E_0$ and $N=0.2$) show a stiffer response before the peak load is reached but then dissipate less plasticity and behave more brittle. The more stable crack growth behaviour observed, relative to the linear brittle case, also relaxes the need for strict tolerance and time-stepping requirements to accurately capture the post-peak regime.\\  

\begin{figure}[H]
  \makebox[\textwidth][c]{\includegraphics[width=0.8\textwidth]{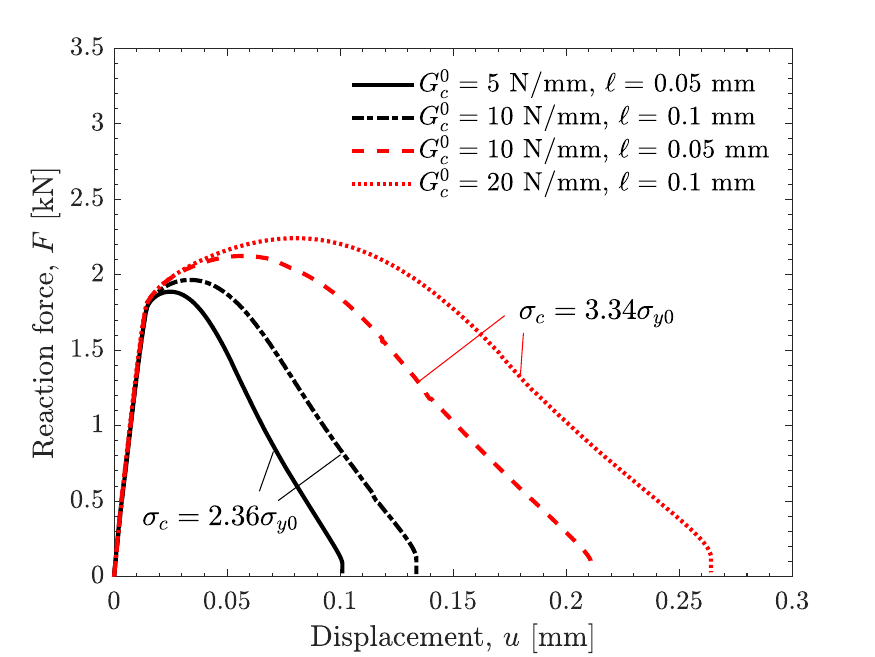}}%
  \caption{Force versus displacement behaviour of an elastic-plastic single-edge tension square plate for various choices of fracture parameters: $G_c^0$ and $\ell$ (which determines $\sigma_c$). The results show more ductility due to greater plastic dissipation with increasing $G_c^0$ and decreasing $\ell$ (increasing $\sigma_c$).}
\label{Fig: Gc0 ductile comparison}  
\end{figure}

We shall now evaluate the influence of varying the fracture parameters, $G_c^0$ and $\ell$. The results are provided in Fig. \ref{Fig: Gc0 ductile comparison} for the linear hardening case, with $H_0=0.03E_0$, and $\beta_p = 0.1$. For a fixed value of $\ell$, increasing the material toughness is seen to produce a more ductile response, in agreement with expectations. The length scale plays a role as it determines the value of the material strength, recall Eq. (\ref{eq:sigmaC}). As discussed in Ref. \cite{kristensen2021assessment}, phase field models can naturally capture both toughness- and strength-dominated failures, and the transition from one to the other. Also, the strength influences the degree of plastic dissipation during crack growth, with higher values of $\sigma_c$ leading to increasing crack growth resistance - see the results in Ref. \cite{Kristensen2020AEmbrittlement} for phase field and in Ref. \cite{Tvergaard1992TheSolids} for cohesive zone models. The results obtained here are consistent with this interpretation; for a fixed $G_c^0=10$ N/mm, decreasing $\ell$ leads to a more ductile behaviour, associated with a higher strength, see Eq. (\ref{eq:sigmaC}), and a larger plastic dissipation during crack growth. In practice, most materials exhibit a strength-toughness trade-off and thus isolating the contributions from $\sigma_c$ and $G_c^0$ is not easy. 

\begin{figure}[H]
  \makebox[\textwidth][c]{\includegraphics[width=1.0\textwidth]{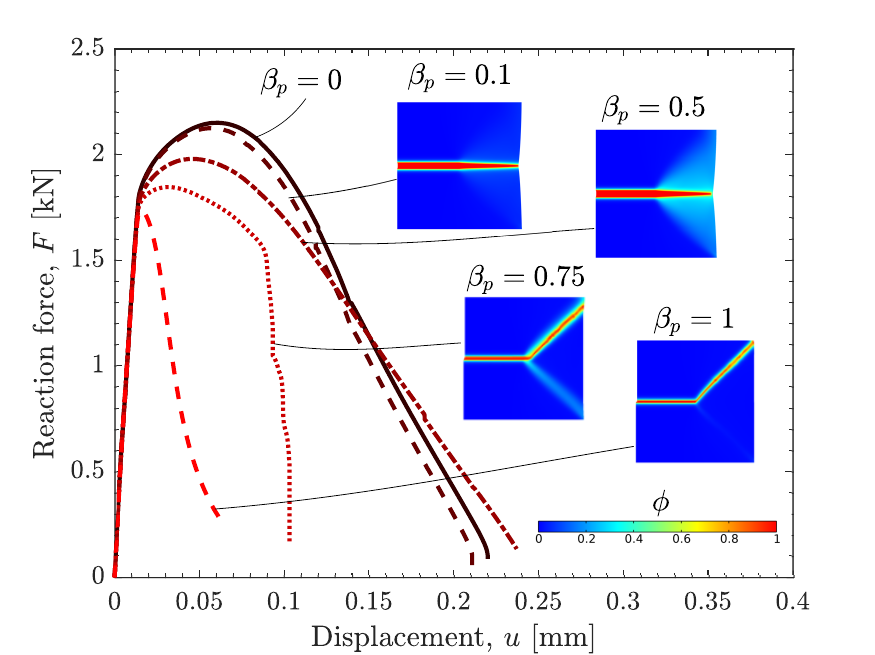}}%
  \caption{Force versus displacement behaviour of an elastic-plastic single-edge tension square plate for various choices of the parameter $\beta_p$, governing the fraction of plastic work involved in the fracture process. Contours of the phase field variable $\phi$ at failure are also superimposed, revealing that failure takes place due to plastic localisation at 45$^\circ$ when the plastic contribution is sufficiently high, versus a straight, mode I crack growth behaviour for lower values of $\beta_p$. Results obtained with $G_c^0=10$ N/mm and $\ell=0.05$ mm.}
\label{Fig: Bp}  
\end{figure}

As discussed in Section \ref{Sec:Theory fracture}, different approaches have been used to weigh the contribution of the plastic work to the crack driving force. For example, Duda \textit{et al.} \cite{Duda2015ASolids} and Kristensen \textit{et al.} \cite{Kristensen2020AEmbrittlement} simulated crack growth in elastic-plastic solids considering the driving force to include only the elastic strain energy density (i.e., $\beta_p = 0$). On the other hand, Borden \textit{et al.} \cite{Borden2016AEffects} fixed $\beta_p = 1$ but considered the previously described plastic threshold $W_0$ to delay plastic damage. While, as discussed in Section \ref{Sec:Theory fracture}, a magnitude of $\beta_p = 0.1$ appears to be the most sensible choice from a physical viewpoint, as experiments show that 90\% of the plastic work dissipates into heat and is thus not available to create new cracks, we here numerically assess the influence of varying $\beta_p$. The results are shown in Fig. \ref{Fig: Bp}, in terms of both the force versus displacement response and the phase field contours, with red colour denoting fully cracked regions ($\phi \approx 1$) and blue colour denoting uncracked material points ($\phi \approx 0$). It can be seen that the higher the value of $\beta_p$, the lower the peak load, as the driving force is greater for the same remote load. More interestingly, the phase field $\phi$ contours reveal a transition from plastic localisation-driven failures for $\beta_p=0.75$ or higher, with damage localising at an angle of 45$^\circ$ relative to the initial crack, to a standard mode I fracture, with the crack growing along the 0$^\circ$ path, for smaller values of $\beta_p$. The $\beta_p=0.5$ case shows that while the crack propagates straight, along the mode I fracture path, a trail of non-zero damage is observed along the new crack due to plastic damage surrounding the advancing front; this trail is not observed for $\beta_p = 0$. In the following simulations, a value of $\beta_p = 0.1$ is chosen, unless otherwise stated, in order to capture a physically-based dissipation.

Now, once insight has been gained on the interplay between plasticity and fracture, we proceed to account for the role of hydrogen. Two values of the hydrogen degradation coefficient are considered, $\chi=0.89$ and $\chi=0.3$, as shown in Figs. \ref{fig:Xi_effect}a and Figs. \ref{fig:Xi_effect}b, respectively. The threshold for the moving chemical boundary conditions is established at $\phi_{th}=0.95$ and three values of the initial and boundary hydrogen concentration are assumed: 0.1, 0.5 and 1 wt ppm. The results show that ductility reduces with increasing hydrogen concentration, with this effect being less significant when $\chi=0.3$, as expected. For the case $\chi=0.89$ and hydrogen contents above 0.5 wt ppm, failure occurs in a very brittle fashion, with very little plastic ductility, if any. For both cases, it appears that increasing the concentration from 0.5 to 1 wt ppm has little effect; as shown in Fig. \ref{Fig: degradation}, the degradation law attains a plateau at around 0.5 wt ppm, for both $\chi=0.3$ and $\chi=0.89$, as the interface is close to being fully covered with hydrogen (for the segregation energy of $\Delta g_b^0 = 30$ kJ/mol, typical of grain boundaries).

\begin{figure}[H]
    \centering
    \makebox[\textwidth][c]{
        \begin{subfigure}[t]{0.55\textwidth}  
            \centering
            \includegraphics[width=1.1\textwidth]{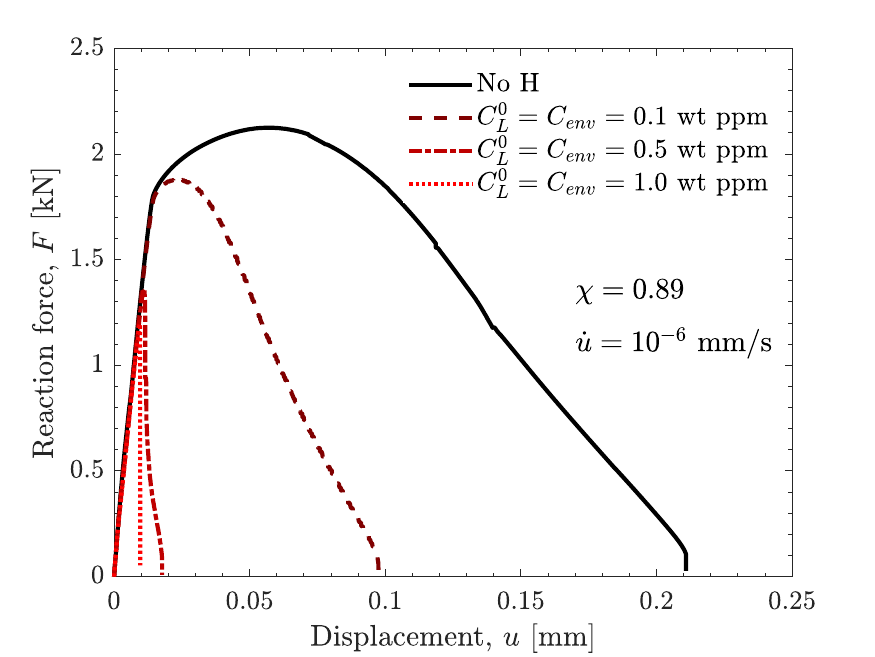}
            \subcaption{}
            \label{Fig: embrittlement ductile Xi 0.89}
        \end{subfigure}
        \begin{subfigure}[t]{0.55\textwidth} 
            \centering
            \includegraphics[width=1.1\textwidth]{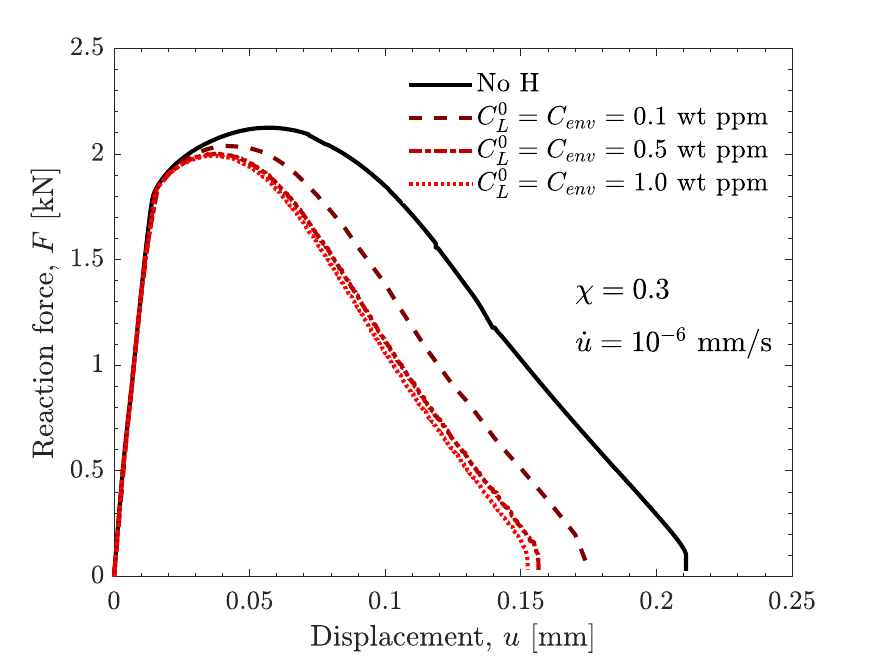}
            \subcaption{}
            \label{Fig: embrittlement ductile Xi 0.89}
        \end{subfigure}
    }
    \caption{Force versus displacement behaviour of an elastic-plastic single-edge tension square plate exposed to a hydrogen-containing environment, considering hydrogen damage coefficients of (a) $\chi=0.89$, as dictated by atomistics for iron-based materials, and (b) $\chi=0.3$. The results are shown for $G_c^0 = 10$ N/mm, a loading rate of $\dot{u} = 10^{-6}$ mm/s, and a segregation energy of $\Delta g_b^0 =$ 30 kJ/mol.}
    \label{fig:Xi_effect}
\end{figure}

\begin{figure}[H]
  \makebox[\textwidth][c]{\includegraphics[width=1.0\textwidth]{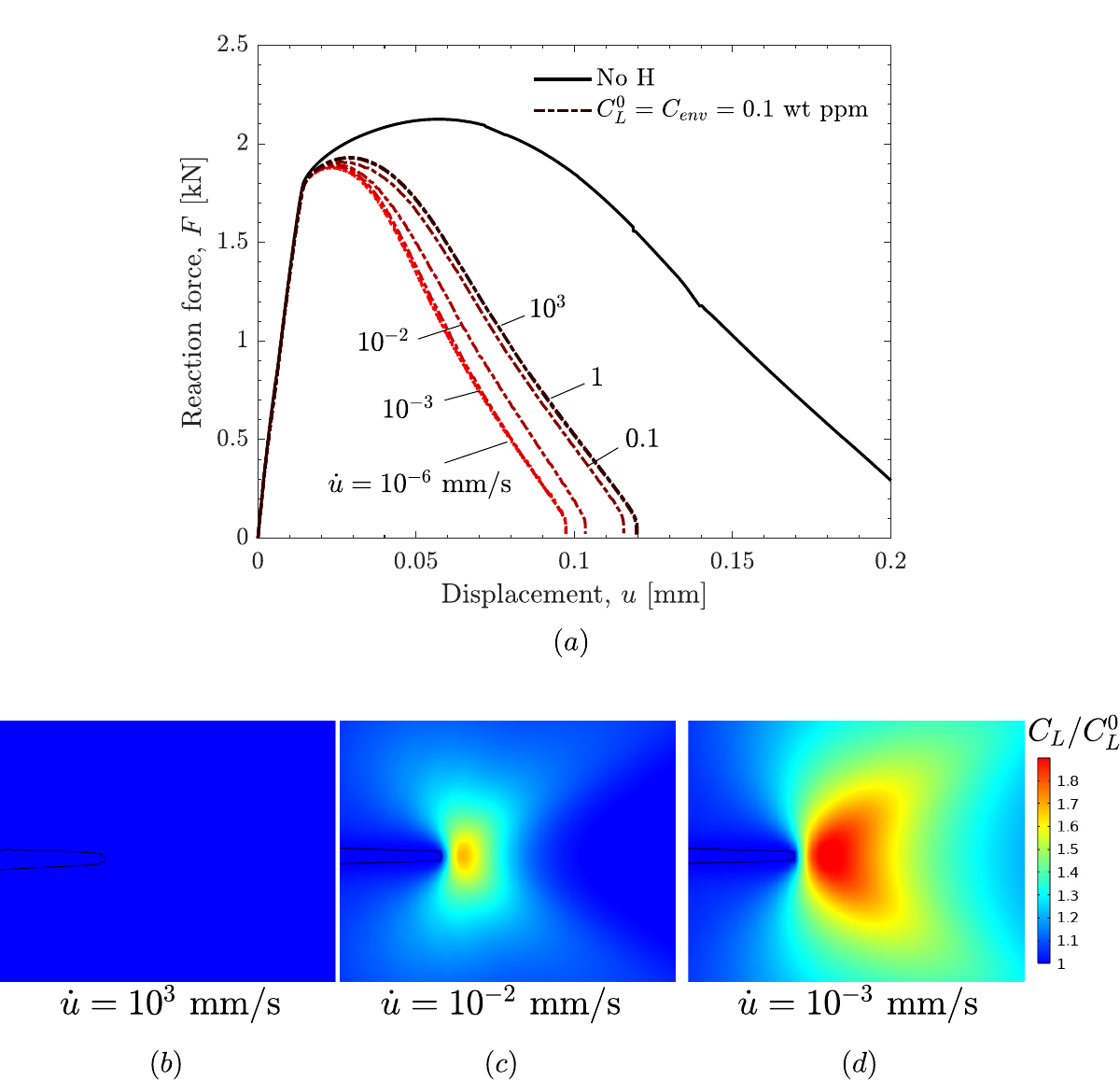}}%
  \caption{Exploring the sensitivity to the loading rate of an elastic-plastic single-edge tension square plate exposed to a hydrogen-containing environment: (a) force versus displacement response for selected choices of the applied displacement rate $\dot{u}$, and crack tip normalised lattice hydrogen concentration contours for (b) $\dot{u}=10^3$ mm/s, (c) $\dot{u}=10^{-2}$ mm/s, and (d) $\dot{u}=10^{-3}$ mm/s, which are representative of the fast, intermediate and slow loading rate regimes, respectively. The $C_L/C_L^0$ contours have been obtained at the point $u=0.04$ mm and a black solid line is used to denote the crack ($\phi=0.95$). Results obtained for $G_c^0 = 10$ N/mm, $\chi = 0.89$ and $\phi_{th}=0.95$.}
\label{fig:StrainRateSENT}  
\end{figure}

Hydrogen accumulates near the crack tip because of the hydrostatic stress effect and therefore a strain rate influence on embrittlement is expected: when the strain rate is high in comparison to the diffusion towards the stressed region, the corresponding $C_L$ peak does not build up and the reduction in fracture energy is only caused by the initial concentration $C_L^0$. On the contrary, for slow strain rates a stationary state can be reached and thus the degradation law is governed by $C_L^0 \text{exp}[\sigma_h\bar{V}_H/(RT)]$. The sensitivity to the loading rate is explored in Fig. \ref{fig:StrainRateSENT}, where the applied displacement rate $\dot{u}$ is varied from $10^{-6}$ to $10^3$ mm/s and the hydrogen concentration is fixed at $C_L^0=C_{env}=0.1$ wt ppm. The hydrogen-free curve is also obtained for comparison. Normalised lattice hydrogen concentration contours $C_L/C_L^0$ are also provided near the crack tip, to rationalise observations. In agreement with expectations, the results reveal three regimes of behaviour. For sufficiently fast loading rates, $\dot{u} > 0.1$ mm/s, hydrogen does not have time to diffuse (see Fig. \ref{fig:StrainRateSENT}b) and an increased fracture resistance is observed, albeit the behaviour is still more brittle than the case without hydrogen due to the role of the initial hydrogen content $C_L^0$ (the influence of pre-charging conditions on the rate susceptibility is addressed in Sections \ref{Subsec: BL hydrogen} and \ref{Sec: R-curves trapping effects}). At the other end are the cases where the loading rate is sufficiently slow $\dot{u} < 10^{-3}$ mm/s such that hydrogen has enough time to accumulate near the crack tip (see Fig. \ref{fig:StrainRateSENT}d), approaching the steady state. For loading rates in between, $10^{-3} < \dot{u} < 0.1$, some hydrogen accumulation is observed (see Fig. \ref{fig:StrainRateSENT}c), bringing in a reduction of ductility and an intermediate force versus displacement response.

\subsection{Boundary Layer: crack growth resistance}
\label{Sec:Boundary Layer}

\begin{figure}[H]
\makebox[\linewidth][c]{%
        \begin{subfigure}[b]{0.4\textwidth}
                \centering
                \includegraphics[scale=0.7]{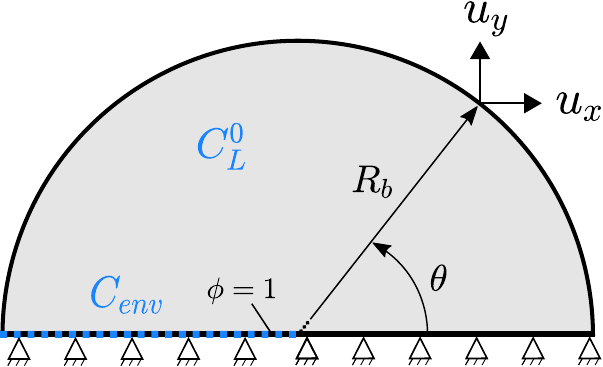}
                \caption{}
                \label{}
        \end{subfigure}
        \begin{subfigure}[b]{0.55\textwidth}
                \raggedleft
                \includegraphics[scale=0.18]{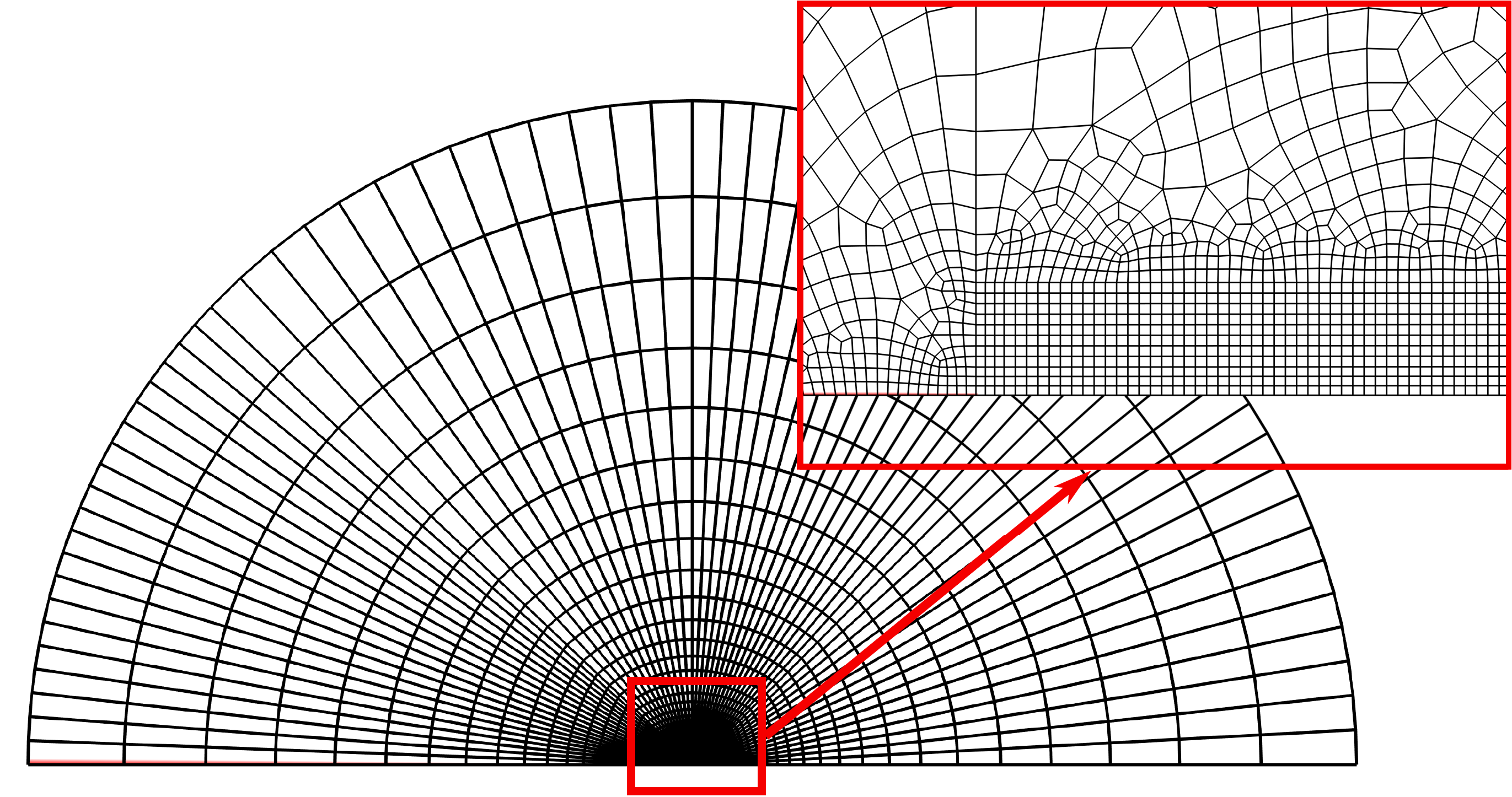}
                \caption{}
                \label{}
        \end{subfigure}}
        \caption{Numerical experiments on a boundary layer geometry where a remote $K_I$ is imposed: (a) scheme of the geometry and the boundary conditions, and (b) finite element mesh.}
        \label{Fig:schemeBL}
\end{figure}

We proceed to gain further insight into the interplay between plasticity, fracture and hydrogen diffusion (and trapping) by conducting virtual fracture experiments. To this end, a boundary layer model is used to apply a remote stress intensity factor $K_I$ and therefore readily obtain crack growth resistance curves, so-called R-curves. The geometry and loading configuration of the boundary layer model are shown in Fig. \ref{Fig:schemeBL}a. The Williams solution is used to prescribe the displacements of the nodes in the outer boundary. Thus, considering a polar coordinate system ($r$, $\theta$) centred at the crack tip, the horizontal and vertical components of the nodes located in the outer periphery ($r=R_b$) read:
\begin{equation}
    u_x(R_b,\theta)=K_I \frac{1+\nu}{E_0}\sqrt{\frac{R_b}{2\pi}}\cos\left(\frac{\theta}{2}\right)\left[2-4\nu+2 \sin^2{\left(\frac{\theta}{2}\right)}\right]
\end{equation}
\begin{equation}
    u_y(R_b,\theta)=K_I \frac{1+\nu}{E_0}\sqrt{\frac{R_b}{2\pi}}\cos\left(\frac{\theta}{2}\right)\left[4-4\nu+2 \cos^2{\left(\frac{\theta}{2}\right)}\right]
\end{equation}
where $K_I$ is the mode I stress intensity factor characterising the crack tip stress state. As in Refs. \cite{Sofronis1989NumericalTip, Krom1999HydrogenTip}, the outer radius is taken to be $R_b=0.15$ m, but results are insensitive to this choice, provided that $R_b$ is much larger than the plastic zone size and the fracture region. The mesh is refined along the crack propagation front, as shown in Fig. \ref{Fig:schemeBL}b, where the characteristic element size fulfils $h_e \le \ell/5$. The mesh consists of a total of 14,002 elements. A cubic discretization is considered for both displacement and damage problems, whereas linear discretization is chosen for hydrogen concentration.

Crack growth resistance curves are obtained in the boundary layer model by registering the applied $K_I$ ramp and the crack advance $\Delta a$ (with the crack front given by $\phi>0.95$). To facilitate the generalisation of our findings all results are normalised. The loading variable $K_I$ is divided by a reference $K_c^0$ that corresponds to the material fracture toughness, which for plane strain is related to $G_c^0$ as,
\begin{equation}
    K_c^0 = \sqrt{\frac{E_0 G_c^0}{1-\nu^2}}
\end{equation}
Similarly, a reference length $R_0$ for the fracture process zone can be defined \cite{Tvergaard1992TheSolids}:
\begin{equation}
    R_0 = \frac{E_0 G_c^0}{3\pi (1-\nu^2)\sigma_{y0}^2}
\end{equation}
which can be used to provide a normalised crack extension: $\Delta a/R_0$.

\subsubsection{R-curves without hydrogen}

\begin{figure}[H]
    \centering
    \makebox[\textwidth][c]{
        \begin{subfigure}[t]{0.55\textwidth}  
            \centering
            \includegraphics[width=1.1\textwidth]{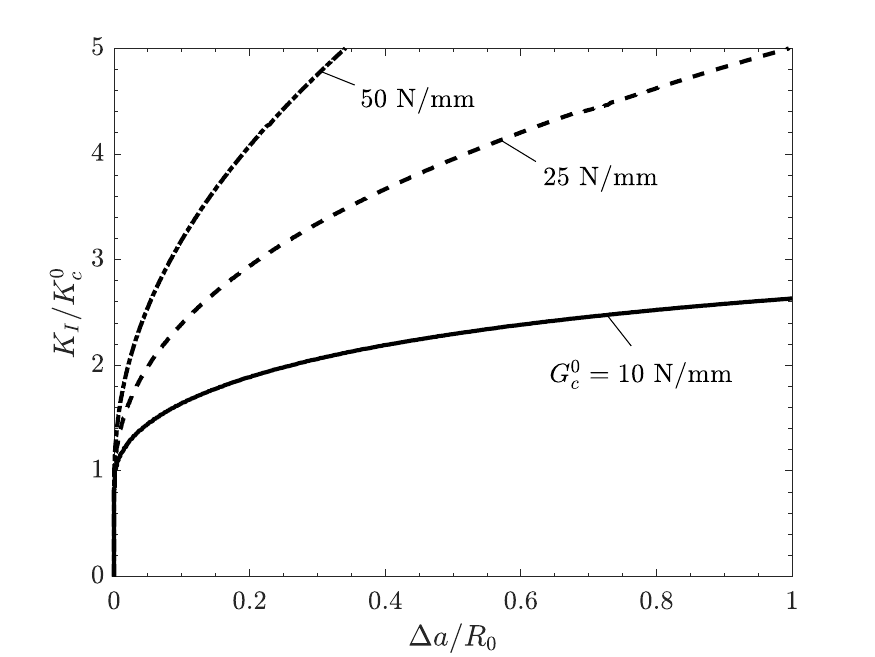}
            \subcaption{}
            \label{Fig: R-curves no H}
        \end{subfigure}
        \begin{subfigure}[t]{0.55\textwidth} 
            \centering
            \includegraphics[width=1.1\textwidth]{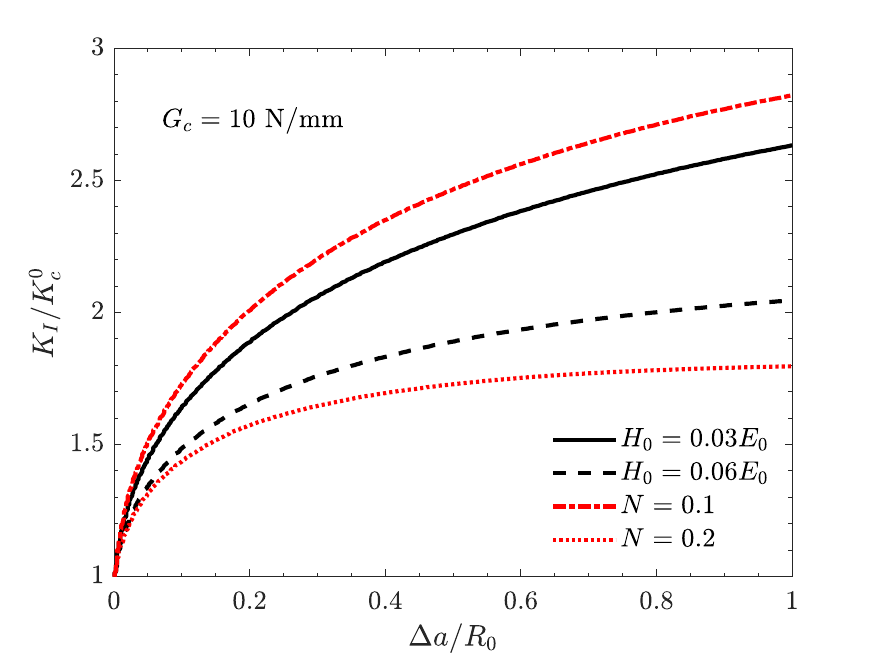}
            \subcaption{}
            \label{Fig: R-curves hardening}
        \end{subfigure}
    }
    \caption{Predictions of crack growth resistance in the absence of hydrogen and with $\ell = 0.05$ mm: (a) influence of the material toughness $G_c$, for $H_0 = 0.03E_0$, and (b) influence of work hardening, as characterised through the power law hardening exponent $n$ and the linear hardening modulus $H_0$.}
    \label{fig:R-curve Gc and hardening}
\end{figure}

We begin by exploring the interplay between plasticity and fracture in the absence of hydrogen. The elastic-plastic phase field fracture model here implemented captures the three characteristic stages of a crack growth resistance curve: crack blunting for $K_I<K_c^0$, initiation at approximately $K_I=K_c^0$, and stable crack propagation for $K_I>K_c^0$. Estimated R-curves are provided in Fig. \ref{fig:R-curve Gc and hardening}. The material properties correspond to those reported in Table \ref{Tab:mat_elastic_plastic}. First, in Fig. \ref{fig:R-curve Gc and hardening}a, we explore the sensitivity of the R-curves to the choice of $G_c^0$ for a material exhibiting linear hardening with hardening modulus $H_0 = 0.03E_0$. It can be seen that in all cases, the initiation of crack growth begins when $K_I$ is approximately equal to $K_c^0$, in agreement with well-established fracture mechanics theory. This result validates the predictive character of the present ductile, elastic-plastic phase field fracture approach in terms of a Griffith-like, energy balance criterion. Hence, Griffith's energy balance can be consistently extended to predict fracture in ductile, elastic-plastic metals, by selecting the material toughness ($G_c^0$) appropriately; i.e., a magnitude that corresponds to the experimentally measured toughness and therefore includes relevant inelastic phenomena, as first proposed by Orowan \cite{orowan1949fracture}. However, as previously discussed by Kristensen \textit{et al.}  \cite{kristensen2021assessment}, to observe the onset of crack growth at $K_I=K_c^0$, or equivalently $G_I = G_c^0$, one must define $\phi=1$ at the crack nodes (i.e., a phase field-induced crack) as geometrically-induced cracks introduce free surfaces that constraint phase field evolution due to the arising natural boundary condition, requiring $\nabla \phi \cdot \mathbf{n} = 0$. 

After the initiation of crack growth, the model predicts a rising R-curve with a decreasing slope, consistent with experiments. This is the result of plastic dissipation. For a purely elastic material, the curve becomes flat after $K_I=K_c^0$ and the fracture process is unstable, absent of any dissipative toughening mechanisms. It can be observed that the higher the magnitude of $G_c^0$, the steeper the rising R-curve. This can be understood from the relationship between $G_c^0$ and the strength $\sigma_c$, see Eq. (\ref{eq:sigmaC}). A higher $G_c^0$ value implies a higher strength and higher crack tip stresses are needed to evolve damage, allowing for the plastic zone to become fully developed. On the contrary, for small $G_c^0$ and $\sigma_c$ values, fracture is triggered more easily, at lower stresses, and plastic dissipation is then small compared to the work of the fracture process. In this way, the model naturally captures a more ductile behaviour for higher toughness values, in agreement with observations.\\

The influence of hardening is explored in Fig. \ref{fig:R-curve Gc and hardening}b. The linear and power law hardening models provide results that are qualitatively similar. In both cases, increasing the work hardening lowers the crack growth resistance, as the behaviour becomes closer to that of a linear elastic solid and less plastic dissipation takes place. The results are consistent with numerical R-curve experiments conducted with cohesive zone models \cite{Tvergaard1992TheSolids,martinez2018crack}.

\subsubsection{Hydrogen effects on crack growth resistance}
\label{Subsec: BL hydrogen}

\begin{figure}[H]
    \centering
    \makebox[\textwidth][c]{
        \begin{subfigure}[t]{0.55\textwidth}  
            \centering
            \includegraphics[width=1.1\textwidth]{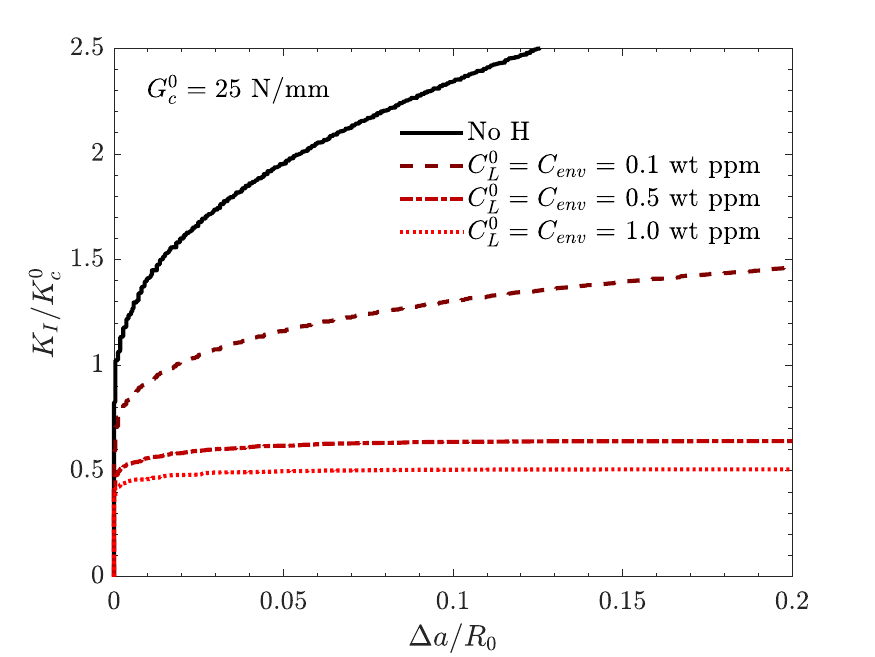}
            \subcaption{}
            \label{Fig: G25 Cinfluence}
        \end{subfigure}
        \begin{subfigure}[t]{0.55\textwidth} 
            \centering
            \includegraphics[width=1.1\textwidth]{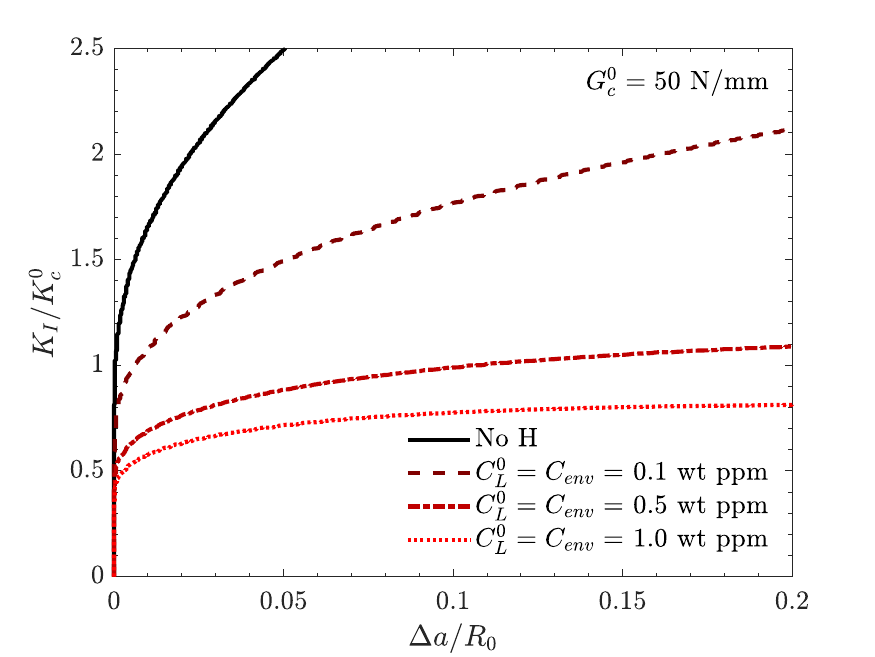}
            \subcaption{}
            \label{Fig: G50 Cinfluence}
        \end{subfigure}
    }
    \caption{Predictions of crack growth resistance in the presence of hydrogen for four concentration conditions (0, 0.1, 0.5 and 1 wt ppm) and two toughness values: (a) $G_c^0 = 25$ N/mm and (b) $G_c^0 = 50$ N/mm. Results obtained with $\chi=0.89$, $H_0 = 0.03E_0$ and $\phi_{th}=0.95$.}
    \label{Fig: R-curves H embrittlment Xi089}
\end{figure}

We shall now explore the effect of hydrogen on crack growth resistance. To this end, three different concentration levels are considered, as in the plate examples, taking the same initial and boundary concentration ($C_L^0=C_{env}$), a load rate of $\dot{K}_I = 0.1$ MPa $\sqrt{\text{m}}$/s and a degradation coefficient $\chi = 0.89$. The predictions are provided in Fig. \ref{Fig: R-curves H embrittlment Xi089}, for two choices of material toughness $G_c^0$: 25 N/mm (a) and 50 N/mm (b). The results show that an increase in hydrogen concentration shifts resistance curves to a brittle behaviour, i.e. the slope is reduced and the propagation initiates for $K_I$ values lower than $K_c^0$. It must be noted that the normalisation variables $K_c^0$ and $R_0$ are calculated using $G_c^0$, since the actual $G_c$ value is not uniform as it depends on the local $C_L$ concentration. Therefore, hydrogen-assisted crack propagation for $K_I<K_c^0$ does not involve a deviation from Griffith's criterion. Small hydrogen concentrations bring a notable reduction in crack growth resistance, due to the initial high slope in the hydrogen degradation law $f(C)$ for $\chi = 0.89$ (see Fig. \ref{Fig: degradation}), but it appears to saturate, as the hydrogen concentration is high enough to reach the plateau region of the degradation curve (Fig. \ref{Fig: degradation}). For a fracture energy of $G_c^0=$ 25 N/mm (Fig. \ref{Fig: R-curves H embrittlment Xi089}a), the highest concentrations (0.5 and 1 wt ppm) result in unstable fracture with almost no plastic toughening. The curves obtained for the higher toughness of $G_c^0=$ 50 N/mm (Fig. \ref{Fig: R-curves H embrittlment Xi089}a) exhibit more inelastic toughening and stable crack growth, appropriately capturing the more ductile behaviour of the material.

\begin{figure}[H]
\makebox[\linewidth][c]{%
        \begin{subfigure}[b]{0.5\textwidth}
                \centering
                \includegraphics[scale=0.6]{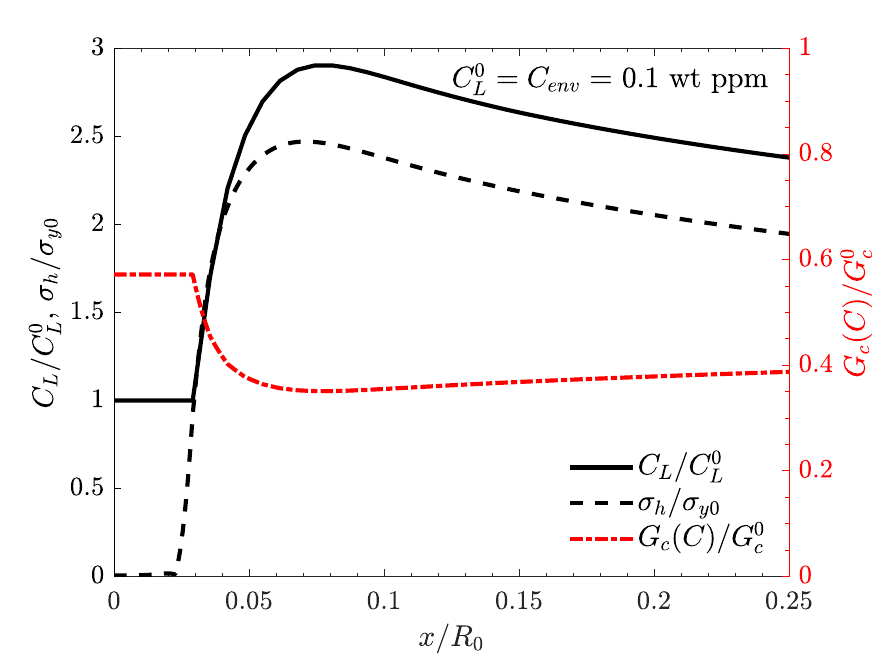}
                \caption{}
                \label{fig: distributions C01}
        \end{subfigure}
        \begin{subfigure}[b]{0.5\textwidth}
                \raggedleft
                \includegraphics[scale=0.6]{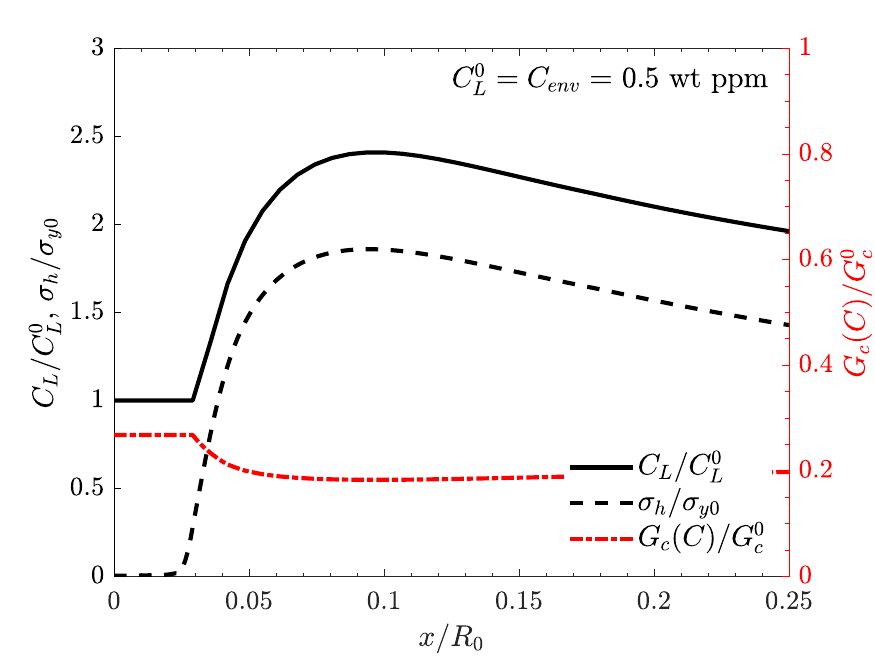}
                \caption{}
                \label{fig: distributions C05}
        \end{subfigure}}
        \begin{subfigure}[b]{1.0\textwidth}
                \centering
                \includegraphics[scale=0.6]{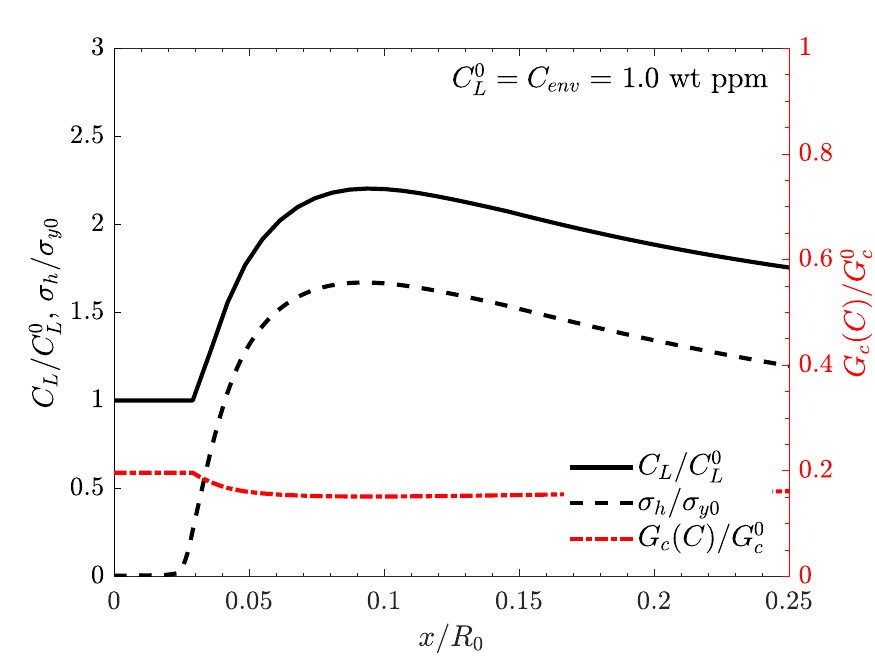}
                \caption{}
                \label{fig: distributions C1}
        \end{subfigure}
        \caption{Crack tip distribution of normalised hydrogen concentration ($C_L/C_L^0$), hydrostatic stress ($\sigma_h/\sigma_y^0$) and fracture energy ($G_c/G_c^0$) at $\Delta a = 0.02R_0$ for different concentrations (a) $C_L^0 = 0.1$ wt ppm; (b) $C_L^0 = 0.5$ wt ppm; (c) $C_L^0 = 1.0$ wt ppm. Results obtained with $\chi=0.89$, $H_0 = 0.03E_0$ and $\phi_{th}=0.95$.}
        \label{Fig: distributions}
\end{figure}

As a consequence of hydrogen embrittlement, the crack tip opening during propagation is reduced. Therefore, higher concentrations result in lower toughnesses and sharper cracks. The local distributions of $G_c$ ahead of an advancing crack are plotted in Fig. \ref{Fig: distributions} (right axis) for different $C_L^0$ values and at $\Delta a = 0.02R_0$, approximately. Distributions of normalised concentration and (damaged) hydrostatic stress are also plotted in Fig. \ref{Fig: distributions}. Despite cracks becoming sharper with increasing hydrogen content, the hydrostatic and lattice hydrogen content peaks decrease, as can be seen by comparing Figs. \ref{Fig: distributions}a and \ref{Fig: distributions}c. This is due to the higher degree of embrittlement and thus lower $K_I$ needed to propagate the crack. The results shown in Figs \ref{Fig: distributions} also reveal that the approach employed to capture the moving chemical boundary is working well. It can be seen that, in the cracked region, $C_L=C_L^0$ (and thus $C_L=C_{env}$). The results also show that the model appropriately captures the drop in stress in the fully cracked regions ($\phi=1$), where $\sigma_h$ vanishes. The crack tip $C_L$ distributions follow those of $\sigma_h$ and the results show that choosing to adopt the damaged hydrostatic stress to drive diffusion leads to consistent trends. Smooth $\sigma_h$ distributions are obtained when a cubic discretization is assigned to the displacement field; a quadratic-order discretization also gives accurate results if the appropriate hydrostatic stress mapping is performed. For a linear discretization of the displacement, a mixed formulation can be used to avoid volumetric locking and spurious hydrostatic stress distribution, as discussed in Part I of this work \cite{PartI}. 

\begin{figure}[H]
  \makebox[\textwidth][c]{\includegraphics[width=0.8\textwidth]{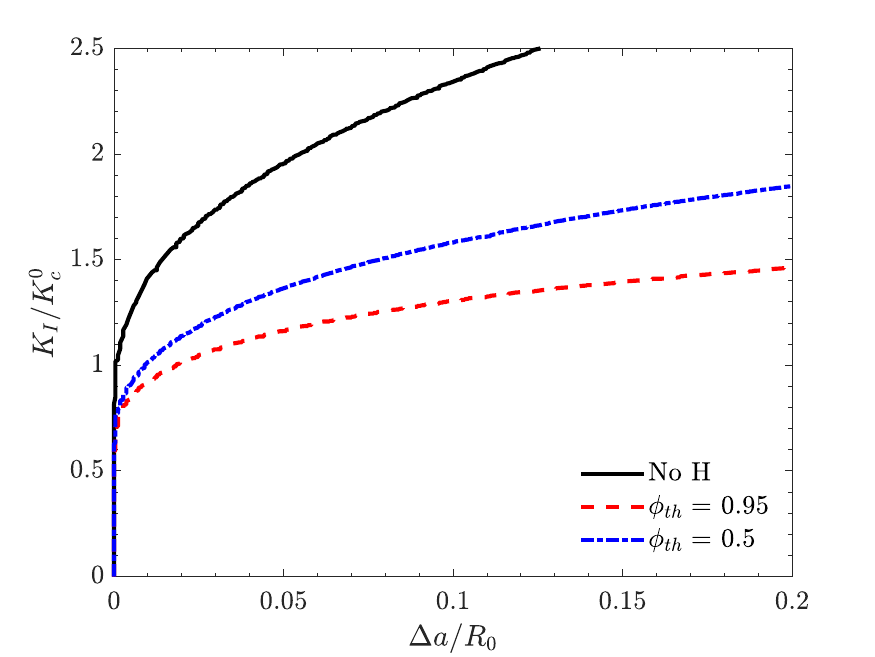}}%
  \caption{Influence of the threshold coefficient $\phi_{th}$ for the prediction of crack growth resistance in materials exposed to hydrogen-containing environments. Results obtained for $G_c^0 = 25$ N/mm, $\chi=0.89$, $H_0 = 0.03E_0$ and $C_L^0 = 0.1$ wt ppm.}
\label{Fig: R-curves movBCs}   
\end{figure}

\begin{figure}[H]
  \makebox[\textwidth][c]{\includegraphics[width=1.0\textwidth]{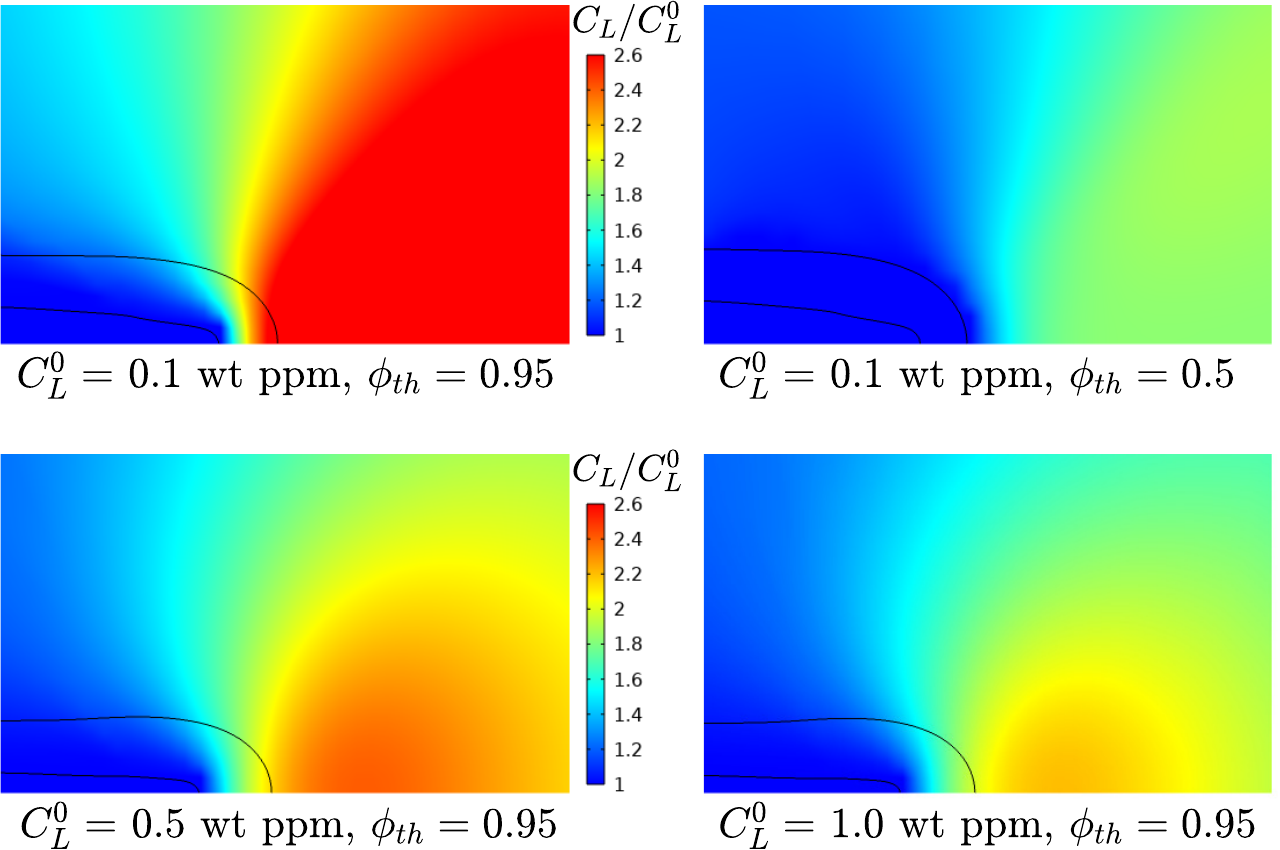}}%
  \caption{Contours of the normalised hydrogen concentration, $C_L/C_L^0$, for three different initial concentrations at $\Delta a =$ 0.05 mm. Two values of the threshold coefficient are considered, $\phi_{th}=0.5$ and $\phi_{th}=0.95$. Black solid lines denote contours for $\phi=0.5$ and $\phi=0.95$. The choice of a lower $\phi_{th}$ translates into the condition $C_L=C_{env}$ (=$C_L^0$) being enforced over a greater region, decreasing the role that $\sigma_h$ plays in elevating $C_L$ ahead of the crack. Results obtained for $G_c^0 = 25$ N/mm, $\chi=0.89$, and $H_0 = 0.03E_0$.}
\label{Fig:CL_Boundary_Layer}   
\end{figure}

So far the threshold coefficient that determines when the hydrogen-containing environment finds its way through a cracked region has been fixed at $\phi_{th}=0.95$. The influence of this choice is assessed in Fig. \ref{Fig: R-curves movBCs} by computing R-curves for two choices of $\phi_{th}$: 0.5 and 0.95. The choice of $\phi_{th}=0.95$ is more conservative and delivers a more brittle response. For the $\phi_{th}=0.5$ case, the enhanced diffusivity approach presented in Section \ref{Sec:theory hydrogen transport} enforces $C_L=C_L^0$ over a greater region, and thus reduces the role of the hydrostatic stress in elevating hydrogen concentrations ahead of cracks. This is shown in Fig. \ref{Fig:CL_Boundary_Layer}, where normalised concentration contours ahead of a crack that has extended 0.05 mm are shown for three selected environments (0.1, 0.5 and 1 wt ppm) and the two thresholds considered ($\phi_{th}=0.5$ and $\phi_{th}=0.95$). The sensitivity to the choice of the threshold coefficient $\phi_{th}$ should decrease with decreasing $\ell$, as the interface (fracture process zone) becomes smaller.   

\begin{figure}[H]
  \makebox[\textwidth][c]{\includegraphics[width=0.8\textwidth]{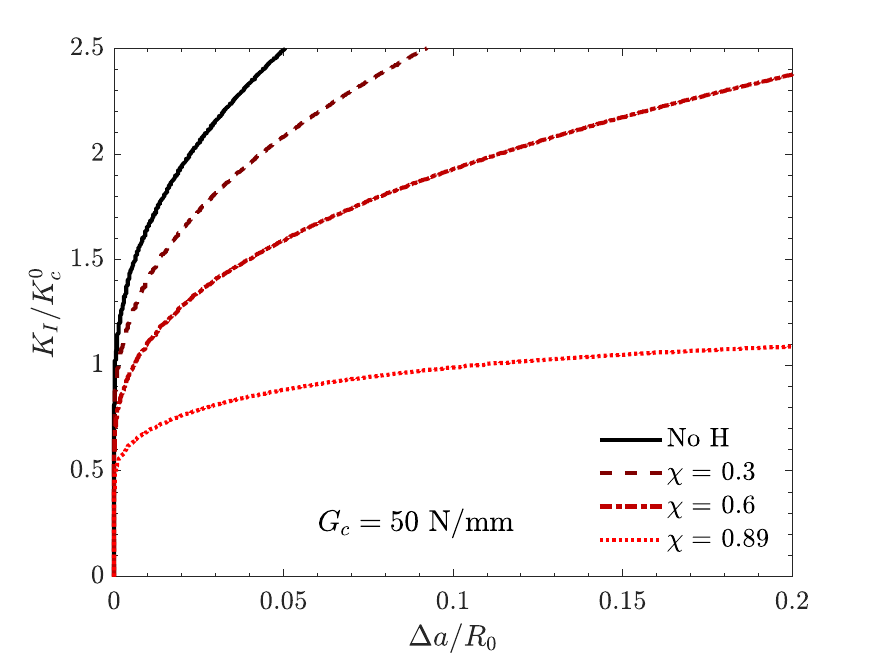}}%
  \caption{Influence of the hydrogen degradation parameter $\chi$ on the prediction of crack growth resistance considering the coverage-based degradation law and $C_L^0 = 0.5$ wt ppm. Results obtained for $G_c^0 = 50$ N/mm, $\phi_{th}=0.95$ and $H_0 = 0.03E_0$.}
\label{Fig: R-curves Xi effect}   
\end{figure}

Next, we proceed to quantify the effect of the hydrogen degradation coefficient, as shown in Fig. \ref{Fig: R-curves Xi effect}. R-curves are computed for three choices of $\chi$: 0.3, 0.6 and 0.89, with the last one corresponding to the atomistically-informed value for iron-based materials. Embrittlement increases with higher $\chi$ values, as expected. Experimental curves could be fitted by a specific coverage-based degradation law, i.e. $\chi$ and $\Delta g_b^0$ values, but the present framework also accommodates any empirical form of $f(C)$.

\begin{figure}[H]
        \begin{subfigure}[h]{1\textwidth}
                \centering                \includegraphics[scale=0.8]{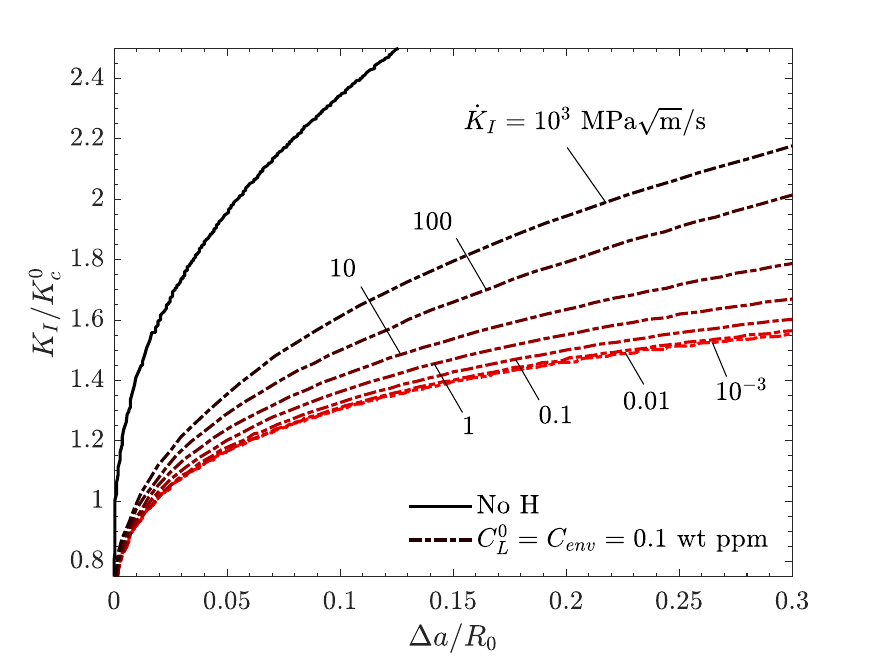}
                \caption{}
                \label{Fig: BL srate}
        \end{subfigure}\\
        \begin{subfigure}[h]{1\textwidth}
                \centering               \includegraphics[scale=0.8]{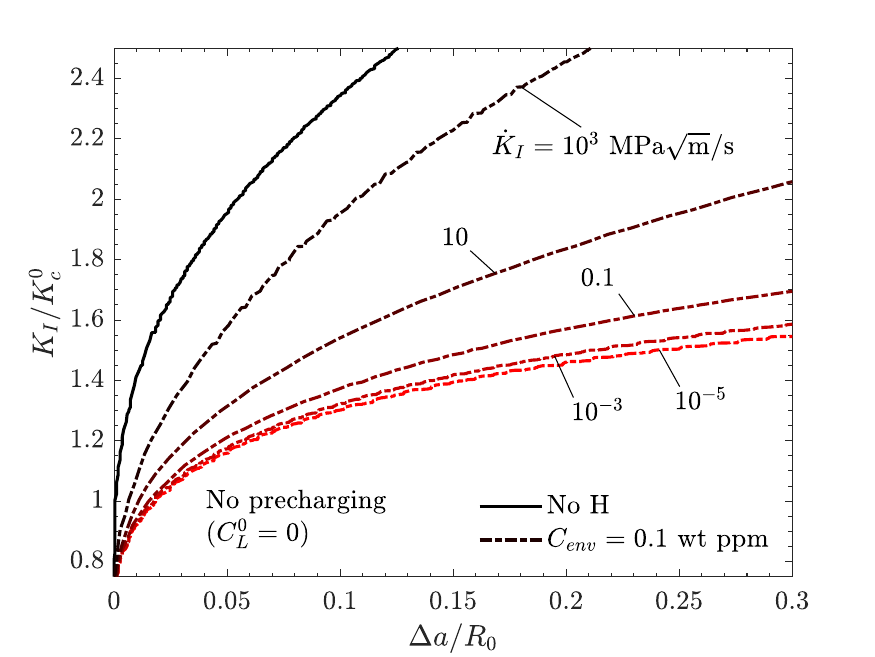}
                \caption{}
                \label{Fig: BL srate noC0}
        \end{subfigure}       
        \caption{Influence of the loading rate on the crack growth resistance of a hydrogen-exposed metal considering (a) a pre-charged condition, where $C_L^0=C_{env}$, and (b) no pre-charging, where $C_L^0=0$. Results obtained for $G_c^0 = 25$ N/mm, $\phi_{th}=0.95$ and $H_0 = 0.03E_0$.}
\label{Fig: R-curves H embrittlment K rate}  
\end{figure}

The sensitivity to the loading rate is another important effect. Hydrogen accumulation near the crack front, as displayed in Fig. \ref{Fig: distributions}, governs hydrogen-assisted cracking and thus transient effects must be taken into account. Crack growth resistance curves are obtained for a wide range of loading rates and two scenarios: a pre-charged sample (Fig. \ref{Fig: R-curves H embrittlment K rate}a), where $C_L^0=C_{env}$, as it has been considered so far, and a scenario where the sample is initially free from hydrogen, $C_L^0=0$ (Fig. \ref{Fig: R-curves H embrittlment K rate}b). In both cases, when the loading rate is slow enough, steady state crack tip $C_L$ distributions are reached and the maximum embrittlement level occurs. On the other hand, for extremely fast loads, the reduction of the resistance curve in comparison to the absence of hydrogen is caused by the pre-charging concentration $C_L^0$. This effect is analogous to the phenomena previously discussed and shown in Fig. \ref{fig:StrainRateSENT} for the single-edge cracked plate. For the boundary layer example, the initially empty condition ($C_L^0=0$) is also assessed (Fig. \ref{Fig: R-curves H embrittlment K rate}b). In this case, the influence of the loading rate is stronger and the R-curves tend to the condition without hydrogen when the loading rate is high enough.  

\subsubsection{Trapping effects}
\label{Sec: R-curves trapping effects}

Finally, we assess the influence of microstructural traps with our phase field-based framework, which combines (for the first time) plasticity and multi-trapping. First, to anticipate the effect of trapping binding energy and density, the ratio $D_L/D_{eff}$ is determined assuming thermodynamic equilibrium:
\begin{equation}
\label{Eq. Deff}
    \frac{D_L}{D_{eff}}=1+\frac{K_T N_T/N_L}{[1+(K_T-1)C_L^0/N_L]^2}
\end{equation}

The inverse of this operational diffusivity corresponding to $C_L^0$ is plotted in Fig. \ref{Fig: Deff for trapping effects. The minimum in $D_{eff}/D_L$ } 
indicates the greatest delay in hydrogen diffusion due to trapping effects for the considered $N_T$ and $E_B$ values. $D_{eff}$ is always lower than $D_L$, but a stronger trap binding energy does not always result in an increased delay: for the right part of the curves, a higher $E_B$ produces the saturation of traps and a reduction in the delay effect. The location of regimes highly depends on the concentration level: a low lattice occupancy shifts the $D_{eff}/D_L$ minimum towards high binding energies. The analysed $C_L^0 = 0.1$ wt ppm is equivalent to a lattice occupation of $\theta_L = 9.22\times10^{-7}$.

\begin{figure}[H]
  \makebox[\textwidth][c]{\includegraphics[width=0.8\textwidth]{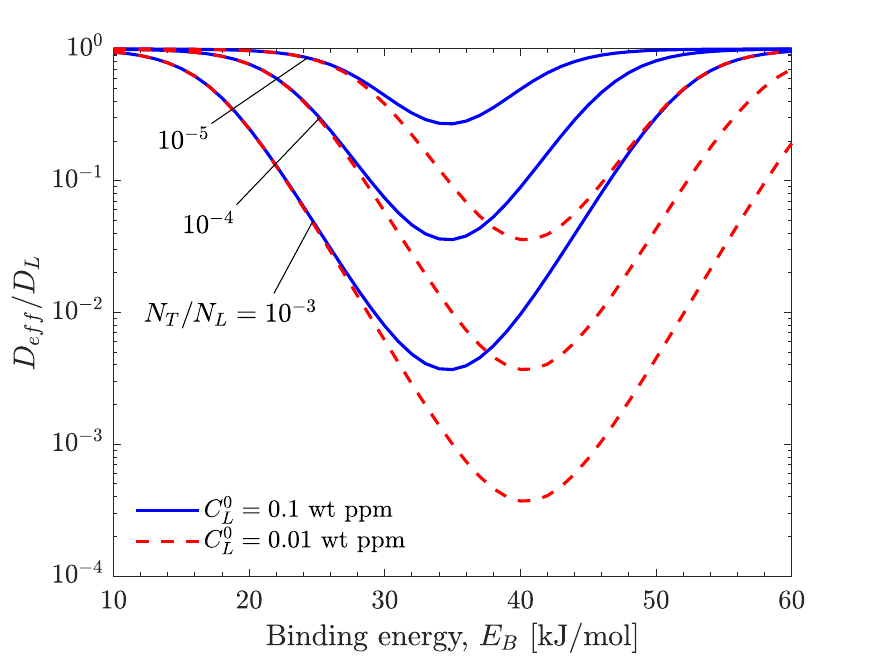}}%
  \caption{Effective diffusivity ratio as a function of binding energy and trap densities for different $C_L^0$ values.}
\label{Fig: Deff for trapping effects}  
\end{figure}

\begin{figure}[H]
  \makebox[\textwidth][c]{\includegraphics[width=0.8\textwidth]{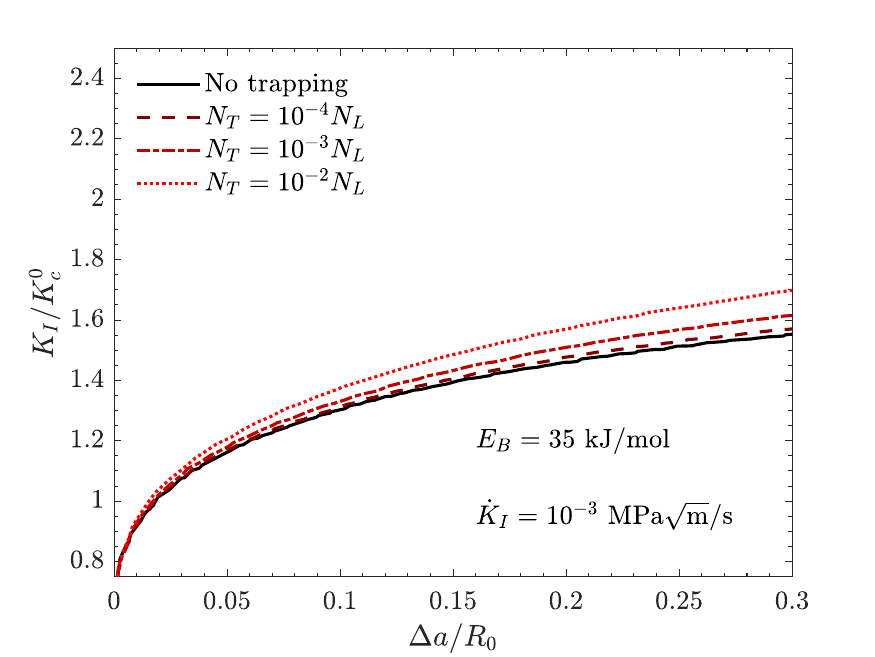}}%
  \caption{Influence of trap density on the crack growth resistance of metals exposed to hydrogen-containing environments. Results obtained for one trap with binding energy of 35 kJ/mol, $G_c^0 = 25$ N/mm, $\phi_{th}=0.95$ and $H_0 = 0.03E_0$.}
\label{Fig: R-curves H embrittlement NT}  
\end{figure}

The impact of varying the trap density on crack growth resistance is evaluated in Fig. \ref{Fig: R-curves H embrittlement NT}, where four values of $N_T$ are considered, including the reference $N_T=0$ case. In all cases, a homogeneous distribution of traps is assumed. As predicted by the $D_{eff}/D_L$ plot, a higher $N_T$ delays diffusion and thus reduces hydrogen embrittlement. The binding energy is fixed as 35 kJ/mol in Fig. \ref{Fig: R-curves H embrittlement NT} since this is the value maximising trapping effects for $C_L^0 = 0.1$ wt ppm. Nevertheless, the effect of trapping appears to be small. 

\begin{figure}[H]
        \begin{subfigure}[h]{1\textwidth}
                \centering                \includegraphics[scale=0.8]{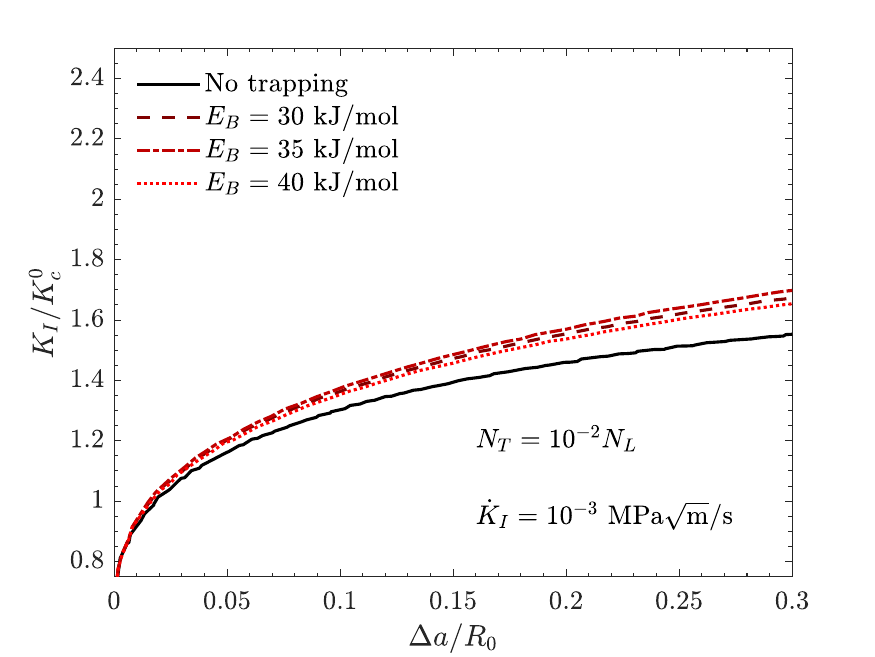}
                \caption{}
                \label{Fig: R-curves H embrittlement EB}
        \end{subfigure}\\
        \begin{subfigure}[h]{1\textwidth}
                \centering               \includegraphics[scale=0.8]{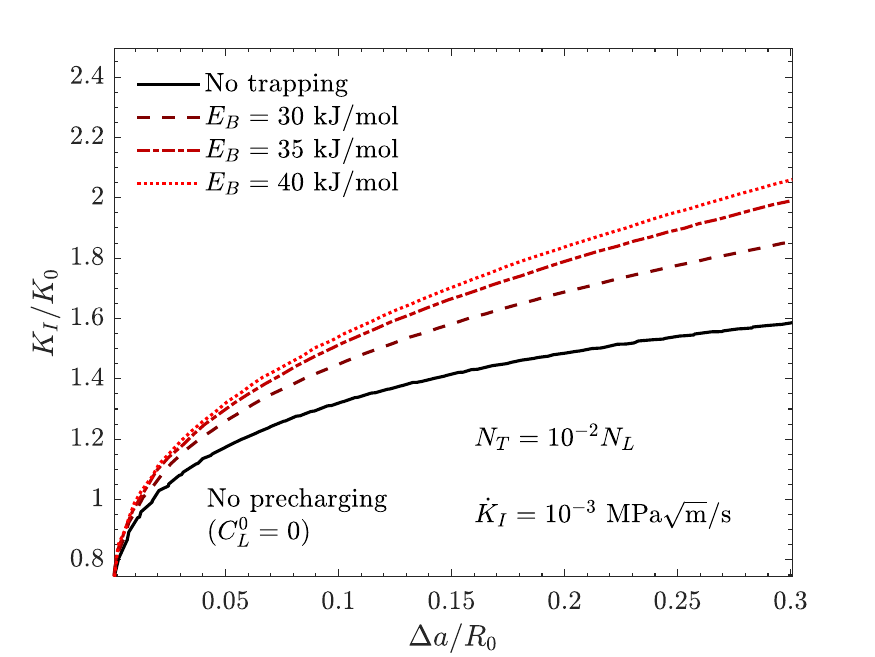}
                \caption{}
                \label{Fig: R-curves H embrittlement EB noC0}
        \end{subfigure}       
        \caption{Influence of binding energy on the crack growth resistance of metals exposed for a fixed trap density $N_T=10^{-2} N_L$. Two scenarios are considered: (a) a sample pre-charged with hydrogen, $C_L^0=C_{env}=0.1$ wt ppm; (b) a sample that has not been previously exposed to hydrogen ($C_L^0=0$, $C_{env}=0.1$ wt ppm). Results are obtained with $G_c^0 = 25$ N/mm, $\phi_{th}=0.95$ and $H_0 = 0.03E_0$.}
\label{Fig: R-curves E_B effect}  
\end{figure}

The effect of the trap binding energy, for a fixed trap density of $N_T = 10^{-3}N_L$, is evaluated in Fig. \ref{Fig: R-curves E_B effect}. Two scenarios are considered: a pre-charged sample, with $C_L^0=C_{env}$ (Fig. \ref{Fig: R-curves E_B effect}a), and a sample with no initial hydrogen, $C_L^0=0$ (Fig. \ref{Fig: R-curves E_B effect}b). Consider first the pre-charged scenario, Fig. \ref{Fig: R-curves E_B effect}a. The results show that increasing $E_B$ from 30 kJ/mol to 35 kJ/mol produces a delayed hydrogen accumulation in the fracture process zone and thus a lower embrittlement effect, shifting the crack growth resistance curves towards a ductile behaviour. However, this effect is maximum for $E_B = 35$ kJ/mol, the value corresponding to the minimum in $D_{eff}/D_L$, but weaker for $E_B = 40$ kJ/mol. The results obtained for the simulations without hydrogen pre-charging, 
Fig. \ref{Fig: R-curves H embrittlement EB noC0}, show that the trapping influence is stronger if pre-charging is not considered. This is explained also by trapping equilibrium at low concentrations, when $D_{eff}$ is lower. In addition, the minimum $D_{eff}/D_L$ is shifted to higher binding energies and therefore the $E_B=$ 40 kJ/mol case results in the most ductile curve in contrast to the pre-charging condition with $C_L^0=$ 0.1 wt ppm.

\begin{figure}[H]
  \makebox[\textwidth][c]{\includegraphics[width=0.8\textwidth]{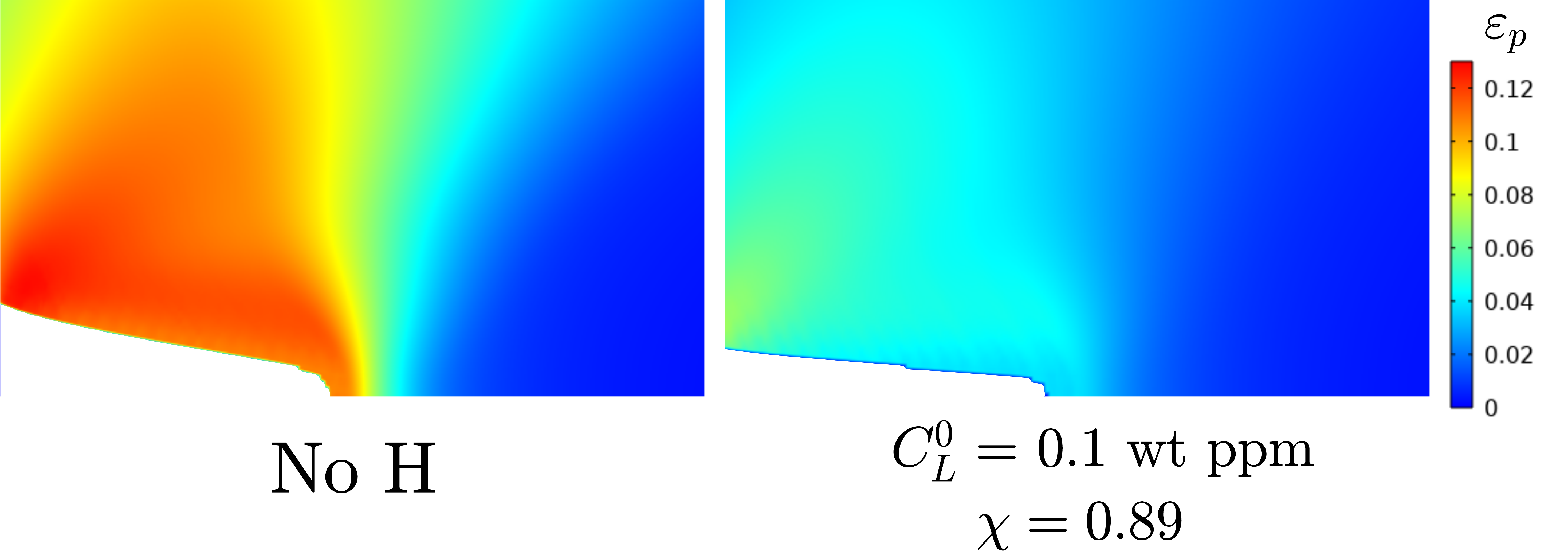}}%
  \caption{Equivalent plastic strain for $G_c^0 =$ 25 N/mm at $\Delta a = 0.18$ mm, without hydrogen (left) and with hydrogen (right). The cracked domain ($\phi>0.95$) has been removed for clarity.}
\label{Fig: Plastic deformation crack front}    
\end{figure}

The influence of non-homogeneous trap density as a function of plastic strain, i.e. following Eq. (\ref{Eq: N_T function of rho}) is also assessed. When considering usual values for dislocation densities, $\rho_0=10^{10}$ m$^{-2}$ and $\gamma=10^{16}$ m$^{-2}$ \cite{Dadfarnia2011HydrogenEmbrittlement}, a negligible influence on R-curves has been obtained. This is caused by the low plastic deformation developed during crack propagation. As observed in Fig. \ref{Fig: Plastic deformation crack front}, the equivalent plastic strain is even lower when hydrogen degradation is implemented and thus trapping effects due to dislocations will be minor during propagation for the analysed fracture energy and length scale. The low plastic deformation is also explained by the assumption made for the ductile phase field model: $\beta_p = 0.1$ produces a plastic degradation function $h(\phi)$ close to the nominal approach, as shown in Fig. \ref{Fig: plastic degradation}, and thus plastic flow is limited damage onset \cite{Marengo2023AFracture}. The maximum value of $\varepsilon_p = 0.12$ corresponds to a $N_T$ equal to $2.33\times 10^{-5}N_L$, resulting in a very slight trapping effect for the $C_L^0$ and $E_B$ values considered.

\begin{figure}[H]
  \makebox[\textwidth][c]{\includegraphics[width=0.8\textwidth]{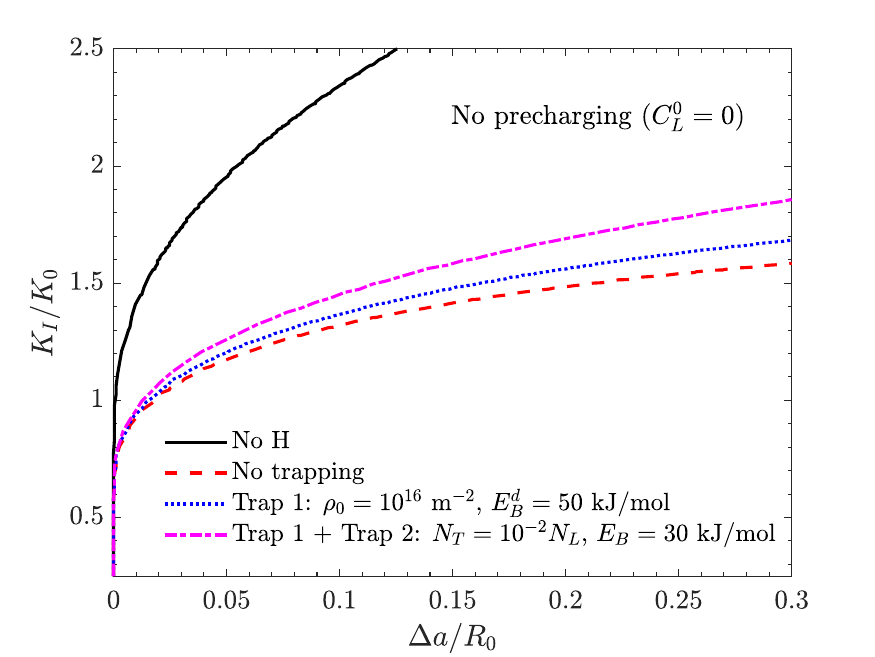}}%
  \caption{Influence of trapping on the crack growth resistance of metals. Results obtained for an initially hydrogen-free metal ($C_L^0=0$) and two trap scenarios: one with a single trap, that evolves in time as is associated with dislocations (with $E_B^d=50$ kJ/mol), and another one with this dislocation trap type and a 'static' trap with $E=30$ kJ/mol and $N_T=10^{-2}N_L$. Results are obtained with $G_c^0 = 25$ N/mm, $\phi_{th}=0.95$ and $H_0 = 0.03E_0$.}
\label{Fig: R-curves H embrittlement multitrap}  
\end{figure}

For the concentration value here simulated, $C_{env} = 0.1$ wt ppm, trapping at dislocations only influences embrittlement when a high initial dislocation density is assumed, e.g. $\rho_0 = 10^{-16}$ m$^{-2}$, as shown in Fig. \ref{Fig: R-curves H embrittlement multitrap}. Different trapping sites can also be considered at the same time by using the multi-trap scheme described in Section \ref{Sec:theory hydrogen transport}. Fig. \ref{Fig: R-curves H embrittlement multitrap} shows the resistance curve corresponding to the combined effects of two traps: the dislocation trapping site with $\rho_0 = 10^{-16}$ m$^{-2}$ and $E_B^d = 50$ kJ/mol (Trap 1) and a defect with a constant trapping density and a lower binding energy, $E_B=30$ kJ/mol (Trap 2). It can be observed that considering only dislocation trapping, the effect on crack growth resistance is minimal, despite assuming an initially hydrogen-free sample. This is due to the low levels of plastic deformation that are attained ahead of the crack tip; while these are relatively larger for the initiation of crack growth, they reduce in relevance as the crack propagates.

\subsection{3D case study: failure of a vessel}
\label{Sec:3D example}

The extension of the present framework to 3D modelling is straightforward: three convection velocity components are incorporated to capture stress-drifted diffusion. Other features of the model (e.g., the phase field implementation) do not need modification. A pressurised vessel with a vent pipe is simulated as a case study to show the capabilities of the present model to predict hydrogen embrittlement in engineering problems. The inner radius of the vessel is equal to 1 m and the thickness equals 90 mm. The side pipe has the same thickness and a 70-mm inner radius. A ramp pressure is applied to all the internal surfaces until unstable fracture occurs. Only half of the pipe is modelled and the corresponding symmetry boundary conditions are applied. To capture the vessel constraint a longitudinal traction equal to $\sigma_l = pR_v/(2t_v)$ is applied at the boundary surface, where $p$ is the applied pressure, $R_v$ is the vessel radius and $t_v$ is the vessel thickness. Material parameters have the same values as the previous 2D examples (Tables \ref{Tab:mat_elastic_plastic} and \ref{Tab:mat_diffusion}). However, the value of the length scale is increased to $\ell=20$ mm for convenience and $G_c^0$ is fixed to result in a material strength $\sigma_c = 2.5\sigma_{y0}$ in the absence of hydrogen. A single-pass staggered scheme is adopted, unlike the previous case studies.\\

\begin{figure}[H]
  \makebox[\textwidth][c]{\includegraphics[width=1\textwidth]{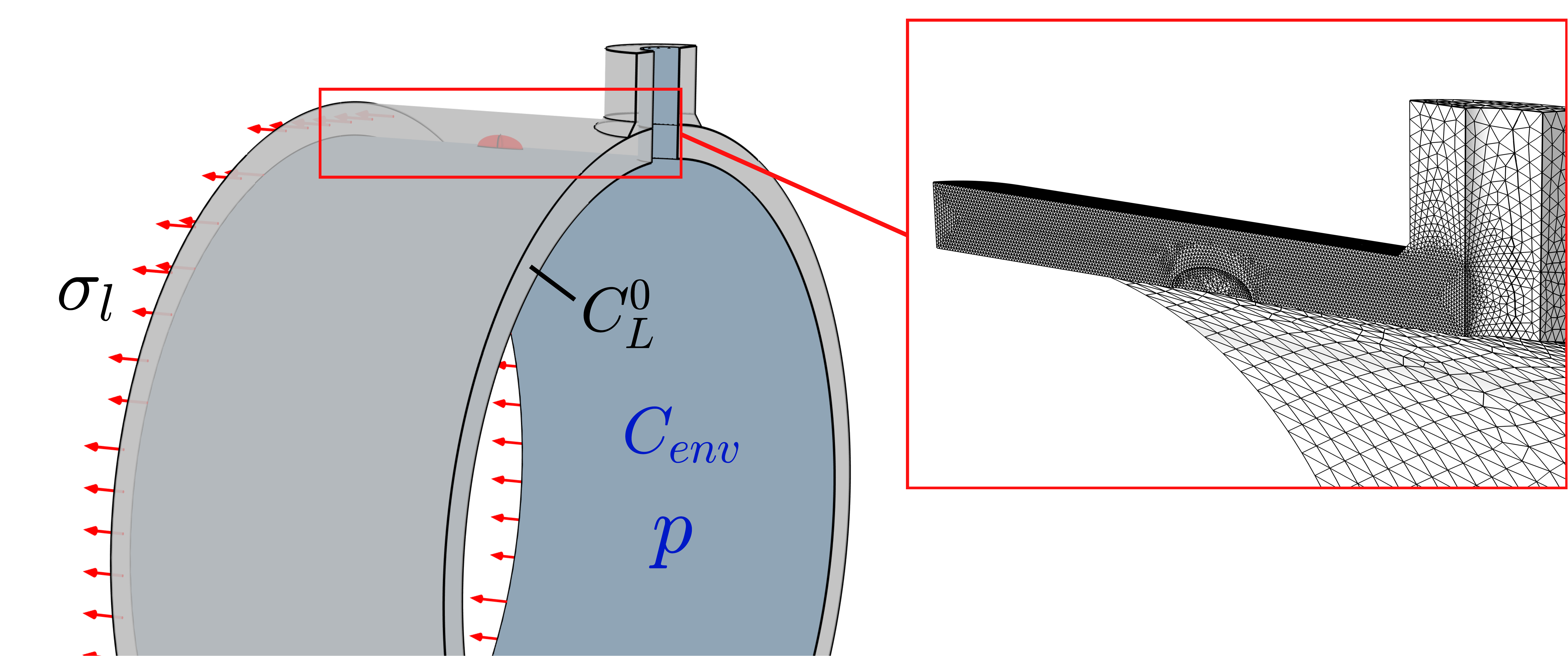}}%
  \caption{Geometry and mesh of the 3D case study reproducing a high-pressure vessel with a vent pipe. An inner ramp pressure $p$ is considered and the corresponding longitudinal traction $\sigma_l$ is applied at one vessel edge. Hydrogen entry from the H$_2$ environment($C_{env}$) is assumed through the inner surfaces and an inital concentration ($C_L^0$), for some cases. Symmetry conditions are applied to simplify the model and the mesh is refined near the expected crack initiation (vent pipe corner or longitudinal crack).}
\label{Fig: 3D concentration}   
\end{figure}

\begin{figure}[H]
  \makebox[\textwidth][c]{\includegraphics[width=1\textwidth]{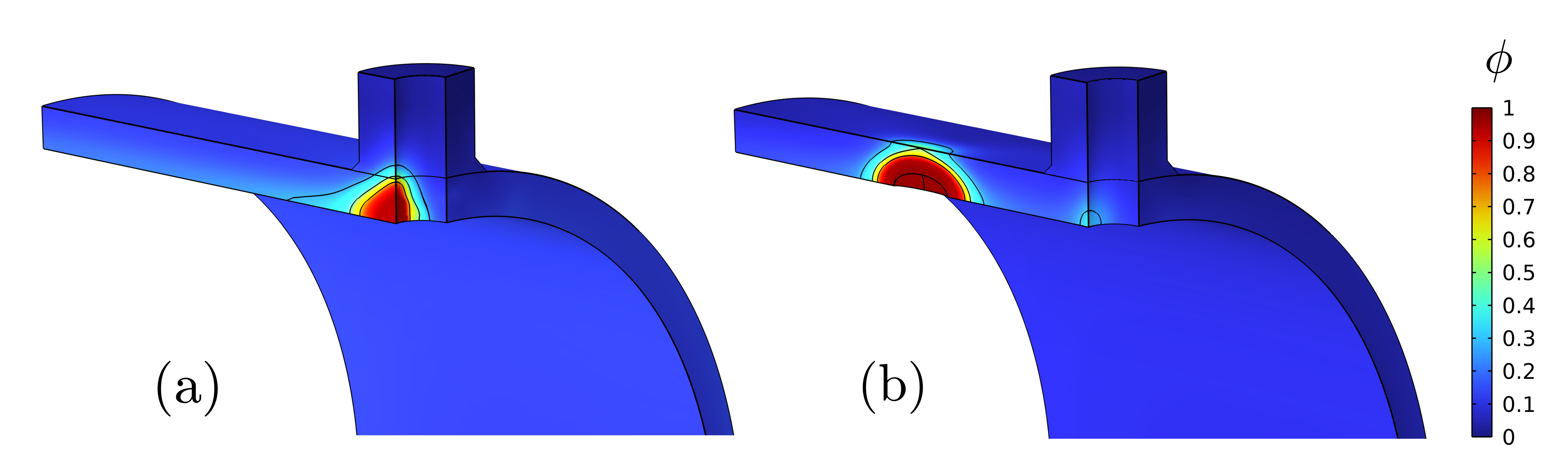}}%
  \caption{3D analysis of component failure, predictions in the absence of hydrogen. The results show contours of damage evolution ($\phi$) and black lines for $\phi=$ 0.25, 0.5 and 0.75, for (a) an uncracked pressurised vessel with a pipe, and (b) for the same vessel containing an elliptical crack.}
\label{Fig: 3D crack vs no crack}   
\end{figure}

First, we begin by exploring the behaviour of the component in the absence of hydrogen, as shown in Fig. \ref{Fig: 3D crack vs no crack}. Two scenarios are considered, one where the component is defect-free (Fig. \ref{Fig: 3D crack vs no crack}a), and one where a longitudinal elliptical crack with depth $a=35$ mm and length $2c = $ 70 mm is present (Fig. \ref{Fig: 3D crack vs no crack}b). In the former case, failure takes place due to the initiation and propagation of damage driven by the hoop stress concentration around the vent pipe. Bending forces due to the inner pressure in the vent pipe also produce damage in the internal surface and failure occurs at a pressure of 92.9 MPa. In the case of the cracked sample, Fig. \ref{Fig: 3D crack vs no crack}b, failure is governed by the growth of this pre-existing flaw, and occurs at a lower pressure (78.1 MPa). 

\begin{figure}[H]
  \makebox[\textwidth][c]{\includegraphics[width=1\textwidth]{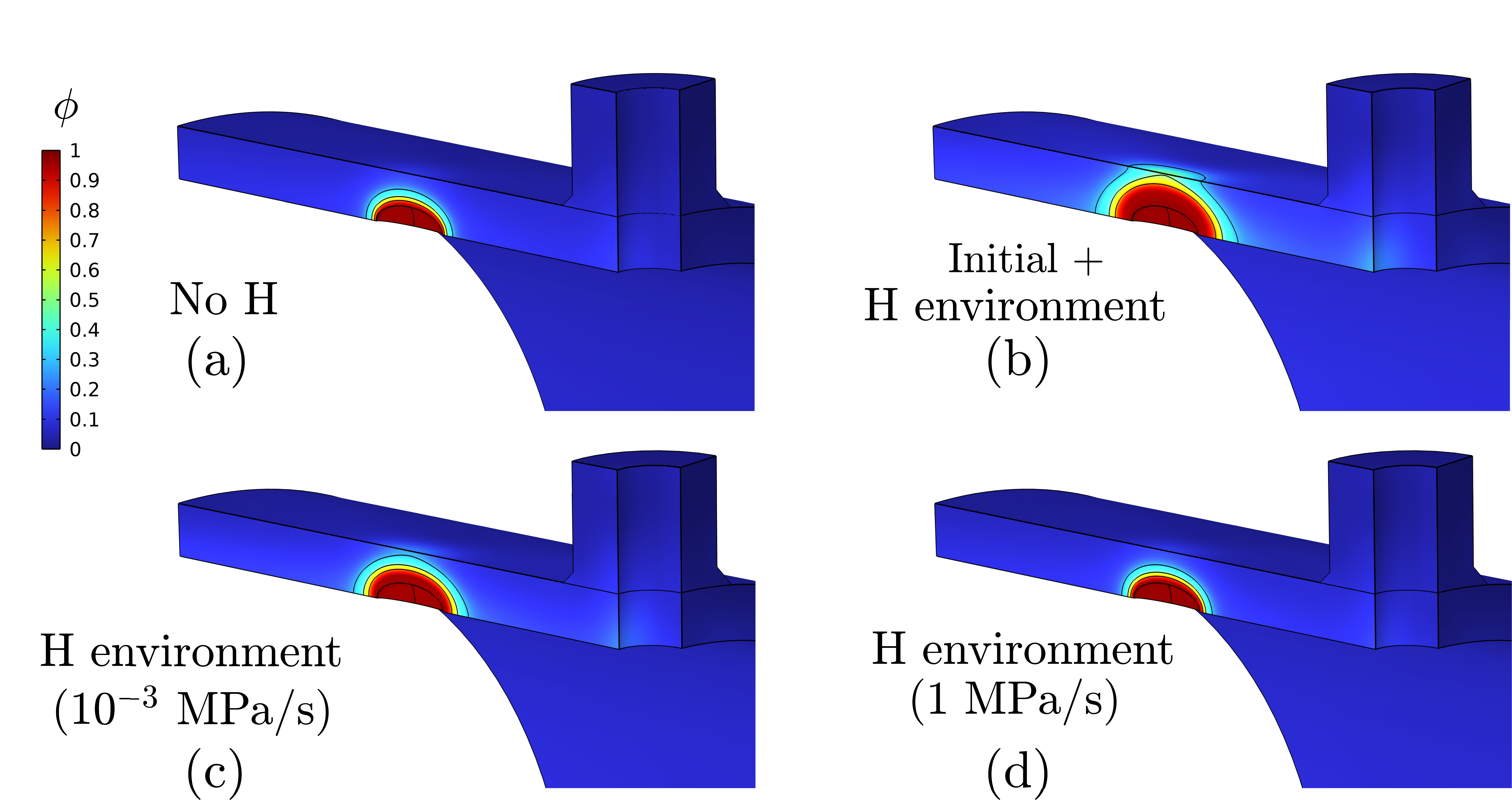}}%
  \caption{3D analysis of component failure, damage propagation ($\phi$) at $p=64.5$ MPa for a pipe with a longitudinal elliptic crack: (a) without hydrogen; (b) with hydrogen uptake from the inner wall and precharged; (c) only with external hydrogen at $10^{-3}$ MPa/s; (c) only with external hydrogen at 1 MPa/s. Black lines for $\phi =$ 0.25, 0.5 and 0.75 are also included.}
\label{Fig: 3D embrittlement}   
\end{figure}

When hydrogen-informed phase field is considered in the 3D model, embrittlement is captured. The effects of $C_L^0=$ 0.1 wt ppm and a degradation coefficient $\chi=0.89$ are assessed for a pressure of $p=64.5$ MPa and four different conditions: (i) no hydrogen, given in Fig. \ref{Fig: 3D embrittlement}a, (ii) exposure to both initial ($C_L^0=$ 0.1 wt ppm) and internal hydrogen ($C_{env}=$ 0.1 wt ppm), with a pressure ramp of $10^{-3}$ MPa/s, shown in Fig. \ref{Fig: 3D embrittlement}b, (iii) internal source of hydrogen ($C_{env}=$ 0.1 wt ppm) with the pressure loading rate being slow ($10^{-3}$ MPa/s), Fig. \ref{Fig: 3D embrittlement}c, and (iv) internal source of hydrogen ($C_{env}=$ 0.1 wt ppm) with the pressure loading rate being fast (1 MPa/s), see Fig. \ref{Fig: 3D embrittlement}d. The concentration at internal surfaces, $C_{env}$, is assumed to be constant and independent of pressure. This simplification has been followed for the sake of simplicity and to compare all cases without hydrogen egress during loading. The damage contours show that 64.5 MPa of internal pressure are sufficient to propagate a crack through the thickness when $C_L^0$ is fixed at the internal surfaces and also as an initial condition (Initial + H Environment).  However, when precharging is not considered, i.e. external hydrogen is only absorbed from the inner surface (Fig. \ref{Fig: 3D embrittlement}c), only some damage propagation is observed. At a very fast pressure rate, 1 MPa/s, hydrogen-induced damage is only caused by the reduction of fracture energy at the crack front, but the loading time is short and hydrogen accumulation is prevented resulting in a delayed crack propagation (Fig. \ref{Fig: 3D embrittlement}d). Therefore, it is demonstrated that the model captures different failure modes, including damage initiation at stress concentrations or the propagation of cracks, and also the transient effects during hydrogen embrittlement. 

For the slower pressure rate, $10^{-3}$ MPa/s, the normalised hydrogen concentration is plotted in Fig. \ref{Fig: 3D concentration}. A peak with $C_L>C_L^0$ is not fully developed due to the low triaxiality and the corresponding low $\sigma_h$ values. However, it can be observed that the damaged stress and the moving boundary conditions reproduce the advance of hydrogen with the crack front.

\begin{figure}[H]
  \makebox[\textwidth][c]{\includegraphics[width=1\textwidth]{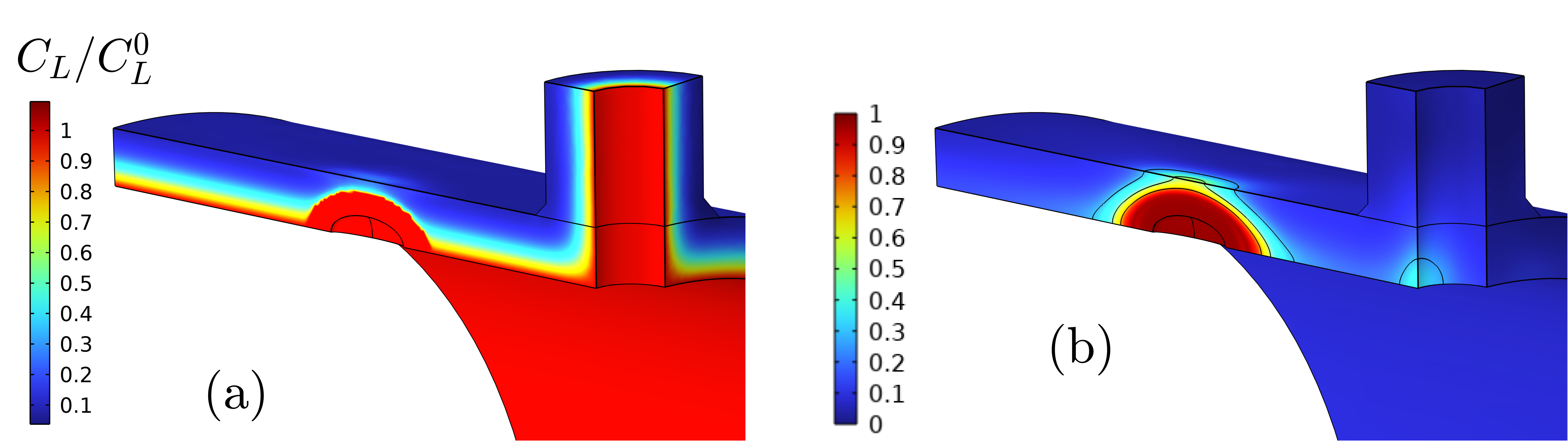}}%
  \caption{3D analysis of component failure: (a) normalized lattice hydrogen concentration, and (b) damage propagation ($\phi$) for a component containing a longitudinal elliptic crack, at a pressure of $p=66.8$ MPa, and with a pressure loading rate of $10^{-3}$ MPa/s.}
\label{Fig: 3D concentration}   
\end{figure}

\section{Conclusions}
\label{Sec:Concluding remarks}

We have presented a generalised, hydrogen-informed elastic-plastic phase field fracture model, which can reproduce both brittle and ductile failures and resolve transient effects related to hydrogen diffusion and trapping. Emphasis has been placed on its implementation into COMSOL Multiphysics, so as to establish a robust tool to predict fracture of metals in the presence of hydrogen. The codes developed can be freely downloaded from \url{https://mechmat.web.ox.ac.uk/codes}. The implementation of the phase field evolution equation through a Helmholtz PDE enables user-defined initial conditions on the phase field variable $\phi$. This approach is validated against literature results and the in-built implementation (see \ref{Appendix: comparison with built-in}). Numerical stability and discretisation aspects are extensively discussed; multi-pass solution schemes increase accuracy and the use of Anderson's acceleration was found to increase efficiency. Hydrogen diffusion and transport have been successfully captured through the implementation of a stress-driven two-level transport model: hydrostatic stress effects are modelled using a convective term whereas trapping is reproduced through a reaction term. 
Three case studies have been addressed, spanning the paradigmatic benchmark of a square notched plate, a boundary layer model to estimate crack growth resistance curves (R-curves), and a 3D analysis of a pressurised vessel with a vent pipe. New insight has been gained into the interplay between plasticity, hydrogen and fracture, and into the predictive abilities of phase field-based models; key findings include:
\begin{itemize}
    \item Changing the weighting of the plastic contribution to the fracture driving force induces a change from mode I crack growth to plastic localisation-driven failures at a 45$^\circ$ angle. The choice of $\beta_p=0.1$ is grounded on physical observations of 90\% of the plastic work being dissipated into heat and delivers consistent results. 
    \item The elastic-plastic phase field model can rigorously capture the three stages involved in crack growth experiments: (i) crack blunting at $K_I < K_c$, (ii) crack initiation at $K_I=K_c$, and (iii) a rising R-curve at $K_I > K_c$, due to plastic dissipation. Likewise, when hydrogen comes into play, its role is naturally captured, predicting a lower $K_I$ at initiation and a more brittle response.
    \item The new artificial diffusivity enhancement proposed is shown to accurately capture how the hydrogen-containing environment follows the crack as it propagates. A threshold value of $\phi_{th}=0.95$ was found to be more conservative.
    \item The model captures the three regimes resulting from varying the loading rate: a slow loading regime, where embrittlement is maximised, a fast loading regime, where steady state conditions are recovered, and an intermediate loading rate-sensitive regime. 
    \item Trapping effects were found to have a small effect on crack growth resistance, unless the sample has not been pre-exposed to hydrogen. 
    \item The analysis of 3D components reveals the ability of the framework to tackle problems of engineering significance, with phenomena such as leak-before-break being a natural outcome of the model.   
\end{itemize}

\section{Acknowledgements}
\label{Acknowledge of funding}

\noindent The authors gratefully acknowledge funding from projects PID2021-124768OB-C21 and TED2021-130413B-I00. This work was also supported by the Regional Government of Castilla y León (Junta de Castilla y León) and by the Ministry of Science and Innovation MICIN and the European Union NextGenerationEU / PRTR through projects H2MetAmo (C17.I01.P01.S21) and MA2TEC (C17.I01). E. Mart\'{\i}nez-Pa\~neda acknowledges financial support from the EPSRC (grant EP/V009680/1), from UKRI’s Future Leaders Fellowship programme [grant MR/V024124/1], and from the UKRI Horizon Europe Guarantee programme (ERC Starting Grant \textit{ResistHfracture}, EP/Y037219/1).


\appendix

\section{Time stepping and discretization}
\label{Appendix: time stepping and discretization}

The time step can influence conditionally stable staggered schemes, particularly under conditions of unstable crack growth. Previous authors have shown that single-pass schemes in phase field fracture modelling are limited and require a small time step to capture brittle crack propagation \cite{mandal2021comparative}. Here, different schemes are evaluated, considering the validation case of Section \ref{Sec:elastic_plate}. The results are shown in Fig. \ref{Fig: Time stepping Plastic}. It is observed that the single-pass scheme predicts a non-instantaneous load decrease when the maximum time increment is fixed as $ \Delta\bar{u}=10^{-3}$. On the other hand, a multi-pass scheme always predicts accurately the unstable crack propagation and the corresponding sudden loss of carrying capacity. Nevertheless, if the time increment is too large, a multi-pass scheme can overshoot the peak load. Importantly, the results reveal that the single-pass scheme is also able to capture unstable crack propagation when the adaptive time increment defined in Eq. (\ref{Eq: adaptive time step}) is applied.
 
\begin{figure}[H]
  \makebox[\textwidth][c]{\includegraphics[width=0.8\textwidth]{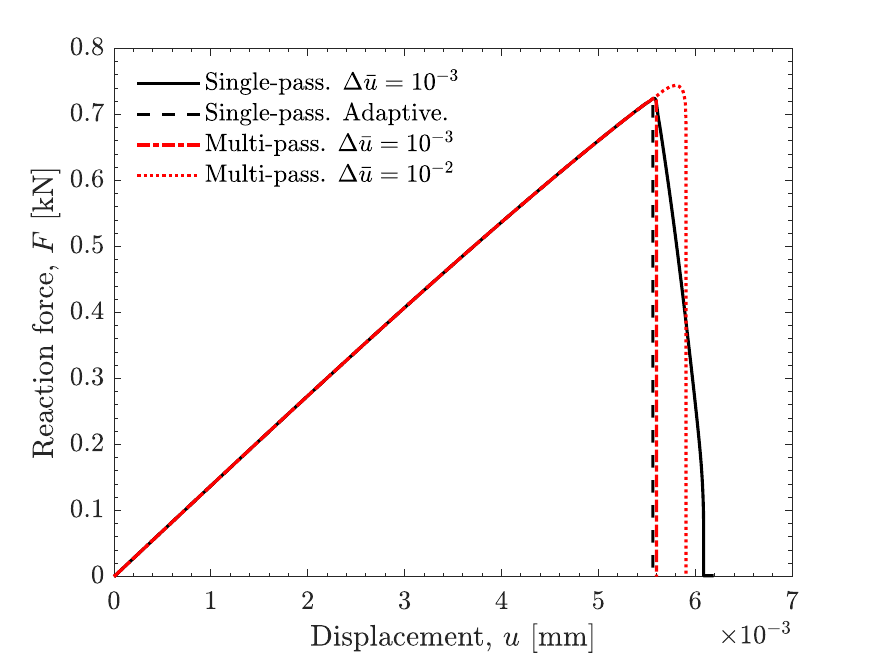}}%
  \caption{Comparison of different solving strategies through the prediction of force versus displacement curves for the notched square plate boundary value problem, considering the elastic-brittle case and in the absence of hydrogen.}
\label{Fig: Time stepping Plastic}  
\end{figure}

Despite these differences, it has been observed that both time-stepping methods yield the same results for the plastic-ductile case. This is illustrated in Fig. \ref{Fig: Time stepping Plastic} for the case of an elastic-plastic single-edge cracked plate exposed to a hydrogen content of $C_L^0=C_{env}=0.1$ wt ppm. This figure also assesses the role of implicit solvers. Two families of implicit solvers are available in COMSOL: \texttt{BDF} and \texttt{Generalised alpha}. \texttt{BDF} is more stable for the analysed coupled problem and circumvents spurious concentration distributions, due to a tighter time step reduction when damage propagation begins. The default \texttt{Generalised alpha} solver results in time steps that are generally bigger and thus simulations are faster.
 
\begin{figure}[H]
  \makebox[\textwidth][c]{\includegraphics[width=0.8\textwidth]{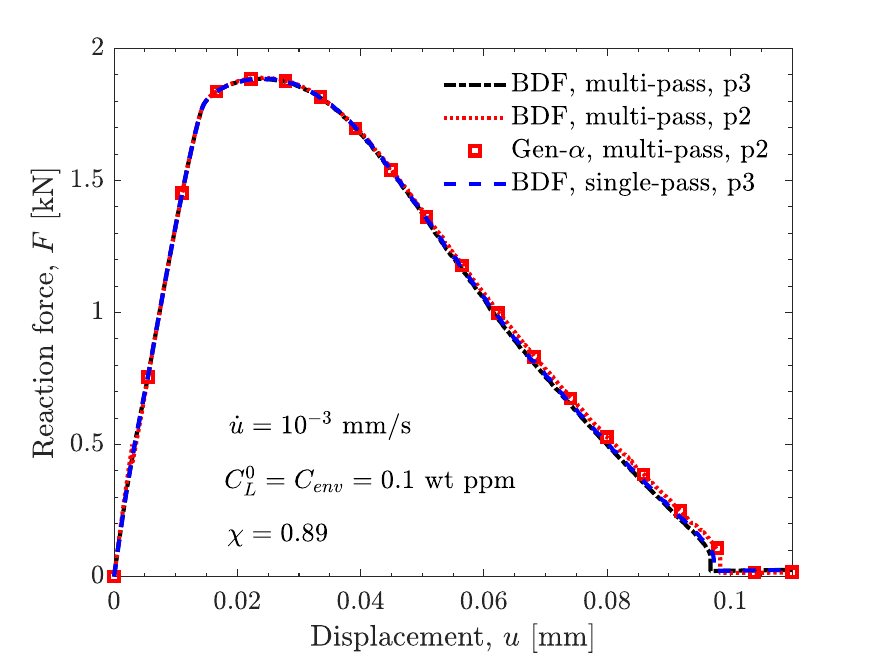}}%
  \caption{Comparison of different solving strategies through the prediction of force versus displacement curves for the notched square plate boundary value problem, considering an elastic-plastic metal exposed to a hydrogen-containing environment.}
\label{Fig: Time stepping Plastic}  
\end{figure}

The influence of discretization for stable crack propagation is low, as demonstrated by the second-order (p2) discretization of the displacement field (\textbf{u}) and phase field ($ \phi$) in comparison to the cubic order (p3) of both. Figure \ref{Fig: Time stepping Plastic} shows that both force-displacement curves are very similar. In the present examples, the deformation of the crack tip is limited and locking effects are not critical. Part I of the present work \cite{PartI} demonstrates the need for appropriate discretization and mapping in order to avoid spurious hydrostatic stress distributions in highly deformed regions, e.g. for high $G_c^0$ values and low embrittlement levels. 

\section{Comparison with built-in phase field damage}
\label{Appendix: comparison with built-in}

The comparison of results from the proposed PDE-based phase field balance and the built-in damage capabilities can only be done with the following two assumptions: (i) the crack must be geometrically introduced, for the built-in model, initial values or Dirichlet boundary conditions cannot be directly assigned to the \texttt{solid.phic} variable; and (ii) the yield expression has to be modified as the yielding criterion in the built-in plasticity uses the undamaged stress values ($|\boldsymbol{\sigma}_0|$); since $h(\phi)$ is not always equal to $g(\phi)$, the yield stress expression reformulated as follows, so the yielding criterion in the built-in approach is equivalent to Eq. (\ref{Eq: Yield criterion}):
    \begin{equation}
    \label{Eq: Yield criterion built-in}
        |\boldsymbol{\sigma}_0|-\frac{h(\phi)}{g(\phi)}\sigma_{f0}=0
    \end{equation}

\begin{figure}[H]
  \makebox[\textwidth][c]{\includegraphics[width=0.8\textwidth]{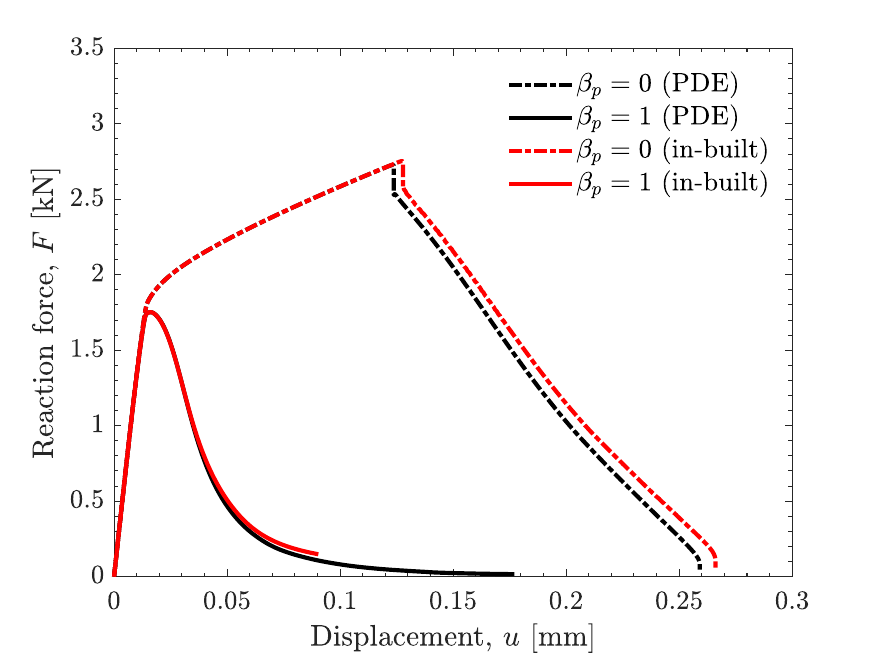}}%
  \caption{Comparison between the in-built and user-defined PDE-based implementation of elastic plastic phase field fracture in COMSOL. The results show the behaviour of a notched square plate in the absence of hydrogen for the conditions where fracture is driven purely by elastic contributions ($\beta_p=0$) and when is driven by both elastic and plastic terms, on equal weights ($\beta_p=1$).}
\label{Fig: Validation built-in}  
\end{figure}

Results are compared in Figure \ref{Fig: Validation built-in} for $\beta_p=0$ and $\beta_p=1$, i.e. with only the \texttt{Elastic strain energy density} or with the \texttt{Total strain energy density} as the \textit{Crack driving force}, in the built-in phase field model, respectively. Agreement between results is adequate and the PDE-based implementation is validated. The small differences observed are attributed to the different integration schemes adopted for the built-in $\phi$ and $\mathcal{H}$ internal variables in comparison to the PDE-based option.  

\end{document}